\documentclass[aps,rmp]{revtex4}

\newcommand{\la}{\lower.5ex\hbox{$\; \buildrel < \over \sim \;$}}
\newcommand{\ga}{\lower.5ex\hbox{$\; \buildrel > \over \sim \;$}}
\newcommand{\OM}{\Omega_{{\rm M}0}}
\newcommand{\OR}{\Omega_{{\rm R}0}}
\newcommand{\OL}{\Omega_{\Lambda 0}}
\newcommand{\OK}{\Omega_{{\rm K}0}}
\newcommand{\ODM}{\Omega_{{\rm DM}0}}
\newcommand{\OB}{\Omega_{{\rm B}0}}

\begin{document}

\preprint{KSUPT-02/3 \hspace{0.5truecm} July 2002}

\title{The Cosmological Constant and Dark Energy}

\author{P.\ J.\ E.\ Peebles}
\affiliation{Joseph Henry Laboratories, Princeton University, Princeton, 
             NJ 08544}
\author{Bharat Ratra}
\affiliation{Department of Physics, Kansas State University, Manhattan, 
             KS 66506}

\begin{abstract}
Physics invites the idea that space contains energy whose
gravitational effect approximates that of Einstein's cosmological
constant, $\Lambda$; nowadays the concept is termed dark energy
or quintessence. Physics also suggests the dark energy could be 
dynamical, allowing the arguably appealing picture that the dark
energy density is evolving to its natural value, zero, and is
small now because the expanding universe is old. This alleviates 
the classical problem of the curious energy scale of order a 
millielectronvolt associated with a constant $\Lambda$. 
Dark energy may have been detected by recent advances in the 
cosmological tests. The tests establish a good scientific case 
for the context, in the relativistic Friedmann-Lema\^\i tre model, 
including the gravitational inverse square law applied to the 
scales of cosmology. We have well-checked evidence that the mean
mass density is not much more than one quarter of the critical
Einstein-de Sitter value. The case for detection of dark
energy is serious but not yet as convincing; we await more checks
that may come out of work in progress. Planned observations might
be capable of detecting evolution of the dark energy density; 
a positive result would be a considerable stimulus to attempts to
understand the microphysics of dark energy. This review presents
the basic physics and astronomy of the subject, reviews the 
history of ideas, assesses the state of the observational
evidence, and comments on recent developments in the search for a
fundamental theory. 
\end{abstract}

\maketitle

\tableofcontents

\section{Introduction}

There is significant observational evidence for the detection of 
Einstein's cosmological constant, $\Lambda$, or a component of
the material content of the universe that varies only slowly with
time and space and so acts like $\Lambda$. We will use the term 
dark energy for $\Lambda$ or a component that acts like it. 
Detection of dark energy would be a new clue to an old puzzle, 
the gravitational
effect of the zero-point energies of particles and fields. The
total with other energies that are close to
homogeneous and nearly independent of time act as dark energy. 
The puzzle has been that the value of the dark energy density 
has to be tiny compared to what is suggested by dimensional
analysis; the startling new evidence is that it may be
different from the only other natural value, zero.  

The main question to consider now has to be whether to accept the
evidence for detection of dark energy. We outline the 
nature of the case in this section. After reviewing the basic
concepts of the relativistic world model, in Sec. II, and in
Sec.~III reviewing the history of ideas, we present
in Sec. IV a more detailed assessment of the state of the
cosmological tests and the evidence for detection of $\Lambda$ or
its analog in dark energy. 

There is little new to report on the big issue for physics ---
why the dark energy density is so small --- since Weinberg's
(1989) review in this Journal\footnote{ 
Sahni and Starobinsky (2000), Carroll (2001), Witten (2001), 
Weinberg (2001), and Ellwanger (2002) present more 
recent reviews.}. But there have been analyses of a simpler
idea: can we imagine the present dark energy density 
is evolving, perhaps approaching zero? Models
are introduced in Secs. II.C and III.E, and recent work is  
summarized in more detail in the Appendix. Feasible advances in
the cosmological tests could detect evolution of the dark energy
density, and maybe its gravitational response to the large-scale
fluctuations in the mass distribution. That would really drive 
the search for a more fundamental physics model for dark energy. 

\subsection{The issues for observational cosmology}

We have to make two points. First, cosmology has a substantial
observational and experimental basis that shows many aspects of
the standard model almost certainly are good  
approximations to reality. Second, the empirical basis is not
nearly as strong as it is for the standard model for particle
physics: in cosmology it is not yet a matter of measuring the 
parameters in a well-established theory.  

To explain the second point we must remind those more accustomed
to experiments in the laboratory than observations in astronomy
of the astronomers' tantalus principle: one can look at distant
objects but never touch them. For example, the observations of 
supernovae in distant galaxies offer evidence for the detection 
of dark energy, under the assumption that distant and nearby 
supernovae are drawn from the same statistical sample (that is, 
that they are statistically similar enough for the purpose of 
this test). There is no direct way to check this, and it is 
easy to imagine differences between distant and nearby
supernovae of the same nominal type. More distant supernovae  
are seen in younger galaxies, because of the light travel time,
and these younger galaxies tend to have more massive rapidly
evolving stars with lower heavy element abundances. How do we
know the properties of the supernovae are not also different?
We recommend Leibundgut's (2001, Sec.~4) discussion of the
astrophysical hazards. Astronomers have checks for this and
other issues of interpretation of the observations used in
the cosmological tests. But it takes nothing 
away from this careful and elegant work to note that the checks
seldom can be convincing, because the astronomy is complicated
and what can be observed is sparse. What is more, we don't know
ahead of time that the physics that is well tested on scales
ranging from the laboratory to the Solar System survives the
enormous extrapolation to cosmology.  

Our first point is that the situation is by no means hopeless, 
because we now have significant cross-checks from the
consistency of results based on independent applications of the
astronomy and of the physics of the cosmological model. If the
physics or astronomy were faulty 
we would not expect consistency from independent lines of
evidence --- apart from the occasional accident, and the
occasional tendency to stop the analysis when it approachs the
``right answer''. We have to demand abundant evidence of
consistency, and that is starting to appear.  

The case for detection of $\Lambda$ or dark energy commences with
the Friedmann-Lema\^\i tre cosmological model. In this 
model the expansion history of the universe is determined by a 
set of dimensionless parameters whose sum is normalized to unity,
\begin{equation}
   \OM + \OR + \OL +\OK = 1.
\label{eq:sum}
\end{equation}
The first, $\OM$, is a measure of the present mean mass
density in nonrelativistic matter, mainly baryons and 
nonbaryonic dark matter. The second, $\OR \sim 1 \times 10^{-4}$, 
is a measure of the present mass in the relativistic 3~K thermal 
cosmic microwave background radiation that almost homogeneously
fills space, and the accompanying low mass 
neutrinos. The third is a measure of $\Lambda$ or the
present value of the dark energy equivalent. The fourth,
$\OK$, is an effect of the curvature of space. We review some 
details of these parameters in the next section, and of their
measurements in Sec.~IV. 

The most direct evidence for detection of dark energy comes from 
observations of supernovae of a type whose intrinsic 
luminosities are close to uniform (after subtle astronomical
corrections, a few details of which are discussed in
Sec.~IV.B.4). The observed brightness as a function of the 
wavelength shift of the radiation 
probes the geometry of spacetime, in what has come to be 
called the redshift-magnitude relation.\footnote{
The apparent magnitude is $m = -2.5\log _{10}f$ plus a constant, 
where $f$ is the detected energy flux density in a chosen wavelength 
band. The standard measure of the wavelength shift, due to the 
expansion of the universe, is the redshift $z$ defined in 
Eq.~(\ref{eq:z}) below.} The measurements agree with the 
relativistic cosmological model with $\OK = 0$, meaning no
space curvature, and $\OL \sim 0.7$, meaning nonzero 
$\Lambda$. A model with $\OL =0$ is two or three standard 
deviations off the best fit, depending on the data set and
analysis technique. This is an important indication, but 2 to 3
$\sigma$ is not convincing, even when we can be sure the
systematic errors are under reasonable control. And we have
to consider that there may be a significant systematic error
from differences between distant, high redshift, and nearby,
low redshift, supernovae.   

There is a check, based on the CDM model\footnote{
The model is named after the nonbaryonic cold dark matter (CDM) 
that is assumed to dominate the masses of galaxies in the present 
universe. There are more assumptions in the CDM model, of course; 
they are discussed in Secs.~III.D and~IV.A.2.}  
for structure formation. The fit of the model to the observations 
reviewed in Sec.~IV.B yields two key constraints. First, the 
angular power spectrum of fluctuations in the temperature of the 
3~K thermal cosmic microwave background  radiation across the sky 
indicates $\OK$ is small. Second, the power spectrum of the
spatial distribution of the galaxies requires $\OM \sim 0.25$.
Similar estimates of $\OM$ follow from independent lines of
observational evidence. The rate of gravitational lensing prefers 
a somewhat larger value (if $\OK$ is small), and some dynamical 
analyses of systems of galaxies prefer lower $\OM$. But the 
differences could all be in the measurement uncertainties. Since 
$\OR$ in Eq.~(\ref{eq:sum}) is small, the conclusion is 
$\OL$ is large, in excellent agreement with what the supernovae 
say. 

Caution is in order, however, because this check depends on the
CDM model for structure formation. We can't see the dark matter,
so we naturally assign it the simplest properties we can get away
with. Maybe it is significant that the model has observational 
problems with galaxy formation, as discussed in Sec.~IV.A.2, or
maybe these problems are only apparent, from the complications of
the astronomy. We are going to have to determine which it is
before we can have a lot of confidence in the role of the CDM
model in the cosmological tests. We will get a strong hint from
the precision measurements in progress of the angular
distribution of the 3~K thermal cosmic microwave background
radiation.\footnote{
At the time of writing the MAP satellite is taking data; the project is
described in http://map.\-gsfc.nasa.gov/.} If the results 
match in all detail the prediction of the relativistic model for 
cosmology and the CDM model for structure formation, with
parameter choices that agree with the constraints from all the
other cosmological tests, it will be strong evidence that
we are approaching a good approximation to reality, and 
the completion of the great program of cosmological tests
that commenced in the 1930s. But all that is to come. 

We emphasize that the advances in the empirical basis for
cosmology already are very real and substantial. 
How firm is the conclusion depends on the issue, of course. Every 
competent cosmologist we know accepts as established beyond
reasonable doubt that the universe is expanding and cooling in a
near homogeneous and isotropic way from a hotter denser state:
how else could space, which is transparent now, have been filled
with radiation that has relaxed to a thermal spectrum? The debate
is about when the expansion commenced or became a meaningful
concept. Some whose opinions and research we respect
question the extrapolation of the gravitational inverse square
law, in its use in estimates of masses in galaxies and systems of
galaxies, and of $\OM$. We agree that this law is one of the
hypotheses to be tested. Our 
conclusion from the cosmological tests in Sec.~IV is that the law
passes significant though not yet complete tests, and that we
already have a strong scientific case, resting on the abundance of
cross-checks, that the matter density parameter $\OM$ is about
one quarter. The case  for detection of $\OL$ is significant too,
but not yet as compelling. 

For the most part the results of the cosmological tests agree
wonderfully well with accepted theory. But the observational
challenges to the tests are substantial: we are drawing profound 
conclusions from very limited information. We have to be liberal
in considering ideas about what the universe is like,
conservative in accepting ideas into the established canon.    

\subsection{The opportunity for physics}

Unless there is some serious and quite unexpected flaw in our
understanding of the principles of physics we can be sure the
zero-point energy of the electromagnetic field at laboratory
wavelengths is real and measurable, as in the Casimir (1948) 
effect.\footnote{
See Bordag, Mohideen, and Mostepanenko (2001) for a recent review.
The attractive Casimir force between two parallel conducting
plates results from the boundary condition that suppresses the
number of modes of oscillation of the electromagnetic field 
between the plates, thus suppressing the energy of the system.
One can  understand the effect at small separation without
reference to the quantum behavior of the electromagnetic field,
as in the analysis of the Van der Waals interaction in
quantum mechanics, by taking account of the term in the particle 
Hamiltonian for the Coulomb potential energy between
the charged particles in the separate neutral objects. But a more
complete  treatment, as discussed by Cohen-Tannoudji, Dupont-Roc, 
and  Grynberg (1992), replaces the Coulomb interaction with the
coupling of the charged particles to the electromagnetic field
operator. In this picture the Van der Waals interaction is
mediated by the exchange of virtual photons. In either way of
looking at the Casimir effect --- the perturbation of the normal
modes or the exchange of virtual quanta of the unperturbed modes
--- the effect is the same, the suppression of the energy of the
system.}
Like all energy, this zero-point energy has to contribute to the
source term in Einstein's gravitational field equation. If, as 
seems likely, the zero-point energy of the electromagnetic field 
is close to homogeneous and independent of the velocity of the  
observer, it manifests itself as a positive contribution to Einstein's 
$\Lambda$, or dark energy. The zero point energies of the
fermions make a negative contribution. Other contributions,
perhaps including the energy densities of fields that interact
only with themselves and gravity, might have either sign. The
value of the sum suggested by dimensional analysis is much larger
than what is 
allowed by the relativistic cosmological model. The only other 
natural value is $\Lambda =0$. If $\Lambda$ really is tiny but
not zero, it adds a most stimulating though enigmatic clue
to physics to be discovered. 

To illustrate the problem we outline an example of a contribution 
to $\Lambda$. The energy density in the 3~K thermal cosmic microwave 
background radiation, 
which amounts to $\OR \sim 5 \times 10^{-5}$ in Eq.~(\ref{eq:sum})
(ignoring the neutrinos) peaks at wavelength $\lambda \sim 2$~mm. 
At this Wien peak the photon occupation number is of order a fifteenth. 
The zero-point energy  amounts to half the energy of a photon at
the given frequency. This means the zero-point energy in the
electromagnetic field at wavelengths $\lambda \sim 2$~mm amounts
to a contribution $\delta\OL \sim 4 \times 10^{-4}$ to the density
parameter in $\Lambda$ or dark energy. The sum over modes scales
as $\lambda ^{-4}$ (as illustrated in Eq.~[\ref{eq:zeropoint}]).
Thus a naive extrapolation to visible wavelengths says the
contribution amounts to $\delta\OL \sim 5 \times 10^{10}$,   
already a ridiculous number.

The situation can be compared to the development of the theory of
the weak interactions. The Fermi point-like interaction model is
strikingly successful for a considerable range of energies, but
it was clear from the start that the model fails at high energy.
A fix was discussed --- mediate the interaction by an intermediate 
boson --- and eventually incorporated into the even more successful 
electroweak theory. General relativity and quantum mechanics are 
strikingly successful on a considerable range of length scales, 
provided we agree not to use the rules of quantum mechanics to 
count the zero-point energy density in the vacuum, even though 
we know we have to count the zero-point energies in all other
situations. 
There are thoughts on how to improve the situation, though they 
seem less focused than was the case for the Fermi model. Maybe a 
new energy component spontaneously cancels the vacuum energy density;
maybe the new component varies slowly with position and here and
there happens to cancel the vacuum energy density well enough to
allow observers like us to flourish. Whatever the nature of the
more perfect theory, it must reproduce the successes of general
relativity and quantum mechanics. That includes the method of 
representing the material content of the observable universe --- 
all forms of mass and energy --- by the stress-energy tensor, and 
the relation between the stress-energy tensor and the curvature of 
macroscopic spacetime. One part has to be adjusted. 

The numerical values of the parameters in Eq.~(\ref{eq:sum}) also
are enigmatic, and possibly trying to tell us something. The
evidence is that the parameters have the approximate values   
\begin{equation}
   \OL \sim 0.7, \qquad \ODM \sim 0.2, \qquad \OB \sim 0.05.
\label{eq:concordance}
\end{equation}
We have written $\OM$ in two parts: $\OB$ measures the density of 
the baryons we know exist and $\ODM$ that of the hypothetical
nonbaryonic cold dark matter we need to fit the cosmological tests.
The parameters $\OB$ and $\ODM$
have  similar values but represent different things --- baryonic
and nonbaryonic matter --- and $\OL$, which is thought to
represent something completely different, is not much larger.
Also, if the parameters were measured when the universe was one
tenth its present size the time-independent $\Lambda$ parameter
would contribute  $\Omega _\Lambda\sim 0.003$. That is, we seem
to have come on the scene just as $\Lambda$ has become an
important factor in the expansion rate. These curiosities surely
are in part accidental, but maybe in part physically 
significant. In particular, one might imagine that the dark energy
density represented by $\Lambda$ is rolling to its natural value, 
zero, and  is very small now because we measure it when the universe
is very old. We will discuss efforts along this line to at least 
partially rationalize the situation.

\subsection{Some explanations}

We have to explain our choice of nomenclature. Basic concepts of
physics say space contains homogeneous zero-point energy, and
maybe also energy that
is homogeneous or nearly so in other forms, real or effective (as
from counter terms in the gravity physics, which make the net
energy density cosmologically acceptable). In the literature this
near homogeneous energy has been termed the vacuum energy, the
sum of vacuum energy and quintessence (Caldwell, Dav\'e, and 
Steinhardt, 1998), and the dark energy (Turner, 1999). We have 
adopted the last term, and we will refer to the dark energy density 
$\rho _\Lambda$ that manifests itself as an effective version of
Einstein's cosmological constant, but one that may vary slowly 
with time and position.\footnote{
The dark energy should of course be distinguished from a
hypothetical gas of particles with velocity dispersion large
enough that the distribution is close to homogeneous.} 

Our subject involves two quite different traditions, in physics
and astronomy. Each has familiar notation, and familiar ideas
that may be ``in the air'' but not in the recent literature. 
Our attempt to take account of these traditions commences with
the summary in Sec. II of the basic notation with brief
explanations. We expect readers will find some of these concepts
trivial and others of some use, and that the useful parts will  
be different for different readers. 

We offer in Sec. III our reading of the history of ideas on 
$\Lambda$ and its generalization to dark energy. This is a  
fascinating and we think edifying illustration of how science may
advance in unexpected directions. It is relevant to an
understanding of the present state of research in cosmology, 
because traditions
inform opinions, and people have had mixed feelings about
$\Lambda$ ever since Einstein (1917) introduced it 85 years ago.
The concept never entirely dropped out of sight in cosmology
because a series of  observations hinted at its presence, and
because to some cosmologists $\Lambda$ fits the formalism too
well to be ignored. The search for the physics of the vacuum, and
its possible relation to $\Lambda$, has a long history too.
Despite the common and strong suspicion that $\Lambda$ must be
negligibly small, because any other acceptable value is absurd,
all this history has made the community well prepared for the
recent observational developments that argue for the detection of
$\Lambda$.    

Our approach in Sec. IV to the discussion of the evidence for
detection of $\Lambda$, from the cosmological tests, also requires
explanation. One occasionally reads that the tests will show us
how the world ends. That certainly seems interesting, but it is
not the main point: why should we trust an extrapolation into the
indefinite future of a theory we can at best show is a good
approximation to reality?\footnote{
Observations may now have detected $\Lambda$, at a characteristic 
energy scale of a millielectronvolt (Eq.~[\ref{eq:lambdaenergy}]). 
We have no guarantee that there is not an even lower energy scale;
such a scale could first become apparent through the cosmological 
tests.} As we remarked in Sec.~I.A, the purpose of the tests is
to check the approximation to reality, by checking the physics
and astronomy of the standard 
relativistic cosmological model, along with any viable alternatives 
that may be discovered. We take our task to be to identify the 
aspects of the standard theory that enter the interpretation of the
measurements and thus are or may be empirically checked or
measured. 

\section{Basic concepts}

\subsection{The Friedmann-Lema\^\i tre model}

The standard world model is close to homogeneous and isotropic  
on large scales, and lumpy on small scales --- the effect
of the mass concentrations in galaxies, stars, people, and all
that. The length scale at the transition from nearly smooth to
strongly clumpy is about 10~Mpc. We use here and throughout the 
standard astronomers' length unit,
\begin{equation}
   1\ \hbox{Mpc} =3.1 \times 10^{24}\ \hbox{cm} = 3.3 \times 10^6\ 
   \hbox{light years}.
   \label{eq:Mpc}
\end{equation}
To be more definite, imagine many spheres of radius 10~Mpc
are placed at random, and the mass within each is measured. At this
radius the rms fluctuation in the set of values of masses is
about equal to the mean value. On smaller scales the departures
from homogeneity are progressively more nonlinear; on larger scales the
density fluctuations are perturbations to the homogeneous model.
From now on we mention these perturbations only when relevant for
the cosmological tests.  

The expansion of the universe means the distance $l(t)$ between
two well-separated galaxies varies with world time, $t$, as 
\begin{equation}
   l(t)\propto a(t),
   \label{eq:scaling}
\end{equation}
where the expansion or scale factor, $a(t)$, is independent of the 
choice of galaxies. It is an interesting exercise, for those who have not
already thought about it, to check that Eq.~(\ref{eq:scaling}) is
required to preserve homogeneity and isotropy.\footnote{
We feel we have to comment on a few details about
Eq.~(\ref{eq:scaling}) to avoid 
contributing to debates that are more intense than 
seem warranted. Think of the world time $t$ as the 
proper time kept by each of a dense set of observers, each moving
so all the others are isotropically moving away, and with the
times synchronized to a common energy density, $\rho (t)$, in the
near homogeneous expanding universe. The distance
$l(t)$ is the sum of the proper distances between neighboring
observers, all measured at time $t$, and along the shortest distance 
between the two observers. The rate of increase of the distance, $dl/dt$, 
may exceed the velocity of light. This is no more problematic in
relativity theory than is the large speed at which the beam of a
flashlight on Earth may swing across the face of the Moon
(assuming an adequately tight beam). Space sections at fixed $t$
may be non-compact, and the total mass of a homogeneous universe 
formally infinite. As far as is known this is not meaningful: we can only
assert that the universe is close to homogeneous and isotropic
over observable scales, and that what can be observed is a finite 
number of baryons and photons.}

The rate of change of the distance in Eq.~(\ref{eq:scaling}) is the 
speed
\begin{equation}
   v = dl/dt = Hl, \qquad H = \dot a/a,
   \label{eq:Hlaw}
\end{equation}
where the dot means the derivative with respect to world time $t$ 
and $H$ is the time-dependent Hubble parameter. When $v$ is small 
compared to the speed of light this is Hubble's law. 
The present value of $H$ is
Hubble's constant, $H_0$. When needed we will use\footnote{
The numerical values in Eq.~(\ref{eq:Ho}) are determined from an
analysis of   
almost all published measurements of $H_0$ prior to mid 1999
(Gott {\em et al.}, 2001). They are a very reasonable summary of 
the current situation. For instance, the Hubble Space Telescope 
Key Project summary measurement value 
$H_0 =  72 \pm 8\ \hbox{km s}^{-1}\hbox{Mpc}^{-1}$ (1 $\sigma$ 
uncertainty, Freedman {\em et al.}, 2001) is in very good agreement
with Eq.~(\ref{eq:Ho}), as is the recent Tammann {\em et al.} (2001)
summary value $H_0 =  60 \pm 6\ \hbox{km s}^{-1}\hbox{Mpc}^{-1}$ 
(approximate 1 $\sigma$ systematic uncertainty). This is an example of 
the striking change in the observational situation over the previous 
5 years: the uncertainty in $H_0$ has decreased by more than a factor of
3, making it one of the better measured cosmological parameters.}
\begin{equation}
   H_0 = 100 h\ \hbox{km s}^{-1}\hbox{Mpc}^{-1} = 67 \pm 7\ 
   \hbox{km s}^{-1}\hbox{Mpc}^{-1} = (15 \pm 2\ \hbox{Gyr})^{-1},
   \label{eq:Ho}
\end{equation}
at two standard deviations. The first equation defines the
dimensionless parameter $h$. 

Another measure of the expansion follows by considering the
stretching of the wavelength of light received from a
distant galaxy. The observed wavelength,
$\lambda _{\rm obs}$, of a feature in the spectrum
that had wavelength $\lambda _{\rm em}$ at emission satisfies
\begin{equation}
   1+z=\lambda _{\rm obs}/\lambda _{\rm em} = 
   a(t_{\rm obs})/a(t_{\rm em}),
   \label{eq:z}
\end{equation} 
where the expansion factor $a$ is defined in Eq.~(\ref{eq:scaling})
and $z$ is the redshift. That is, the wavelength of freely
traveling radiation stretches 
in proportion to the factor by which the universe 
expands. To understand this, imagine a large part of the universe
is enclosed in a cavity with perfectly reflecting walls. The
cavity expands with the general expansion, the widths
proportional to $a(t)$. Electromagnetic radiation is a sum of
the normal modes that fit the cavity. At interesting wavelengths 
the mode frequencies are much larger than the rate of expansion
of the universe, so adiabaticity says a photon
in a mode stays there, and its wavelength thus must vary as 
$\lambda\propto a(t)$, as stated in Eq.~(\ref{eq:z}). The cavity
disturbs the long wavelength part of the radiation, but the
disturbance can be made exceedingly small by
choosing a large cavity.

Equation (\ref{eq:z}) defines the redshift $z$. The redshift is a
convenient label for epochs in the early universe, where $z$
exceeds unity. A good exercise for the student is to check 
that when $z$ is small Eq.~(\ref{eq:z}) reduces to Hubble's law,
where $\lambda z$ is the first-order Doppler shift in the
wavelength $\lambda$, and Hubble's parameter $H$ is given by
Eq.~(\ref{eq:Hlaw}). Thus Hubble's law may be written as
$cz = Hl$ (where we have put in the speed of light).  

These results follow from the symmetry of the cosmological model
and conventional local physics; we do not need general
relativity theory. When $z\ga 1$ we need the relativistic theory
to compute the relations among the redshift and other
observables. An example is the relation between redshift and
apparent 
magnitude used in the supernova test. Other cosmological tests 
check consistency among these relations, and this checks the
world model.

In general relativity the second time derivative of the expansion
factor satisfies
\begin{equation}
   {\ddot a\over a} = -{4\over 3}\pi G (\rho + 3p).
   \label{eq:f1}
\end{equation}
The gravitational constant is $G$. Here and throughout we choose units 
to set the velocity of light to unity. The mean mass density, $\rho (t)$,
and the pressure, $p(t)$, counting all contributions including
dark energy, satisfy the local energy conservation law,  
\begin{equation}
   \dot\rho = -3{\dot a\over a}(\rho + p).
   \label{eq:energy}
\end{equation}
The first term on the right-hand side represents the decrease of mass 
density due to the expansion that more broadly disperses the matter. 
The $pdV$ work in the second term is a familiar local concept, and
meaningful in general relativity. But one should note that energy
does not have a general global meaning in this theory. 

The first integral of Eqs.~(\ref{eq:f1}) and~(\ref{eq:energy}) is 
the Friedmann equation 
\begin{equation}
   \dot a^2 = {8\over 3}\pi G\rho a^2 + {\rm constant}.
   \label{eq:f}
\end{equation}
It is conventional to rewrite this as 
\begin{equation}
   \left({\dot a\over a}\right)^2 = H_0^2 E(z)^2 
   = H_0^2\left(\OM (1+z)^3 + \OR (1+z)^4 + 
   \OL +  \OK (1+z)^2\right) .
   \label{eq:f2}
\end{equation}
The first equation defines the function $E(z)$ that is introduced
for later use. The second equation assumes constant $\Lambda$; 
the time-dependent dark energy case is reviewed in Secs.~II.C and
III.E. The first term in the last part of Eq.~(\ref{eq:f2}) 
represents non-relativistic matter with negligibly small 
pressure; one sees from Eqs.~(\ref{eq:z}) and~(\ref{eq:energy}) that the
mass density in this 
form varies with the expansion of the universe as $\rho_{\rm M}\propto 
a^{-3}\propto (1+z)^3$. The second term represents radiation and
relativistic matter,  
with pressure $p_{\rm R} = \rho_{\rm R}/3$, whence $\rho_{\rm R} \propto 
(1+z)^4$. The third term is the effect of Einstein's cosmological constant, or 
a constant dark energy density. The last term, discussed in more detail 
below, is the constant of integration in Eq.~(\ref{eq:f}). The four 
density parameters 
$\Omega_{i0}$ are the fractional contributions to the square of
Hubble's constant, $H_0^2$, that is, 
$\Omega_{i0}(t) = 8 \pi G  \rho_{i0}/(3 H_0^2)$. At the present
epoch, $z=0$, the present value of $\dot a/a$ is $H_0$, 
and the $\Omega_{i0}$ sum to unity (Eq.~[\ref{eq:sum}]).  

In this notation, Eq.~(\ref{eq:f1}) is 
\begin{equation}
   {\ddot a\over a} = -H_0^2\left( \OM (1+z)^3/2 + \OR (1+z)^4 - \OL \right) .
\label{eq:f2p}
\end{equation}

The constant of integration in Eqs.~(\ref{eq:f})
and~(\ref{eq:f2}) is related to the geometry of 
spatial sections at constant world time. Recall that in general relativity 
events in spacetime are labeled by the four coordinates $x^\mu$ of time and
space. Neighboring events 1 and 2 at separation $dx^\mu$ have
invariant separation $ds$ defined by the line element 
\begin{equation}
   ds^2 = g_{\mu\nu}dx^\mu dx^\nu. 
\label{eq:lineelement}
\end{equation}
The repeated indices are summed, and the metric tensor $g_{\mu\nu}$
is a function of position in spacetime. If $ds^2$ is positive
then $ds$ 
is the proper (physical) time measured by an observer who moves
from event 1 to 2; if negative, $|ds|$ is the proper distance
between events 1 and 2 measured by an observer who is moving so
the events are seen to be simultaneous. 

In the flat spacetime of special relativity one can choose
coordinates so the metric tensor has the Minkowskian form
\begin{equation}
   \eta _{\mu\nu} = \left(\matrix{1 & 0 & 0 & 0 \cr 
                                  0 & -1 & 0 & 0 \cr
                                  0 & 0 & -1 & 0 \cr 
                                  0 & 0 & 0 & -1 \cr }\right) .
\label{eq:minkowski}
\end{equation}
A freely falling, inertial, observer can choose locally Minkowskian
coordinates, such that along the path of the observer 
$g_{\mu\nu}=\eta _{\mu\nu}$ and the first derivatives of $g_{\mu\nu}$ vanish.   

In the homogeneous world model we can choose coordinates so the
metric tensor is of the form that results in the line element 
\begin{eqnarray}
   ds^2 &=& dt^2 - a(t)^2\left( {dr^2\over 1 + Kr^2} + r^2(d\theta ^2 
   + \sin ^2\theta d\phi ^2)\right) \nonumber \\
  &=& dt^2 - K^{-1}a(t)^2\left( d\chi ^2 + \sinh ^2\chi (d\theta ^2 
   + \sin ^2\theta d\phi ^2)\right) .
  \label{eq:rw}
\end{eqnarray}
In the second expression, which assumes $K>0$, the radial
coordinate is $r=K^{-1/2}\sinh\chi$. The expansion factor 
$a(t)$ appears in Eq.~(\ref{eq:scaling}). If $a$ were constant  
and the constant $K$ vanished this would represent the flat spacetime 
of special relativity in polar coordinates. The key point for now is 
that $\OK$ in Eq.~(\ref{eq:f2}), which represents the constant of
integration in Eq.~(\ref{eq:f}), is related to the constant $K$:
\begin{equation}
   \OK = K/(H_0 a_0)^2,
\label{eq:ok}
\end{equation}
where $a_0$ is the present value of the expansion factor
$a(t)$. Cosmological tests that are sensitive to the geometry of
space constrain the value of the parameter $\OK$,
and $\OK$ and the other density parameters $\Omega_{i0}$ in
Eq.~(\ref{eq:f2}) determine the expansion history of the
universe.     

It is useful for what follows to recall that the metric tensor 
in Eq.~(\ref{eq:rw}) satisfies Einstein's field equation, a
differential equation we can write as 
\begin{equation}
   G_{\mu\nu} = 8\pi G T_{\mu\nu}.
\label{eq:einsteineq}
\end{equation}
The left side is a function of $g_{\mu\nu}$ and its first two
derivatives and represents the geometry of spacetime. The 
stress-energy tensor $T_{\mu\nu}$ represents the material
contents of the universe, including particles, radiation, fields,
and zero-point energies. An observer in a homogeneous and
isotropic universe, moving so the universe is observed to be
isotropic, would measure the stress-energy tensor to be 
\begin{equation}
   T_{\mu\nu} = \left(\matrix{\rho & 0 & 0 & 0 \cr 
                              0 & p & 0 & 0 \cr
                              0 & 0 & p & 0 \cr 
                              0 & 0 & 0 & p\cr }\right) .
\label{eq:idealfluid}
\end{equation}
This diagonal form is a consequence of the symmetry; the diagonal
components define the pressure and energy density. With
Eq.~(\ref{eq:idealfluid}), the differential  
equation~(\ref{eq:einsteineq}) yields the expansion rate 
equations~(\ref{eq:f2}) and~(\ref{eq:f2p}).

\subsection{The cosmological constant}

Special relativity is very successful in laboratory physics. Thus
one might guess any inertial observer would see the same
vacuum. A freely moving inertial observer represents spacetime in
the neighborhood by locally Minkowskian coordinates, with the 
metric tensor $\eta _{\mu\nu}$  given in Eq.
(\ref{eq:minkowski}). A Lorentz transformation 
to an inertial observer with another velocity does not change
this Minkowski form. The same must be true of the stress-energy
tensor of the vacuum, if all observers see the same vacuum, 
so it has to be of the form
\begin{equation} 
   T^\Lambda _{\mu\nu} =\rho _\Lambda g_{\mu\nu}, 
   \label{eq:TLambda}
\end{equation}
where $\rho _\Lambda$ is a constant, in a general coordinate
labeling. On writing this 
contribution to the stress-energy 
tensor separately from all the rest, we bring the field
equation~(\ref{eq:einsteineq}) to 
\begin{equation}
   G_{\mu\nu} = 8\pi G(T_{\mu\nu} + \rho _\Lambda g_{\mu\nu}).
   \label{eq:einstein2}
\end{equation}
This is Einstein's (1917) revision of the field equation of
general relativity, where
$\rho _\Lambda$ is proportional to his cosmological constant
$\Lambda$; his reason for writing down this equation is discussed
in Sec.~III.A. In many dark energy scenarios $\rho _\Lambda$
is a slowly varying function of time and its stress-energy
tensor differs slightly from Eq.~(\ref{eq:TLambda}), so the
observed properties of the vacuum do depend on the
observer's velocity.   

One sees from Eqs.~(\ref{eq:minkowski}), (\ref{eq:idealfluid}), 
and~(\ref{eq:TLambda}) that the new component in the stress-energy 
tensor looks like an ideal fluid with negative pressure  
\begin{equation}
   p_\Lambda = -\rho _\Lambda .
   \label{eq:plambda}
\end{equation}
This fluid picture is of limited use, but the following 
properties are worth noting.\footnote{These arguments have been
familiar, in some circles, for a long time, though in our
experience discussed more often in private than the
literature. Early statements of elements are in  
Lema\^{\i}tre (1934) and McCrea (1951); see Kragh (1999, pp. 397-8)
for a brief historical account.} 

The stress-energy tensor of an ideal fluid with four-velocity
$u^\mu$ generalizes from Eq.~(\ref{eq:idealfluid}) to 
$T^{\mu\nu} = (\rho + p)u^\mu u^\nu - p g^{\mu\nu}$. The
equations of fluid dynamics follow from the vanishing of the 
divergence of $T^{\mu\nu}$.  Let us consider the simple case of 
locally Minkowskian coordinates, meaning free fall, and a fluid 
that is close to homogeneous. By the latter we mean the fluid
velocity ${\vec v}$ --- the space part 
of the four-velocity $u^\mu$ --- and the density fluctuation
$\delta\rho$ from homogeneity may be treated in linear
perturbation theory. Then the equations of energy and momentum
conservation are 
\begin{equation}
   \delta\dot\rho + (\langle\rho\rangle + \langle p\rangle ) 
      \nabla \cdot {\vec v} = 0 ,\qquad
   (\langle\rho\rangle + \langle p\rangle ) \dot {\vec v} 
      + c_s^2 \nabla \delta\rho = 0 ,
\label{eq:fluid}
\end{equation}
where $c_s^2=dp/d\rho$ and the mean density and pressure are
$\langle\rho\rangle$ 
and $\langle p\rangle$. These combine to
\begin{equation}
   \delta\ddot\rho = c_s^2\nabla ^2\delta\rho .
\label{eq:waveeq}
\end{equation}
If $c_s^2$ is positive this is a wave equation, and $c_s$ is the
speed of sound. 

The first of Eqs.~(\ref{eq:fluid}) is the local energy
conservation law, as in Eq.~(\ref{eq:energy}). If $p=-\rho$,
the $pdV$ work cancels the $\rho dV$ part: the work done to
increase  the volume cancels the effect of the 
increased volume. This has to be so for a Lorentz-invariant 
stress-energy tensor, of course, where all
inertial observers see the same vacuum. Another way to see this
is to note that the energy flux density in Eqs.~(\ref{eq:fluid})
is $(\langle \rho \rangle + \langle p \rangle ){\vec v}$. This vanishes 
when $p=-\rho$:
the streaming velocity loses meaning. When $c_s^2$ is negative
Eq.~(\ref{eq:waveeq}) says the fluid is unstable, in general.
But when $p=-\rho$ the vanishing divergence of $T^{\mu\nu}$ becomes
the condition seen in Eq.~(\ref{eq:fluid}) that 
$\rho =\langle\rho\rangle + \delta\rho$ is constant. 

There are two measures of gravitational interactions with a
fluid: the passive gravitational mass density determines how the
fluid streaming velocity is affected by an applied gravitational
field, and the active gravitational mass  density determines the
gravitational field produced by the fluid. When the fluid
velocity is nonrelativistic the expression for the former in
general relativity is $\rho +p$, as one sees by writing out the
covariant divergence of $T^{\mu\nu}$. This vanishes when
$p=-\rho$, consistent with the loss of meaning of the streaming
velocity. 
The latter is $\rho +3p$, as one sees from Eq.~(\ref{eq:f1}).
Thus a fluid with $p=-\rho /3$, if somehow kept homogeneous and
static, would produce no gravitational field.\footnote{
Lest we contribute to a wrong problem for the 
student we note that a fluid with $p=-\rho /3$ held in a
container would have net positive gravitational mass, from the
pressure in the container walls required for support against the
negative pressure of the contents. We have finessed the walls by 
considering a homogeneous situation. We believe Whittaker (1935)
gives the first derivation of the relativistic active
gravitational mass density. Whittaker also presents an example of
the general proposition that the active gravitational mass of an
isolated stable object is the integral of the time-time part of
the stress-energy tensor in the locally Minkowskian rest frame. 
Misner and Putman (1959) give the general demonstration.}
In the model in Eqs.~(\ref{eq:TLambda}) and~(\ref{eq:plambda})
the active gravitational mass density is  
negative when $\rho_\Lambda$ is positive. When this positive
$\rho_\Lambda$ dominates the stress-energy tensor 
$\ddot a$ is positive: the rate of expansion of the universe
increases. In the language of Eq.~(\ref{eq:einstein2}), this
cosmic repulsion is a gravitational effect of the negative active
gravitational mass density, not a new force law. 

The homogeneous active mass represented by $\Lambda$ changes the
equation of relative motion of freely moving test particles in
the nonrelativistic limit to 
\begin{equation}
   {d^2\vec r\over dt^2} = \vec g + \OL H_0^2\vec r,
   \label{eq:desscattering}
\end{equation}
where $\vec g$ is the relative gravitational acceleration
produced by the distribution of ordinary 
matter.\footnote{
This assumes the particles are close enough for
application of the ordinary operational definition of proper
relative position. The parameters in the last term follow from
Eqs.~(\ref{eq:f1}) and~(\ref{eq:plambda}).} 
For an illustration of the size of the last term
consider its effect on our motion in a nearly circular orbit
around the center of the Milky Way galaxy. The Solar System is
moving at speed $v_c=220$ km~s$^{-1}$ at radius $r=8$~kpc. The
ratio of the acceleration $g_\Lambda$ produced by $\Lambda$ to 
the total gravitational acceleration $g=v_c^2/r$ is
\begin{equation}
    g_\Lambda /g = \OL H_0 ^2r^2/v_c^2\sim 10^{-5},
\end{equation}
a small number. Since we are towards the edge of the luminous
part of our galaxy, a search for the effect of $\Lambda$ on the 
internal dynamics of galaxies like the Milky Way does not look
promising. The precision of celestial dynamics in the Solar System
is much greater, but the effect of $\Lambda$ is very much smaller;
for the orbit of the Earth, $g_\Lambda /g \sim 10^{-22}$.

One can generalize Eq.~(\ref{eq:TLambda}) to a variable $\rho_\Lambda$, 
by taking $p_\Lambda$ to be negative but different from $-\rho_\Lambda$. 
But if the dynamics were that of a fluid, with pressure a function of  
$\rho_\Lambda$, stability would require $c_s^2 = dp_\Lambda/d\rho_\Lambda >0$, 
from Eq.~(\ref{eq:waveeq}), which seems quite contrived. A viable working 
model for a dynamical $\rho _\Lambda$ is the dark energy of a scalar field 
with self-interaction potential chosen to make the variation of the
field energy acceptably slow, as discussed next.

\subsection{Inflation and dark energy}

The negative active gravitational mass density associated with a
positive cosmological constant is an early precursor
of the inflation picture of the early universe; 
inflation in turn is one precursor of the idea that $\Lambda$
might generalize into evolving dark energy. 

To begin, we review some aspects of causal relations between
events in spacetime. Neglecting space curvature, a light ray 
moves proper distance $dl = a(t)dx = dt$ in time interval $dt$,  
so the integrated coordinate displacement is 
\begin{equation}
 x = \int dt/a(t).
 \label{eq:horizon}
\end{equation}
If $\OL =0$ this integral converges in the past --- we see
distant galaxies that at the time of observation cannot have seen us
since the singular start of expansion at $a=0$. This ``particle 
horizon problem" is curious: how could distant galaxies in different 
directions in the sky know to look so similar? The inflation idea 
is that in the early universe the  
expansion history approximates that of de Sitter's (1917)
solution to Einstein's field equation for $\Lambda > 0$ and
$T_{\mu\nu}=0$ in Eq.~(\ref{eq:einstein2}). We can choose the
coordinate labels in this de Sitter spacetime so space curvature
vanishes. Then Eqs.~(\ref{eq:f2}) and~(\ref{eq:f2p}) 
say the expansion parameter is 
\begin{equation}
   a\propto e^{H_\Lambda t},
   \label{eq:des}
\end{equation}
where $H_\Lambda$ is a constant. As one sees by working the integral 
in Eq.~(\ref{eq:horizon}), here everyone can have seen everyone else 
in the past. The details need not concern us; for the following
discussion two concepts are important. First, the early universe acts 
like an approximation to de~Sitter's solution because it is
dominated by a large effective cosmological ``constant'', or
dark energy density. Second, the dark energy is modeled as 
that of a near homogeneous field, $\Phi$. 

In this scalar field model, motivated by grand unified
models of very high energy particle physics,
the action of the real scalar field, $\Phi$ 
(in units chosen so Planck's constant $\hbar$ is unity) is
\begin{equation}
   S = \int d^4x \sqrt{-g} \left[ {1 \over 2} 
   g^{\mu\nu}\partial_\mu \Phi \partial_\nu \Phi 
   - V(\Phi ) \right] .
   \label{eq:action}
\end{equation}
The potential energy density $V$ is a function of the field $\Phi$, 
and $g$ is the determinant of the metric tensor. When the field is 
spatially homogeneous (in the line element of Eq.~[\ref{eq:rw}]), and 
space curvature may be neglected, the field equation is
\begin{equation}
   \ddot\Phi + 3{\dot a\over a}\dot\Phi + {dV\over d\Phi } = 0.
   \label{eq:fieldeq}
\end{equation}
The stress-energy tensor of this homogeneous field is diagonal
(in the rest frame of an observer moving so the universe is
seen to be isotropic), with time and space parts along the diagonal
\begin{eqnarray}
   \rho_\Phi & = & {1\over 2} \dot\Phi^2 + V(\Phi), \nonumber \\
      p_\Phi & = & {1\over 2} \dot\Phi^2 - V(\Phi).
   \label{eq:rhophi}
\end{eqnarray}
If the scalar field varies slowly in time, so that 
$\dot\Phi^2 \ll V$, the 
field energy approximates the effect of Einstein's cosmological
constant, with $p_\Phi\simeq - \rho_\Phi$.  

The inflation picture assumes the near exponential expansion of
Eq.~(\ref{eq:des}) in the early universe lasts long enough that
every bit of the present observable universe has seen every other
bit, and presumably has discovered how to relax to almost exact
homogeneity. The field $\Phi$ may then start varying rapidly
enough to produce the entropy of our universe, and the field or the 
entropy may produce the baryons, leaving $\rho_\Phi$ small or zero. But
one can imagine the late time evolution of $\rho_\Phi$ is slow.
If slower than the evolution in the mass density in matter,
there comes a time when $\rho_\Phi$ again dominates, and the 
universe appears to have a cosmological constant. 

A model for this late time evolution assumes a potential of the form 
\begin{equation}
   V = \kappa /\Phi^\alpha ,
   \label{eq:powerlawV}
\end{equation}
where the constant $\kappa$ has dimensions of mass raised to the power
$\alpha + 4$. For simplicity let us suppose the 
universe after inflation but at high redshift is dominated by
matter or radiation, with mass density $\rho$, that drives power law
expansion, $a\propto t^n$. Then the power law solution to the
field equation~(\ref{eq:fieldeq}) with the potential in
Eq.~(\ref{eq:powerlawV}) is
\begin{equation}
   \Phi\propto t^{2/(2+\alpha )}, 
   \label{eq:phioft}
\end{equation}
and the ratio of the mass densities in the scalar field and
in matter or radiation is 
\begin{equation}
   \rho_\Phi /\rho \propto t^{4/(2+\alpha)}.
   \label{eq:densityratio}
\end{equation}
In the limit where the parameter $\alpha$ approaches zero, 
$\rho _\Phi$ is constant, and this model is equivalent to 
Einstein's $\Lambda$. 

When $\alpha > 0$ the field $\Phi$ in this model grows
arbitrarily large at large time, so $\rho _\Phi\rightarrow 0$,
and the universe approaches the Minkowskian spacetime of special
relativity. This is within a simple model, of course. It is easy
to imagine that in other models $\rho _\Phi$ approaches a 
constant positive value at large time, and spacetime approaches
the de Sitter solution, or $\rho _\Phi$ passes through zero and 
becomes negative, causing spacetime to collapse to a Big Crunch.  

The power law model with $\alpha >0$ has two properties that seem 
desirable. First, the solution in Eq.~(\ref{eq:phioft}) is said
to be an attractor (Ratra and 
Peebles, 1988) or tracker (Steinhardt, Wang, and Zlatev, 1999),
meaning it is the asymptotic solution for a broad range of
initial conditions at high redshift. That includes relaxation to
a near homogeneous energy distribution even when gravity  
has collected the other matter into nonrelativistic clumps.
Second, the energy density in the attractor solution decreases
less rapidly than that of matter and radiation. This allows
us to realize the scenario: after inflation but at high redshift
the field energy density $\rho_\Phi$ is small so it does not
disturb the standard model for the origin of the light elements,
but eventually $\rho_\Phi$ dominates and the universe acts as if it
had a cosmological constant, but one 
that varies slowly with position and time. We comment on details
of this model in  Sec~III.E.   

\section{Historical remarks}

These comments on what people were thinking are gleaned from the 
literature and supplemented by private discussions and our own 
recollections. More is required for a satisfactory history of the  
subject, of course, but we hope we have captured the main themes 
of the discussion and the way the themes have evolved to
the present appreciation of the situation.  

\subsection{Einstein's thoughts}

Einstein disliked the idea of an island universe in asymptotically 
flat spacetime, because a particle could leave the island and 
move arbitrarily far from all the other matter in the universe, 
yet preserve all its inertial properties, which he considered 
a violation of Mach's idea of the relativity of inertia. Einstein's 
(1917) cosmological model accordingly assumes the universe is
homogeneous and isotropic, on average, thus removing the
possibility of arbitrarily isolated particles. Einstein had no
empirical support for this assumption, yet it agrees with modern
precision tests. There is no agreement on whether this is more
than a lucky guess.  

Motivated by the observed low velocities of the then known stars, 
Einstein assumed that the large-scale structure of the universe
is static. He introduced the cosmological constant to reconcile
this picture with his general relativity theory. In the notation
of Eq.~(\ref{eq:f2p}), one sees that a positive value of $\OL$ 
can balance the positive values of $\OM$ and $\OR$ for consistency 
with $\ddot a =0$. The balance is unstable: a small perturbation 
to the mean mass density or the mass distribution causes expansion 
or contraction of the whole or parts of the universe. One sees this in
Eq.~(\ref{eq:desscattering}): the mass distribution can be chosen
so the two terms on the right hand side cancel, but the balance
can be upset by redistributing the mass.\footnote{To
help motivate the introduction of $\Lambda$,  
Einstein (1917) mentioned a modification of Newtonian gravity 
physics that might make the theory well defined when the mass
distribution is homogeneous. In Einstein's example, similar to
what was considered by Seeliger and Neumann in the mid 1890s, the
modified field equation for the gravitational potential $\varphi$ is 
$\nabla ^2\varphi -\lambda\varphi = 4\pi G\rho_{\rm M}$. This allows 
the nonsingular homogeneous static solution $\varphi = -4\pi 
G\rho_{\rm M} /\lambda$. In this example the potential for an
isolated point mass is  
the Yukawa form, $\varphi \propto e^{-\sqrt{\lambda } r}/r$. 
Trautman (1965) points out that this is not the nonrelativistic
limit of general relativity with the cosmological term. 
Rather, Eq.~(\ref{eq:desscattering}) follows 
from  $\nabla ^2\varphi = 4\pi G(\rho_{\rm M} -2\rho_\Lambda)$,
where the active gravitational mass density of the $\Lambda$ term
is $\rho _\Lambda + 3p_\Lambda = -2\rho _\Lambda$. 
Norton (1999) reviews the history of ideas of this 
Seeliger-Neumann Yukawa-type potential in gravity physics.}

Einstein did not consider the cosmological constant to be part of
the stress-energy term: his form for the field
equation (in the streamlined notation of
Eq.~[\ref{eq:einsteineq}]) is 
\begin{equation}
   G_{\mu\nu} - 8\pi G \rho_\Lambda g_{\mu\nu} = 8\pi G T_{\mu\nu}.
   \label{eq:einstein3}
\end{equation}
The left hand side contains the metric tensor and its
derivatives; a new constant of nature, $\Lambda$, appears 
in the addition to Einstein's original field equation. 
One can equally well put Einstein's new term on the right hand 
side of the equation, as in Eq.~(\ref{eq:einstein2}), and count  
$\rho _\Lambda g_{\mu\nu}$ as part of the source term in
the stress-energy tensor. 
The distinction becomes interesting when $\rho_\Lambda$ takes
part in the dynamics, and the field equation is properly written
with $\rho _\Lambda$, or its generalization, as part of the
stress-energy tensor. One would then say that the differential 
equation of gravity physics has not been changed from Einstein's
original form; instead there is a new component in the content of
the universe.

Having taken it that the universe is static, Einstein did not
write down the differential equation for $a(t)$, and so did not
see the instability. Friedmann (1922, 1924) found the evolving
homogeneous solution, but had the misfortune to do it before the
astronomy became suggestive. Slipher's measurements of the
spectra of the 
spiral nebulae --- galaxies of stars --- showed most are
shifted toward the red, and Eddington (1924, pp. 161-2)
remarked that that might be a manifestation of the second,
repulsive, term  in Eq.~(\ref{eq:desscattering}). Lema\^\i tre
(1927) introduced the relation between  
Slipher's redshifts and a homogeneous matter-filled expanding 
relativistic world model. He may have been influenced by Hubble's
work that was leading to the publication in 1929 of the 
linear redshift-distance relation (eq.~[\ref{eq:Hlaw}]): as a
graduate student at MIT he attended a lecture by Hubble.

In Lema\^\i tre's (1927) solution the expanding universe traces
asymptotically back to Einstein's static case. Lema\^\i tre then
turned to what he called the primeval atom, and we would term a
Big Bang model. This solution expands from densities so large
as to require some sort of quantum treatment, passes through  
a quasi-static approximation to Einstein's solution, and then
continues expanding to de Sitter's (1917) empty space solution. To
modern tastes this ``loitering'' model requires incredibly 
special initial conditions, as will be discussed. Lema\^\i tre
liked it because the loitering epoch allows the expansion time to
be acceptably long for Hubble's (1929) estimate of $H_0$, which is an 
order of magnitude high.   

The record shows Einstein never liked the $\Lambda$ term. His
view of how general relativity might fit Mach's principle was
disturbed by de Sitter's (1917) solution to
Eq.~(\ref{eq:einstein3}) for empty 
space ($T_{\mu\nu}=0$) with $\Lambda >0$.\footnote{
North (1965) reviews the confused early history of ideas on the
possible astronomical significance of de Sitter's solution for an
empty universe with $\Lambda >0$; we add a few comments on
the physics that contributed to the discovery of the expanding
world model. Suppose an observer in de Sitter's spacetime 
holds a string tied to a source of light, so the source 
stays at fixed physical distance $r\ll H_\Lambda ^{-1}$. The
source is much less massive than the observer, the gravitational
frequency shift due to the observer's mass may be neglected, and
the observer is moving freely. Then the observer receives light  
from the source shifted to the red by 
$\delta\lambda /\lambda = -(H_\Lambda r)^2/2$. The observed
redshifts of particles moving on geodesics depend on the
initial conditions. Stars in the outskirts of our galaxy are held
at fixed mean distances from us by their transverse motions. The
mean shifts of the spectra of light from these stars include this
quadratic de Sitter term as well as the much larger Doppler and
ordinary gravitational shifts. The prescription for initial
conditions that reproduces the linear redshift-distance relation
for distant galaxies follows Weyl's (1923) principle: the world
particle geodesics trace back to a near common position in the
remote past, in the limiting case of the Friedmann-Lema\^\i tre 
model at $\OM\rightarrow 0$. This spatially homogeneous coordinate
labeling of de Sitter's spacetime, with space sections with
negative curvature, already appears in de Sitter (1917, Eq.
[15]), and is repeated in Lanczos (1922). This line element is
the second expression in our Eq.~(\ref{eq:rw}) with
$a\propto\cosh H_\Lambda t$. Lema\^\i tre (1925) and
Robertson (1928) present the coordinate labeling for the
spatially-flat case, where the line element is 
$ds^2 = dt^2 - e^{2H_\Lambda t}(dx^2+dy^2+dz^2)$ (in the
choice of symbols and signature in Eqs.~[\ref{eq:rw}]
and~[\ref{eq:des}]). Lema\^\i tre (1925) and Robertson (1928)
note that 
particles at rest in this coordinate system  present a linear
redshift-distance relation, $v = H_\Lambda r$, at small $v$.
Robertson (1928) estimated $H_\Lambda$, and Lema\^\i tre (1927) 
its analog for the Friedmann-Lema\^\i tre model, from
published redshifts and Hubble's galaxy distances. Their
estimates are not far off Hubble's (1929) published 
value.}${^{,}} $\footnote{To the present way of thinking the
lengthy debate about the singularity in de Sitter's 
static solution, chronicled in North (1965), seems surprising, 
because de Sitter (1917) and Klein (1918) had presented de
Sitter's solution as a sphere embedded in 4 plus 1 dimensional
flat space, with no physical singularity.}
Pais (1982, p. 288) points out
that Einstein in a letter to Weyl in 1923 comments on the effect
of $\Lambda$ in Eq.~(\ref{eq:desscattering}):
``According to
De Sitter two material points that are sufficiently far apart,
continue to be accelerated and move apart. If there is no 
quasistatic world, then away with the cosmological term.'' 
We do not know whether at this time Einstein was influenced by
Slipher's redshifts or Friedmann's expanding world model. 

The earliest published comments we have found on Einstein's
opinion of $\Lambda$ within the evolving world model (Einstein,
1931; Einstein and de Sitter, 1932) make the point that, since
not all the terms in the expansion rate Eq.~(\ref{eq:f2}) are
logically required, and the  matter term surely is present and
likely dominates over radiation at low redshift, a reasonable
working model drops $\OK$ and $\OL$ and ignores $\OR$. This
simplifies the expansion rate equation to what has come to be
called the Einstein-de Sitter model, 
\begin{equation}
   {\dot a^2\over a^2} = {8\over 3}\pi G\rho_{\rm M} ,
   \label{eq:edes}
\end{equation}
where $\rho_{\rm M}$ is the mass density in non-relativistic matter;
here $\Omega_{\rm M} = 8 \pi G \rho_{\rm M}/(3H^2)$ is unity. 
The left side is a measure of the kinetic
energy of expansion per unit mass, and the right-hand side a
measure of the negative of the gravitational potential energy. In
effect, this model universe is expanding with escape velocity.

Einstein and de Sitter point out that 
Hubble's estimate of $H_0$ and de Sitter's estimate of the
mean mass density in galaxies are not inconsistent with
Eq.~(\ref{eq:edes}) (and since both quantities scale with
distance in the same way, this result is not affected by the
error in the distance scale that affected Hubble's initial 
measurement of $H_0$). But the evidence now is that the
mass density is about one quarter of what is predicted by this
equation, as we will discuss. 

Einstein and de Sitter (1932) remark that the curvature term in
Eq.~(\ref{eq:f2}) is ``essentially determinable, 
and an increase in the precision of the data derived from
observations will enable us in the future to fix its sign and
determine its value.'' This is happening, 70 years later. 
The cosmological constant term is measurable in principle, too,
and may now have been detected. But 
Einstein and de Sitter say only that the theory of an expanding
universe with finite mean mass density ``can be reached without
the introduction of" $\Lambda$. 

Further to this point, in the appendix of the second edition of 
his book, {\em The Meaning of Relativity}, Einstein (1945, p. 127)
states that the ``introduction of the `cosmologic member'{\,}" ---
Einstein's terminology for $\Lambda$ --- ``into the equations of 
gravity, though possible from the point of view of relativity, is 
to be rejected from the point of view of logical economy'', and 
that if ``Hubble's expansion had been discovered at the time of 
the creation of the general theory of relativity, the 
cosmologic member would never have been introduced. It seems now
so much less justified to introduce such a member into the field
equations, since its introduction loses its sole original
justification,---that of leading to a natural solution of the
cosmologic problem.'' 
Einstein knew that without the cosmological constant the 
expansion time derived from Hubble's estimate of $H_0$ is
uncomfortably short compared to estimates of the ages of the
stars, and opined that that might be a problem with the star
ages. The big error, the value of $H_0$, was corrected by 
1960 (Sandage, 1958, 1962).

Gamow (1970, p. 44) recalls that ``when I was discussing
cosmological problems with Einstein, he remarked that the
introduction of the cosmological term was the biggest blunder he 
ever made in his life.'' This certainly is consistent with all 
of Einstein's written comments we have seen on the cosmological 
constant {\it per se}; we do not know whether Einstein was also
referring to the missed chance to predict the evolution of the
universe.  

\subsection{The development of ideas}

\subsubsection{Early indications of $\Lambda$}

In the classic book, {\it The Classical Theory of Fields}, Landau
and Lifshitz (1951, p. 338) second Einstein's opinion of the
cosmological constant $\Lambda$, stating there is ``no basis
whatsoever'' for adjustment of the 
theory to include this term. The empirical side of cosmology is
not much mentioned in this book, however (though there is a
perceptive comment on the limited empirical support for the
homogeneity assumption; p. 332). In the Supplementary Notes to
the English translation of his book, {\em Theory of Relativity},
Pauli (1958, p. 220) also endorses Einstein's position. 

Discussions elsewhere in the literature on how one might find
empirical constraints on the values of the cosmological
parameters usually take account of $\Lambda$.    
The continued interest was at least in part driven by indications
that $\Lambda$  might be needed to reconcile theory and
observations. Here are three examples. 

First, the expansion time is uncomfortably short if $\Lambda =0$.
Sandage's recalibration of the distance scale in the 1960s indicates 
$H_0 \simeq 75$ km s$^{-1}$ Mpc$^{-1}$. If $\Lambda = 0$ this says
the time of expansion from densities too high for stars to have
existed is $< H_0^{-1} \simeq 13$~Gyr, maybe less than the ages
of the oldest stars, then estimated to be greater than about 15~Gyr.
Sandage (1961a) points out that the problem is removed by adding a
positive $\Lambda$. The present estimates reviewed below (Sec. IV.B.3)
are not far from these numbers, but still too uncertain for a
significant case for $\Lambda$.  

Second, counts of quasars as a function of redshift show a peak at 
$z\sim 2$, as would be produced by the loitering epoch in 
Lema\^\i tre's $\Lambda$ model (Petrosian, Salpeter, and
Szekeres, 1967;  Shklovsky, 1967; Kardashev, 1967). The peak
is now well established, centered at $z\sim 2.5$ (Croom {\em et
al.}, 2001; Fan {\em et al.}, 2001). It usually is interpreted as
the evolution in the rate of violent activity in the nuclei of
galaxies, though in the absence of a loitering epoch the
indicated sharp variation in quasar activity with time is
curious (but certainly could be a consequence of astrophysics
that is not well understood).   

The third example is the redshift-magnitude relation. Sandage's
(1961a) analysis indicates this is a promising method of 
distinguishing world models. The Gunn and Oke (1975) measurement
of this relation for giant elliptical galaxies, with Tinsley's
(1972) correction for evolution of the star population from
assumed formation at high redshift, indicates curvature away from 
the linear relation in the direction that, as Gunn and Tinsley 
(1975) discuss, could only be produced by $\Lambda$ (within 
general relativity theory). The new application of the 
redshift-magnitude test, to Type Ia supernovae (Sec. IV.B.4), is 
not inconsistent with the Gunn-Oke measurement; we
do not know whether this agreement of the measurements is
significant, because Gunn and Oke were worried about galaxy 
evolution.\footnote{
Early measurements of the redshift-magnitude relation were meant 
in part to test the Steady State cosmology of Bondi and Gold 
(1948) and Hoyle (1948). Since the Steady State cosmology assumes 
spacetime is independent of time its line 
element has to have the form of the de Sitter solution with 
$\OK = 0$ and the expansion parameter in Eq.~(\ref{eq:des}). The
measured curvature of the redshift-magnitude relation is in the
direction predicted by the Steady State cosmology. But this
cosmology fails other tests discussed in Sec. IV.B.} 

\subsubsection{The coincidences argument against $\Lambda$}

An argument against an observationally interesting value 
of $\Lambda$, from our distrust of accidental coincidences, has
been in the air for decades, and became very influential in the
early 1980s with the introduction of the inflation scenario for
the very early universe. 

If the Einstein-de Sitter model in Eq.~(\ref{eq:edes}) were a
good approximation at the present epoch, an observer measuring  
the mean mass density and Hubble's constant when the age of the
universe was one tenth the present value, or at ten times the
present age, would reach the same conclusion, that the
Einstein-de Sitter model is a good approximation. That is, we
would flourish at a time that is not special in the course of 
evolution of the universe. If on the other hand two or more 
of the terms in the expansion rate equation~(\ref{eq:f2}) made
substantial contributions to the present value of the expansion
rate, it would mean we are present at a special epoch, because
each term in Eq.~(\ref{eq:f2}) varies with the expansion factor 
in a different way. To put this in more detail, we imagine that the 
physics of the very early universe, when the relativistic 
cosmological model became a good approximation, set the values 
of the cosmological parameters. The initial values of the 
contributions to the expansion rate equation had to have been
very different from each other, and had to have been exceedingly 
specially fixed, to make two of the $\Omega_{i0}$'s have
comparable values. This would be a most remarkable and
unlikely-looking coincidence. The   
multiple coincidences required for the near vanishing of 
$\dot a$ and $\ddot a$ at a redshift not much larger than unity 
makes an even stronger case against Lema\^\i tre's coasting
model, by this line of argument.  

The earliest published comment we have found on this point 
is by Bondi (1960, p. 166), in the second edition of his book {\it
Cosmology}. Bondi notes the ``remarkable property'' of the
Einstein-de Sitter model: the dimensionless parameter we now call
$\Omega_{\rm M}$ is independent of the time at which it is computed
(since it is unity). The coincidences argument follows and extends
Bondi's comment. It is presented in McCrea (1971, p. 151). 
When Peebles was a postdoc, in the early 1960s, in 
R. H. Dicke's gravity research group, the coincidences argument
was discussed, but published much later
(Dicke, 1970, p. 62; Dicke and Peebles, 1979). We do not know its
provenance in Dicke's group, whether from Bondi, McCrea, Dicke,
or someone else. We would not  
be surprised to learn others had similar thoughts. 

The coincidences argument is sensible but not a proof, of course.
The discovery of the 3~K thermal cosmic microwave background radiation 
gave us a term in the expansion rate equation that is down from the 
dominant one by four orders of magnitude, not such a large factor by
astronomical standards. This might be  
counted as a first step away from the argument. The evidence from
the dynamics of galaxies that $\OM$ is less than unity is
another step (Peebles, 1984, p. 442; 1986). And yet another is
the development of the evidence that the $\Lambda$ and dark
matter terms  differ by only a factor of three
(Eq.~[\ref{eq:concordance}]).
This last is the most curious, but the community has 
come to accept it, for the most part. The precedent makes
Lema\^{\i}tre's coasting model more socially acceptable. 

A socially acceptable value of $\Lambda$ cannot be such as to
make life impossible, of course.\footnote{
If $\Lambda$ were negative and the magnitude too large there 
would not be enough time
for the emergence of life like us. If $\Lambda$ were positive and
too large  the universe would expand too rapidly to allow galaxy
formation. Our existence, which requires something resembling the
Milky Way galaxy to contain and recycle heavy elements, thus
provides an upper bound on the value of $\Lambda$. 
Such anthropic considerations are discussed by Weinberg (1987, 2001),  
Vilenkin (2001), and references therein.} 
But perhaps the most productive interpretation of the coincidences 
argument is that it demands a search for a more fundamental 
underlying model. This is discussed  further in Sec.~III.E and
the Appendix. 

\subsubsection{Vacuum energy and $\Lambda$}

Another tradition to consider is the relation between $\Lambda$  
and the vacuum or dark energy density. In one approach to the
motivation for the Einstein field equation, taken by McVittie (1956)
and others, $\Lambda$ appears as a constant of integration (of the 
expression for local conservation of energy and momentum). McVittie
(1956, p. 35) emphasizes that, as a constant of integration,
$\Lambda$ ``cannot 
be assigned any particular value on {\it a priori} grounds.''
Interesting variants of this line of thought are still under
discussion (Weinberg, 1989; Unruh, 1989, and references therein). 

The notion of $\Lambda$ as a constant of integration may be
related to the issue of the zero point of energy. In laboratory
physics one measures 
and computes energy differences. But the net energy matters for
gravity physics, and one can 
imagine $\Lambda$ represents the difference between the true
energy density and the sum one arrives at by laboratory physics.  
Eddington (1939) and Lema\^\i tre (1934, 1949) make this point.

Bronstein (1933)\footnote{
Kragh (1996, p. 36) describes Bronstein's motivation and history. We 
discuss this model in more detail in Sec.~III.E, and comment on why 
decay of dark energy into ordinary matter or radiation would be
hard to reconcile with the thermal spectrum of the 3~K cosmic
microwave background radiation. Decay into the dark sector may be
interesting.}   
carries the idea further, allowing
for transfer of energy between ordinary matter and that represented
by $\Lambda$. In our notation, Bronstein expresses this picture
by generalizing Eq.~(\ref{eq:energy}) to 
\begin{equation}
   \dot\rho _\Lambda = -\dot\rho - 3{\dot a\over a}(\rho + p),
   \label{eq:bronstein}
\end{equation}
where $\rho$ and $p$ are the energy density and pressure of 
ordinary matter and radiation. Bronstein goes on to propose a
violation of local energy conservation, a thought that no longer
seems interesting. North (1965, p. 81) finds Eddington's (1939)
interpretation of the zero of energy also somewhat hard to defend. 
But for our purpose the important point is that the idea of $\Lambda$ 
as a form of energy has been in the air, in at least some circles, for
many years.  

The zero-point energy of fields contributes to the dark energy
density. To make 
physical sense the sum over the zero-point mode energies must be
cut off at a short distance or a high 
frequency up to which the model under consideration is valid.
The integral of the zero-point energy ($k/2$) of normal modes (of 
wavenumber $k$) of a massless real bosonic scalar field ($\Phi$), 
up to the wavenumber cutoff $k_c$, gives the vacuum energy density
quantum-mechanical expectation value\footnote{
Eq.~(\ref{eq:zeropoint}), which usually figures in discussions
of the vacuum energy puzzle, gives a helpful indication of the 
situation: the zero-point energy of each mode is real and the
sum is large. The physics is seriously incomplete,
however. The elimination of spatial momenta with magnitudes $k>k_c$
only makes sense if there is a preferred reference frame in which
$k_c$ is defined. Magueijo and Smolin (2002) mention a related issue:
In which reference frame is the Planck momentum of a virtual 
particle at the threshold for new phenomena? In both cases one
may implicitly choose the rest frame
for the large-scale distribution of matter and radiation. It
seems strange to think the microphysics cares about large-scale 
structure, but maybe it happens in a sea of interacting
fields. The cutoff in Eq.~(\ref{eq:zeropoint}) might be
applied at fixed comoving wavenumber $k_c\propto a(t)^{-1}$, or
at a fixed physical value of $k_c$. The first prescription can
be described by an action written as a sum of terms 
$\dot\Phi_{\vec k}^2/2 + k^2\Phi _{\vec k}^2/(2a(t)^2)$ for the
allowed modes. The zero-point energy of each mode scales with the
expansion of the universe as $a(t)^{-1}$, and the sum over modes
scales as $\rho _\Phi\propto a(t)^{-4}$, consistent with
$k_c\propto a(t)^{-1}$. In the limit of exact spatial
homogeneity, an equivalent approach uses the spatial average of
the standard expression for the field stress-energy tensor.
Indeed, DeWitt (1975) and Akhmedov (2002) show that the 
vacuum expectation value of the stress-energy tensor, expressed
as an integral cut off at $k=k_c$, and computed in the preferred
coordinate frame, is diagonal with space part 
$p_\Phi = \rho _\Phi /3$, for the massless field we are
considering. That is, in this prescription the vacuum zero point
energy acts like a homogeneous sea of radiation. This defines a
preferred frame of motion, where the stress-energy tensor is
diagonal, which is not unexpected because we need a preferred
frame to define $k_c$. It is unacceptable as a model for
the properties of dark energy, of course. For example, if the
dark energy density were normalized to the value now under
discussion, and varied as $\rho _\Lambda\propto a(t)^{-4}$, it
would quite mess up the standard model for the origin of the
light elements. We get a more acceptable model for the behavior
of $\rho _\Lambda$ from the second prescription, with the cutoff
at a fixed physical momentum. If we also want to satisfy local
energy conservation we must take the pressure to be 
$p_\Phi =-\rho _\Phi$. This does not contradict the derivation of
$p_\Phi$ in the first prescription, because the second situation
cannot be described by an action: the pressure must be
stipulated, not derived. What is worse, the known fields at
laboratory momenta certainly do not allow this stipulation; they
are well described by analogs of the action in the first
prescription. This quite unsatisfactory situation illustrates how
far we are from a theory of the vacuum energy.}   
\begin{equation}
   \rho_\Phi = \int_0^{k_c} {4\pi k^2 dk \over (2\pi)^3} {k \over 2}
   = {k_c^4 \over 16 \pi^2} .
   \label{eq:zeropoint}
\end{equation}

Nernst (1916) seems to have been the first to write down this
equation, in connection with the idea that the zero-point energy
of the electromagnetic field fills the vacuum, as a light aether,
that could have physically significant properties.\footnote{
A helpful discussion of Nernst's ideas on cosmology is in Kragh 
(1996, pp. 151-7).} 
This was before Heisenberg and Schr\"odinger: Nernst's
hypothesis is that each degree of freedom, which classical
statistical mechanics assigns energy $kT/2$, has 
``Nullpunktsenergie'' $h\nu /2$. This would mean the ground
state energy of a one-dimensional harmonic oscillator is $h\nu$, twice  
the correct value. Nernst's expression for the energy density
in the electromagnetic field thus differs from
Eq.~(\ref{eq:zeropoint}) by a factor of two (after taking account
of the two polarizations), which is 
wonderfully close. For a numerical example, Nernst noted that if 
the cutoff frequency were $\nu =10^{20}$~Hz, or 
$\sim 0.4$~MeV, the energy density of 
the ``Licht\"ather'' (light aether) would be 
$10^{23}$ erg~cm$^{-3}$, or about 100 g~cm$^{-3}$. 

By the end of the 1920s Nernst's hypothesis was replaced with the
demonstration that in quantum mechanics the zero-point energy of
the vacuum is as real as any other. W. Pauli, in unpublished
work in the 1920s,\footnote{
This is discussed in Enz and Thellung (1960), Enz (1974),  
Rugh and Zinkernagel (2000, pp. 4-5), and Straumann (2002).} 
repeated Nernst's calculation, with the correct factor of 2, taking 
$k_c$ to correspond to the
classical electron radius. Pauli knew the value of $\rho_\Lambda$
is quite unacceptable: the radius of the static Einstein universe
with this value of $\rho_\Lambda$ ``would not even reach to the
moon" (Rugh and Zinkernagel, 2000, p. 5).\footnote{
In an unpublished letter in 1930, G. Gamow considered the gravitational  
consequences of the Dirac sea (Dolgov, 1989, p. 230). We thank A. Dolgov
for helpful correspondence on this point.} 
The modern version of this ``physicists' cosmological constant problem" is 
even more acute, because a natural value for $k_c$ is thought to
be much larger than what Nernst or Pauli used.\footnote{
In terms of an energy scale $\epsilon _\Lambda$ defined by 
$\rho_\Lambda =\epsilon _\Lambda ^4$, the Planck energy $G^{-1/2}$ is
about 30 orders of magnitude larger than the ``observed" value of
$\epsilon _\Lambda$.  
This is, of course, an extreme case, since a lot of the theories of 
interest break down well below the Planck scale. Furthermore, in 
addition to other contributions, one is allowed to add a counterterm 
to Eq.~(\ref{eq:zeropoint}) to predict any value of $\rho_\Lambda$.
With reference to this point, it is interesting to note that while Pauli
did not publish his computation of $\rho_\Lambda$, he remarks in his
famous 1933 {\em Handbuch der Physik} review on quantum mechanics that 
it is more consistent to ``exclude a zero-point energy for each degree 
of freedom as this energy, evidently from experience, does not interact
with the gravitational field" (Rugh and Zinkernagel, 2000, p. 5).
Pauli was fully aware that one must take account of zero-point
energies in the binding energies of molecular structure, for
example (and we expect he was aware that what contributes to the  
energy contributes to the gravitational mass). He chose to drop
the section with the above comment 
from the second (1958) edition of the review (Pauli, 1980, pp.
iv-v). In a globally supersymmetric field theory there are equal 
numbers of bosonic and fermionic degrees of freedom, and the net 
zero-point vacuum energy density $\rho_\Lambda$ vanishes (Iliopoulos 
and Zumino, 1974; Zumino, 1975). However, supersymmetry is not a 
symmetry of low energy physics, or even at the electroweak unification
scale. It must be broken at low energies, and the proper setting for
a discussion of the zero-point $\rho_\Lambda$ in this case is locally 
supersymmetric supergravity. Weinberg (1989, p. 6) notes ``it is very 
hard to see how any property of supergravity or superstring theory 
could make the effective cosmological constant sufficiently small".
Witten (2001) and Ellwanger (2002) review more recent developments 
on this issue in the superstring/M theory/branes scenario.}

While there was occasional discussion of this issue in the middle of the 
20$^{\rm th}$ century (as in the quote from N. Bohr in Rugh and
Zinkernagel, 2000, p. 5), 
the modern era begins with the paper by Zel'dovich (1967) that 
convinced the community to consider the possible
connection between the vacuum energy density of quantum physics 
and Einstein's cosmological constant.\footnote{
For subsequent more detailed discussions 
of this issue, see Zel'dovich (1981), Weinberg (1989), Carroll, Press, 
and Turner (1992), Sahni and Starobinsky  (2000), Carroll (2001),
and Rugh and Zinkernagel (2000).}

If the physics of the vacuum looks the same to any inertial
observer its contribution to the stress-energy tensor is the
same as Einstein's cosmological constant
(Eq.~[\ref{eq:TLambda}]). Lema\^\i tre (1934) notes this: 
``in order that absolute motion, i.e., motion relative to the
vacuum, may not be detected, we must associate a pressure 
$p=-\rho c^2$ to the energy density $\rho c^2$ of 
vacuum''. Gliner (1965) goes further, presenting the relation
between the metric tensor and the stress-energy tensor of a
vacuum that appears the same to any inertial observer. But it was
Zel'dovich (1968) who presented the argument clearly enough and
at the right time to catch the attention of the community.

With the development of the concept of broken symmetry in the now
standard model for particle physics came the idea that the
expansion and cooling of the universe is accompanied by a
sequence of first-order phase transitions accompanying the
symmetry breaking. Each first-order transition has a latent heat
that appears as a contribution to an effective time-dependent
$\Lambda (t)$ or dark energy density.\footnote{ 
Early references to this point are Linde (1974), Dreitlein (1974), 
Kirzhnitz and Linde (1974), Veltman (1975), Bludman and Ruderman (1977), 
Canuto and Lee (1977), and Kolb and Wolfram (1980).} 
The decrease in value of the dark energy density at 
each phase transition is much larger than an acceptable present
value (within relativistic cosmology); the natural presumption is
that the dark energy is negligible now. This final
condition seems bizarre, but the picture led to the very
influential concept of inflation. We discussed the basic elements
in connection with  Eq.~(\ref{eq:des}); we turn now to some
implications.

\subsection{Inflation}

\subsubsection{The scenario}

The deep issue inflation addresses is the origin
of the large-scale homogeneity of the observable universe. 
In a relativistic model with positive pressure we can see distant   
galaxies that have not been in causal contact with each other
since the singular start of expansion (Sec.~II.C,
Eq.~[\ref{eq:horizon}]); they are said
to be outside each other's particle horizon. Why do apparently
causally unconnected parts of space look so
similar?\footnote{
Early discussions of this question are 
reviewed by Rindler (1956); more recent examples are Misner
(1969), Dicke and Peebles (1979), and Zee (1980).}  
Sato (1981a, 1981b), Kazanas (1980), and Guth (1981) make the key 
point: if the early universe were dominated by the energy density
of a relatively flat real scalar field (inflaton) potential 
$V(\Phi)$ that acts like $\Lambda$, 
the particle horizon could spread beyond 
the universe we can see. This would allow for the possibility that
microphysics during inflation could smooth inhomogeneities sufficiently
to provide an explanation of the observed large-scale
homogeneity. (We are unaware of a definitive demonstration
of this idea, however.)

In the inflation scenario the field $\Phi$ rolls down its
potential until eventually $V(\Phi)$ steepens enough to terminate
inflation. Energy in the scalar field is supposed to 
decay to matter and radiation, heralding the usual Big Bang
expansion of the universe. With the modifications of Guth's
(1981) scenario by Linde (1982)  
and Albrecht and Steinhardt (1982), the community quickly accepted this 
promising and elegant way to understand the origin of our homogeneous
expanding universe.\footnote{Aspects of the present state of the
subject are  reviewed by Guth (1997), Brandenberger (2001), and
Lazarides (2002).}

In Guth's (1981) picture the inflaton kinetic energy density is
subdominant during inflation, $\dot\Phi^2 \ll V(\Phi)$, so from
Eqs.~(\ref{eq:rhophi}) the pressure $p_\Phi$ is very close to the
negative of the mass density $\rho_\Phi$, and the 
expansion of the universe approximates the de Sitter solution, 
$a \propto \exp (H_\Lambda t)$ (Eq.~[\ref{eq:des}]). 

For our comments on the spectrum of mass density fluctuations
produced by inflation and the properties of solutions of the dark 
energy models in Sec.~III.E we will find it useful to have another 
scalar field model. Lucchin and Matarrese (1985a, 1985b) consider 
the potential 
\begin{equation}
   V(\Phi) = {A \over G^2} \exp \left[- \Phi\sqrt{8\pi qG}\right] ,
   \label{eq:exppot}
\end{equation}
where $q$ and $A$ are parameters.\footnote{
Similar exponential potentials appear in some higher-dimensional
Kaluza-Klein models. For an early discussion see Shafi and Wetterich 
(1985).}
They show that the scale factor and the 
homogeneous part of the scalar field evolve in time as
\begin{eqnarray}
     a(t) & = & a_0 [ 1 + N t ]^{2/q} , \nonumber \\
  \Phi(t) & = & {1 \over \sqrt{2\pi qG}} \ln [ 1 + N t ] ,
   \label{eq:exppotsoln}
\end{eqnarray}
where $N = 2q \sqrt{\pi A}/\sqrt{G(6-q)}$. If $q < 2$ this model
inflates. Halliwell (1987) and Ratra and Peebles (1988) show that the 
solution~(\ref{eq:exppotsoln}) of the homogeneous equation of
motion has the attractor property\footnote{
Ratra (1989, 1992a) shows that spatial inhomogeneities do not destroy 
this property, that is, for $q < 2$ 
the spatially inhomogeneous scalar field perturbation has no growing mode.}
mentioned in connection with Eq.~(\ref{eq:powerlawV}). This exponential
potential is of historical interest: it provided the first clear
illustration of an attractor solution. We return to this point in Sec.~III.E.

A signal achievement of inflation is that it offers a theory for
the origin of the departures from homogeneity. Inflation
tremendously stretches length scales, so cosmologically 
significant lengths now correspond to extremely short lengths
during inflation. On these tiny length scales quantum mechanics
governs: the wavelenghts of zero-point field fluctuations generated
during inflation are stretched by the inflationary expansion,\footnote{
The strong curvature of spacetime during inflation makes the 
vacuum state quite different from that of Minkowski spacetime
(Ratra 1985). This is somewhat analogous to how the Casimir metal  
plates modify the usual Minkowski spacetime vacuum state.}
and these fluctuations are converted to classical density fluctuations 
in the late time universe.\footnote{
For the development of these ideas see Hawking (1982), Starobinsky 
(1982), Guth and Pi (1982), Bardeen, Steinhardt,
and Turner (1983), and Fischler, Ratra, and Susskind (1985).}
 
The power spectrum of the fluctuations depends on the model for 
inflation. If the expansion rate during inflation is close to
exponential (Eq.~[\ref{eq:des}]), the zero-point fluctuations are
frozen into primeval mass density fluctuations 
with power spectrum  
\begin{equation}
    P(k) = \langle|\delta(k, t)|^2\rangle = A k T^2(k) .
    \label{eq:scaleinvariantpk}
\end{equation} 
Here $\delta (k, t)$ is the Fourier transform at wavenumber $k$
of the mass density contrast 
$\delta (\vec{x}, t) = \rho(\vec{x}, t)/\langle \rho(t) \rangle - 1$,
where $\rho$ is the mass density and $\langle\rho\rangle$ the mean value.  
After inflation, but at very large redshifts, the spectrum in
this model is $P(k)\propto k$ on all interesting length scales.
This means the curvature fluctuations produced by the mass
fluctuations diverge only as $\log k$. The form $P(k)\propto k$
thus need not be cut off anywhere near observationally
interesting lengths, and in this sense it is 
scale-invariant.\footnote{
The virtues of a spectrum that is scale-invariant in this sense
were noted before inflation, by Harrison (1970), Peebles and Yu
(1970), and Zel'dovich (1972).}
The transfer function $T(k)$ accounts for the effects of
radiation pressure and the dynamics of nonrelativistic matter on  
the evolution of $\delta (k, t)$, computed in linear perturbation
theory, at redshifts $z\la 10^4$. The constant $A$ is
determined by details of the chosen inflation model we need 
not get into. 

The exponential potential model in Eq.~(\ref{eq:exppot}) produces
the power spectrum\footnote{
This is discussed by Abbott and
Wise (1984), Lucchin and Matarrese (1985a, 1985b), and Ratra (1989, 1992a).}
\begin{equation}
    P(k) = A k^n T^2(k), \qquad n = (2 - 3q)/(2 - q).
    \label{eq:exponentialV}
\end{equation} 
When $n \neq 1 (q \neq 0)$ the power spectrum is said to be tilted. This
offers a parameter $n$ to be adjusted to fit the observations of
large-scale structure, though as we will discuss the simple
scale-invariant case $n=1$  is close to the best fit to the
observations. 

The mass fluctuations in these inflation models are 
said to be adiabatic, because they are what you get by
adiabatically compressing or decompressing parts of an exactly 
homogeneous universe. This means the initial conditions for the
mass distribution are described by one function of position,
$\delta (\vec x, t)$. This function is a realization of a
spatially stationary random Gaussian process, because it
is frozen out of almost free quantum field fluctuations.
Thus the single function of position is
statistically prescribed by its power spectrum, as in
Eqs.~(\ref{eq:scaleinvariantpk}) and~(\ref{eq:exponentialV}).
More complicated models for inflation produce density
fluctuations that are not Gaussian, or do not have simple
power law spectra, or have parts that break adiabaticity, as
gravitational waves (Rubakov, Sazhin, and Veryaskin, 1982) or 
magnetic fields (Turner and Widrow, 1988; Ratra, 1992b) or new
hypothetical fields. All these extra features may be invoked to 
fit the observations, if needed. It may be significant that none 
seem to be needed to fit the main cosmological structure constraints 
we have now. 

\subsubsection{Inflation in a low density universe}

We do need an adjustment from the simplest case --- an
Einstein-de Sitter cosmology --- to account for
the measurements of the mean mass density. In the two models that
lead to Eqs.~(\ref{eq:scaleinvariantpk}) and~(\ref{eq:exponentialV})
the enormous expansion factor during inflation suppresses the
curvature of space sections, making $\OK$ negligibly small. If $\Lambda = 0$,
this fits the Einstein-de Sitter model (Eq. [\ref{eq:edes}]),
which in the absence of data clearly is the elegant choice. But
the high mass density in this model was already seriously
challenged by the data available in 1983, on the low streaming
flow of the nearby galaxies toward the nearest known large mass 
concentration, in the Virgo cluster of galaxies, 
and the small relative velocities of galaxies outside the rich 
clusters of galaxies.\footnote{
This is discussed in Davis and Peebles (1983a, 1983b) and Peebles (1986). 
Relative velocities of galaxies in rich clusters are large, but the masses 
in clusters are known to add up to a modest mean mass density. Thus most 
of the Einstein-de Sitter mass would have to be outside the dense parts of 
the clusters, where the relative velocities are small.} 
A striking and long familiar example of the latter is that the galaxies 
immediately outside the Local Group of galaxies, at distances of a
few megaparsecs, are moving away from us in a good approximation
to Hubble's homogeneous flow, despite the very clumpy
distribution of galaxies on this scale.\footnote{
The situation a half century ago is illustrated by the compilation of 
galaxy redshifts in Humason, Mayall, and Sandage (1956). In this 
sample of 806 galaxies, 14 have negative redshifts (after correction 
for the rotation of the Milky Way galaxy and for the motion of the 
Milky Way toward the other large galaxy in the Local Group, the
Andromeda Nebula), indicating motion toward us. Nine are members 
of the Local Group, at distances $\la 1$~Mpc. Four are in the direction 
of the Virgo cluster, at redshift 
$\sim 1200$ km~s$^{-1}$ and distance $\sim 20$~Mpc. Subsequent 
measurements indicate two of these four really have negative redshifts, 
and plausibly are members of the Virgo cluster on the tail of the 
distribution of peculiar velocities of the cluster members.
(Astronomers use the term peculiar velocity to denote the deviation
from  the uniform Hubble expansion velocity.) The last of the 14, NGC 
3077, is in the M~81 group of galaxies at 3~Mpc distance. It is
now known to have a small positive redshift.} The options (within 
general relativity) are that the mass density is low, so its
clumpy distribution has 
little gravitational effect, or the mass density is high and the
mass is more smoothly distributed than the galaxies. We comment
on the first option here, and the second in connection with the
cold dark matter model for structure formation in Sec.~III.D. 

Under the first option we have two choices: introduce a
cosmological constant, or space curvature, or maybe even both.  
In the conventional inflation picture space curvature is 
unacceptable, but there is another line of thought that leads to
a universe with open space sections. Gott's (1982) scenario 
commences with a 
large energy density in an inflaton at the top of its
potential. This behaves as Einstein's cosmological
constant and produces a near de Sitter universe expanding
as $a \propto \exp (H_\Lambda t)$, with sufficient inflation to 
allow for a microphysical explanation of the large-scale
homogeneity of the observed universe. As the 
inflaton gradually rolls down the potential it reaches a point  
where there is a small bump in the potential. The inflaton
tunnels through this bump by nucleating a bubble. Symmetry 
forces the interior of the bubble to have open spatial sections
(Coleman and De Luccia, 1980), and the 
continuing presence of a non-zero $V(\Phi)$ inside the bubble acts like 
$\Lambda$, resulting in an open inflating universe. The potential
is supposed to steepen, bringing the second limited epoch of
inflation to an end before space curvature has been completely 
redshifted away. The region inside the open bubble at the end of
inflation is a radiation-dominated Friedmann-Lema\^{\i}tre open model, 
with $0< \OK < 1$ (Eq.~[\ref{eq:ok}]). This can fit the dynamical evidence 
for low $\OM$ with $\Lambda =0$.\footnote{Gott's scenario is
resurrected by Ratra and  Peebles (1994, 1995). See Bucher and
Turok (1995), Yamamoto, Sasaki, and Tanaka (1995), and Gott (1997), for
further discussions of this  model. In this case spatial curvature provides
a second cosmologically-relevant length scale (in addition to that set 
by the Hubble radius $H^{-1}$), so there is no natural preference 
for a power law power spectrum (Ratra, 1994; Ratra and  Peebles, 1995).}  

The decision on which scenario, spatially-flat or open, is
elegant, if either, depends ultimately on which Nature has
chosen, if either.\footnote{
At present, high energy physics considerations do not provide
a compelling specific inflation model, but there are strong indications 
that inflation happens in a broad range of models, so it might not 
be unreasonable to think that future advances in high energy physics
could give us a compelling and observationally successful model of 
inflation, that will determine whether it is flat or open.}   
But it is natural to make judgments in advance of the evidence. 
Since the early 1980s there have been occasional explorations 
of the open case, but the community generally has favored the 
flat case, $\OK =0$, without or, more recently, with a cosmological 
constant, and indeed the evidence now is that space sections are
close to flat. The earlier preference for the Einstein-de Sitter 
case with $\OK =0$ and $\OL =0$ led to considerable interest in the 
picture of biased galaxy formation in the cold dark matter model, as 
we now describe.  

\subsection{The cold dark matter model}

Some of the present cosmological tests were understood in the
1930s; others are based on new ideas about structure formation. A 
decade ago a half dozen models for structure formation were under 
discussion\footnote{
A scorecard is given in Peebles and Silk (1990). Structure formation 
models that assume all matter 
is baryonic, and those that augment baryons with hot dark matter
such as low mass neutrinos, were already seriously challenged a
decade ago. Vittorio and Silk (1985) show that the Uson and
Wilkinson (1984) bound on the small-scale anisotropy of the 3~K
cosmic microwave background temperature rules out a baryon-dominated 
universe with adiabatic  initial conditions. This is because the 
dissipation of the baryon density  
fluctuations by radiation drag as the primeval plasma combines to 
neutral hydrogen (at redshift $z\sim 1000$) unacceptably
suppresses structure formation on the scale of galaxies. Cold
dark matter avoids this problem by eliminating radiation drag.
This is one of the reasons attention turned to the hypothetical 
nonbaryonic cold dark matter. There has not been a thorough
search for more baroque initial conditions that might save the
baryonic dark matter model, however.};
now the known viable models have been winnowed to one class: cold 
dark matter (CDM) and variants. We comment on the present state of
tests of the CDM model in Sec.~IV.A.2, and in connection with the 
cosmological tests in Sec. IV.B. 

The CDM model assumes the mass of the universe now is dominated by 
dark matter that is nonbaryonic and acts like a gas of massive, weakly 
interacting particles with negligibly small primeval velocity
dispersion.
Structure is supposed to have formed by the gravitational growth of 
primeval departures from homogeneity that are adiabatic,
scale-invariant, and Gaussian. The early discussions also assume
an Einstein-de Sitter universe. These features all are
naturally implemented in simple models for inflation, and the
CDM model may have been inspired in part by the developing ideas
of inflation. But the motivation in writing down this 
model was to find a simple way to show that the observed
present-day mass fluctuations can agree with the growing evidence
that the anisotropy of the 3~K thermal cosmic microwave background 
radiation is very small (Peebles, 1982). The first steps toward 
turning this picture into a model for structure formation were
taken by Blumenthal {\em et al.} (1984). 

In the decade commencing about 1985 the standard cosmology for
many active in research in this subject was the Einstein-de
Sitter model, and for good reason: it 
eliminates the coincidences problem, it avoids the curiosity of
nonzero dark energy, and it fits the condition from conventional
inflation that space sections have zero curvature. But unease 
about the astronomical problems with the high mass density of
the Einstein-de Sitter model led to occasional discussions of a 
low density universe with or without a cosmological constant,
and the CDM model played an important role in these considerations,
as we now discuss. 

When the CDM model was introduced it was known that the
observations disfavor the high mass density of the Einstein-de
Sitter model, unless the mass is more smoothly distributed than
the visible matter (Sec.~III.C). The key papers showing that this 
wanted biased distribution of visible galaxies relative to the 
distribution of all of the mass can follow in a natural way in 
the CDM theory are Kaiser (1984) and Davis {\em et al.} (1985). 
In short, where the 
mass density is high  enough to lead to the gravitational
assembly of a large galaxy the mass density tends to be high
nearby, favoring the formation of neighboring large galaxies

The biasing concept is important and certainly had to be
explored. But in 1985 there was little empirical evidence for the 
effect and there were significant arguments against it, mainly 
the empty state of the voids between the concentrations of large 
galaxies.\footnote{
The issue is presented in Peebles (1986); the data and history 
of ideas are reviewed in Peebles (2001).} 
In the biasing picture the voids contain most of the mass of an
Einstein-de Sitter universe, but few of the galaxies, because
galaxy formation there is supposed to have been suppressed. But
it is hard to see how galaxy formation could be entirely 
extinguished: the CDM model would be expected to predict a void 
population of irregular galaxies, that show signs of a difficult
youth. Many irregular galaxies are observed, but they avoid the
voids. The straightforward reading of the observations thus is
that the voids are empty, and that the dynamics of the motions
of the visible galaxies therefore say $\OM$ is well below unity,
and that the mass is not more smoothly distributed than the 
visible galaxies. 

In a low density open universe, with $\OL =0$ and positive $\OK$, 
the growth of mass clustering is suppressed at $z\la\OM^{-1} -
1$. Thus to agree with the observed low redshift mass
distribution density fluctuations at high redshift must be larger
in the open  model than in the Einstein-de Sitter case. This
makes it harder to understand the small 3~K cosmic microwave
background anisotropy. In a low density spatially-flat universe
with $\OK =0$ and a cosmological constant, the transition from
matter-dominated expansion to $\Lambda$-dominated expansion is
more recent than the transition  from matter-dominated expansion
to space-curvature-dominated expansion 
in an open universe with $\Lambda = 0$ , as one sees from
Eq.~(\ref{eq:f2}). This makes density fluctuations grow almost as
much as in the Einstein-de Sitter model, thus allowing smaller
peculiar velocities in the flat-$\Lambda$ case, a big help
in understanding the observations.\footnote{
The demonstration that the suppression of peculiar velocities 
is a lot stronger than the suppression of the growth of structure 
is in Peebles (1984) and Lahav {\em et al.} (1991).}  

An argument for low $\OM$, with or without $\Lambda$, developed
out of the characteristic length scale for structure in the CDM model. 
In the Friedmann-Lema\^\i tre cosmology the mass distribution is
gravitationally unstable. This simple statement has a profound
implication: the early universe has to have been very close to
homogeneous, and the growing departures from homogeneity at high
redshift are well described by 
linear perturbation theory. The linear density fluctuations may
be decomposed into Fourier components (or generalizations for
open or closed space sections). At high enough redshift the 
wavelength of a mode is much longer than the time-dependent 
Hubble length $H^{-1}$, and gravitational instability makes 
the mode amplitude grow. Adiabatic fluctuations remain adiabatic,
because different regions behave as if they were parts of
different homogeneous universes. When the Hubble length becomes
comparable to the mode proper wavelength, the baryons 
and radiation, strongly coupled by Thomson scattering at high
redshift, oscillate as an 
acoustic wave and the mode amplitude for the cold dark matter stops 
growing.\footnote{
At high redshift the dark matter mass density is less than that of the 
radiation. The radiation thus fixes the expansion rate, which is too 
rapid for the self-gravity of the dark matter to have any effect on 
its distribution. Early discussions of this effect are in Guyot and 
Zel'dovich (1970) and M\'esz\'aros (1972).}  
The mass densities in dark matter and radiation are equal at 
redshift $z_{\rm eq}=2.4\times 10^4\OM h^2$; thereafter the dark
matter mass density dominates and the fluctuations in its
distribution start to grow again. The suppressed growth of
density fluctuations within the Hubble length at 
$z>z_{\rm eq}$ produces a bend in the power spectrum of the dark mass 
distribution, from $P(k)\propto k$ at long wavelengths, if scale-invariant 
(Eq.~[\ref{eq:scaleinvariantpk}]), to $P(k)\propto k^{-3}$ at
short wavelengths.\footnote{
That is, the transfer function in Eq.~(\ref{eq:scaleinvariantpk}) goes from 
a constant at small $k$ to $T^2(k) \propto k^{-4}$ at large $k$.}
This means that at small scales (large $k$) the contribution to the
variance of the mass density per logarithmic interval of
wavelength is constant, and at small $k$ the contribution to the
variance of the Newtonian gravitational potential per logarithmic
interval of $k$ is constant. 

The wavelength at the break in the spectrum is set by the Hubble 
length at equal radiation and matter mass densities, 
$t_{\rm eq}\propto z_{\rm eq}^{-2}$. This characteristic break 
scale grows by the factor $z_{\rm eq}$ to 
$\lambda _{\rm break}\sim z_{\rm eq}t_{\rm eq}$ at the present
epoch. The numerical value is (Peebles, 1980a, Eq.~[92.47])
\begin{equation}
   \lambda _{\rm break}\sim 50\OM^{-1}h^{-2}\hbox{ Mpc}.
   \label{eq:lambdabreak}
\end{equation}
If $\Lambda$ is close to constant, or $\Lambda =0$ and space
sections are curved, it does not appreciably affect
the expansion rate at redshift $z_{\rm eq}$, so this
characteristic length is the same in flat and open cosmological
models. In an Einstein-de Sitter model, with $\OM=1$,
the predicted length scale at the break in the power spectrum of
CDM mass fluctuations is uncomfortably small relative to
structures such as are observed in clusters of clusters of galaxies 
(superclusters), and relative to the measured galaxy two-point
correlation function. That is, more power is observed on scales 
$\sim 100$ Mpc than is predicted in the CDM model with $\OM = 1$. 
Since $\lambda _{\rm break}$ scales as 
$\OM ^{-1}$, a remedy is to go to a universe with small
$\OM$, either with  $\Lambda =0$ and open space sections
or $\OK =0$ and a nonzero cosmological constant. The latter case
is now known as  $\Lambda$CDM.\footnote{
The scaling of $\lambda _{\rm break}$ with $\OM$
was frequently noted in the 1980s. The earliest discussions
we have seen of the significance for large-scale structure are in
Silk and Vittorio (1987) and Efstathiou, Sutherland, and Maddox
(1990), who consider a spatially-flat universe, Blumenthal,
Dekel, and Primack (1988), who consider the open case, and
Holtzman (1989), who considers both.
For the development of tests of the open case see Lyth and 
Stewart (1990), Ratra and Peebles (1994), Kamionkowski {\em et al.} 
(1994a), and G\'orski {\em et al.} (1995).
Pioneering steps in the analysis of the anisotropy of the 3~K cosmic
microwave background temperature 
in the $\Lambda$CDM model include Kofman and Starobinsky (1985) and 
G\'orski, Silk, and Vittorio (1992). We review developments after
the COBE detection of the anisotropy (Smoot {\em et al.}, 1992) in
Secs. IV.B.11 and 12.}
  
\subsection{Dark energy}

The idea that the universe contains close to homogeneous dark energy 
that approximates a time-variable cosmological ``constant'' arose
in particle physics, through the discussion of phase transitions in
the early universe and through the search for a dynamical 
cancellation of the vacuum energy density;
in cosmology, through the discussions of how to reconcile a
cosmologically flat universe with the small mass density
indicated by galaxy peculiar velocities; and on both
sides by the thought that $\Lambda$ might be very small now
because it has been rolling toward zero for a very long time.\footnote{
This last idea is similar in spirit to Dirac's (1937, 1938) attempt
to explain the large dimensionless numbers of physics. 
He noted that the gravitational force between two
protons is much smaller than the electromagnetic force, 
and that that might be because the gravitational constant $G$
is decreasing in inverse proportion to the world time. This is the 
earliest discussion we know of what has come to be called the 
hierarchy problem, that is, the search for a mechanism that might
be responsible for the large ratio between a possibly more fundamental
high energy scale, for example, that of grand unification or the
Planck scale  
(where quantum gravitational effects become significant) and a
lower possibly less fundamental energy scale, for example that of
electroweak unification (see, for example, Georgi, Quinn, and
Weinberg, 1974). The hierarchy problem in particle physics 
may be rephrased as a search for a mechanism to prevent the
light electroweak symmetry breaking Higgs scalar field mass from being 
large because of a quadratically divergent quantum mechanical 
correction (see, for example, Susskind, 1979). In this sense it
is similar in spirit to the physicists' cosmological constant
problem of Sec.~III.B.} 

The idea that the dark energy is decaying by emission of matter 
or radiation is now strongly constrained by the condition that 
the decay energy must not significantly disturb the spectrum of 
the 3~K cosmic microwave background radiation. But the history of 
the idea is interesting, and decay to dark matter still a
possibility, so we comment on both here. The picture of
dark energy in the form of defects in cosmic fields has not
received much attention in recent years, in part because the
computations are difficult, but might yet prove to be
productive. Much discussed nowadays is dark energy in a slowly
varying scalar field. The idea is reviewed at some length here
and in even more detail in the Appendix. We begin with another
much discussed approach: prescribe the dark energy
by parameters in numbers that seem fit for the quality of the
measurements.  

\subsubsection{The XCDM parametrization}

In the XCDM parametrization
the dark energy interacts only with itself and gravity,  
the dark energy density $\rho_{\rm X}(t) > 0$ is approximated as a
function of world time alone, and the pressure is written as 
\begin{equation}
  p_{\rm X} = w_{\rm X} \rho_{\rm X},
  \label{eq:wx}
\end{equation}
an expression that has come to be known as the cosmic 
equation of state.\footnote{
Other parametrizations of dark energy are discussed by Hu (1998)
and Bucher and Spergel (1999). The name, XCDM, for the case
$w_{\rm X} < 0$ in Eq.~(\ref{eq:wx}), was 
introduced by Turner and White (1997). There is a long history in
cosmology of applications of such an equation of state, and the
related evolution of $\rho _\Lambda$; examples are 
Canuto {\em et al.} (1977), Lau (1985), Huang (1985), Fry (1985), 
Hiscock (1986), \"Ozer and Taha (1986), and Olson and Jordan (1987).
See Ratra and Peebles (1988) for references to other early work 
on a time-variable $\Lambda$ and 
Overduin and Cooperstock (1998) and Sahni and Starobinsky (2000)
for reviews. More recent discussions of this and related models may 
be found in John and Joseph (2000), Zimdahl {\em et al.} (2001), 
Dalal {\em et al.} (2001), Gudmundsson and Bj\"ornsson (2002), Bean 
and Melchiorri (2002), Mak, Belinch\'on, and Harko (2002), and
Kujat {\em et al.} (2002), through which other recent work may be traced.}
Then the local energy conservation
equation~(\ref{eq:energy}) is  
\begin{equation}
{d\rho _{\rm X}\over dt} = - 3{\dot a\over a}\rho _{\rm X}
	(1 + w_{\rm X}).
\end{equation}
If $w_{\rm X}$ is constant the dark energy density scales with
the expansion factor as 
\begin{equation}
  \rho_{\rm X} \propto a^{-3(1+w_{\rm X})}.
  \label{eq:xcdm}
\end{equation} 

If $w_{\rm X} < -1/3$ the dark energy makes a positive contribution
to $\ddot a/a$ (Eq.~[\ref{eq:f1}]). If  
$w_{\rm X} = -1/3$ the dark energy has no effect on $\ddot a$, and 
the energy density varies as $\rho_{\rm X} \propto 1/a^2$, the same as 
the space curvature term in $\dot a^2/a^2$ (Eq.~[\ref{eq:f2}]).
That is, the expansion time histories are the same in an open
model with no dark energy and in a spatially-flat model with 
$w_{\rm X} =-1/3$, although the spacetime geometries   
differ.\footnote{
As discussed in Sec. IV, it 
appears difficult to reconcile the case $w_{\rm X} = -1/3$
with the Type Ia supernova apparent magnitude data (Garnavich {\em et al.},
1998; Perlmutter {\em et al.}, 1999a).}
If $w_{\rm X} < -1$ the dark energy density is increasing.\footnote{
This is quite a step from the thought that the dark energy
density is small because it has been rolling to zero for a long
time, but the case has found a context (Caldwell, 2002; Maor 
{\em et al.}, 2002). Such models were first discussed in the context
of inflation (e.g., Lucchin and Matarrese, 1985b), where it was shown that
the $w_{\rm X} < -1$ component could be modeled as a scalar field
with a negative kinetic energy density (Peebles, 1989a).}

Equation~(\ref{eq:xcdm}) with constant $w_{\rm X}$ has the great
advantage of simplicity. An appropriate generalization for the
more precise measurements to come might be guided by the idea that
the dark energy density is close to homogeneous, spatial
variations rearranging themselves at close to the speed of light,
as in the scalar field models discussed below. Then for most of
the cosmological tests we have an adequate general description of 
the dark energy if we let $w_{\rm X}$ be a free function of
time.\footnote{The availability of a free function greatly
complicates the search for tests as opposed to curve fitting!
This is clearly illustrated by Maor {\em et al.} (2002). For more
examples see Perlmutter, Turner, and White (1999b) and Efstathiou
(1999).} In scalar field pictures $w_{\rm X}$ is derived from
the field model; it can be a complicated function of time 
even when the potential is a simple function of the scalar field.  

The analysis of the large-scale anisotropy of the 3~K cosmic microwave
background radiation requires a prescription for how the spatial
distribution of the dark energy is gravitationally related to 
the inhomogeneous distribution of other matter and radiation
(Caldwell {\em et al.}, 1998). In XCDM this requires at least one 
more parameter, an effective speed of sound, with $c^2_{s{\rm X}}>0$
(for stability, as discussed in Sec.~II.B), in addition to
$w_{\rm X}$.  

\subsubsection{Decay by emission of matter or radiation}

Bronstein (1933) introduced the idea that the dark energy density
$\rho _\Lambda$ is decaying by the emission of matter or
radiation. The continuing discussions of this and the associated 
idea of decaying dark matter (Sciama, 2001, and references
therein) are testimony to the appeal. Considerations in the decay
of dark energy include the
effect on the formation of light elements at $z\sim 10^{10}$, the
contribution to the $\gamma$-ray or  optical extragalactic
background radiation, and the perturbation to the spectrum of 
the 3~K cosmic microwave background radiation.\footnote{These
considerations generally are phenomenological: the evolution  of
the dark energy density, and its related coupling to matter or
radiation, is assigned rather than derived from an action
principle. Recent discussions include 
Pollock (1980), Kazanas (1980), Freese {\em et al.} (1987),
Gasperini (1987), Sato, Terasawa, and Yokoyama (1989), 
Bartlett and Silk (1990), Overduin, Wesson, and Bowyer (1993),
Matyjasek (1995), and Birkel and Sarkar (1997).}

The effect on the 3~K cosmic microwave background was of 
particular interest a
decade ago, as a possible explanation of indications of a
significant departure from a Planck spectrum. Precision
measurements now show the spectrum is very close to thermal.
The measurements and their interpretation are discussed by Fixsen
{\em et al.} (1996). They show that the allowed addition to the
3~K cosmic microwave background energy density $\rho_{\rm R}$ 
is limited to just 
$\delta\rho_{\rm R}/\rho_{\rm R} \la 10^{-4}$ since redshift 
$z\sim 10^5$, when the interaction between matter and radiation 
was last strong enough for thermal relaxation. The bound on 
$\delta\rho_{\rm R}/\rho_{\rm R}$ is not inconsistent with what
the galaxies are thought to produce, but it is well below an
observationally interesting dark energy density.  

Dark energy could decay by emission of dark matter, cold or hot,
without disturbing the spectrum of the 3~K cosmic microwave background
radiation. For example, let us suppose the dark energy
equation of state is $w_{\rm X}=-1$, and hypothetical
microphysics causes the dark energy density to decay as 
$\rho _\Lambda\propto a^{-n}$ by the production of
nonrelativistic dark matter. Then Bronstein's
Eq.~(\ref{eq:bronstein}) says the dark matter density varies with
time as  
\begin{equation}
  \rho _{\rm M}(t) = {A\over a(t)^3} + 
	{n\over 3 - n}\rho _\Lambda (t),
  \label{eq:DDE}
\end{equation}
where $A$ is a constant and $0<n<3$. In the late time limit the
dark matter density is a fixed fraction of the dark energy. But
for the standard interpretation of the measured anisotropy of the
3~K background we would have to suppose the first term on the
right hand side of Eq.~(\ref{eq:DDE}) is not much smaller than
the second, so the coincidences issue discussed in Sec.~III.B.2
is not much relieved. It does help relieve the problem with the small
present value of $\rho _\Lambda$ (to be discussed in connection
with Eq.~[\ref{eq:lambdaenergy}]). 

We are not aware of any work on this decaying dark energy
picture. Attention instead has turned to the idea that the dark
energy density evolves without emission, as illustrated in
Eq.~(\ref{eq:xcdm}) and the two classes of physical models to be
discussed next.  

\subsubsection{Cosmic field defects}

The physics and cosmology of topological defects produced at
phase transitions in the early universe are reviewed by Vilenkin
and Shellard (1994). An example of dark energy is a tangled web 
of cosmic string, with fixed mass per unit length, which 
self-intersects without reconnection. In Vilenkin's (1984) 
analysis\footnote{
The string flops at speeds comparable to light, making the coherence 
length comparable to the expansion time $t$. Suppose a string 
randomly walks across a region of physical size $a(t)R$ in
$N$ steps, where $aR\sim N^{1/2}t$. The total length of this
string within the region $R$ is $l\sim Nt$. Thus the mean
mass density of the string scales with time as 
$\rho _{\rm string}\propto l/a^3\propto (ta(t))^{-1}$. One
randomly walking string does not fill space, but we can imagine
many randomly placed strings produce a nearly smooth mass
distribution. Spergel and Pen (1997) compute the 3~K cosmic microwave 
background radiation
anisotropy in a related model, where the string network is
fixed in comoving coordinates so the mean mass density
scales as $\rho _{\rm string}\propto a^{-2}$.} 
the mean mass density in strings scales as 
$\rho _{\rm string}\propto (ta(t))^{-1}$. When ordinary matter is
the dominant contribution to $\dot a^2/a^2$, the ratio of mass
densities is $\rho _{\rm string}/\rho\propto t^{1/3}$. Thus at late
times the string mass dominates. In this limit, 
$\rho _{\rm string}\propto a^{-2}$, $w_{\rm X} = -1/3$ for the XCDM 
parametrization of Eq.~(\ref{eq:xcdm}), and the universe expands as
$a\propto t$. Davis (1987) and Kamionkowski and Toumbas (1996) 
propose the same behavior for a texture model.
One can also imagine domain walls fill space densely enough not 
to be dangerous. If the domain walls are fixed in comoving
coordinates the domain wall energy density scales as 
$\rho_{\rm X}\propto a^{-1}$ (Zel'dovich, Kobzarev, and Okun,
1974; Battye, Bucher, and Spergel, 1999). The corresponding
equation of state parameter is $w_{\rm X} = -2/3$, which is
thought to be easier to reconcile 
with the supernova measurements than  $w_{\rm X} = -1/3$
(Garnavich {\em et al.}, 1998; Perlmutter {\em et al.}, 1999a).
The cosmological tests of defects models for the dark energy have
not been very thoroughly explored, at least in part because an
accurate treatment of the behavior of the dark energy is
difficult (as seen, for example, in Spergel and Pen, 1997; Friedland, 
Muruyama, and Perelstein, 2002), but this class of models is worth 
bearing in mind.    
 
\subsubsection{Dark energy scalar field}

At the time of writing the popular picture for dark energy is a 
classical scalar field with a self-interaction potential $V(\Phi)$ 
that is shallow enough that the field energy density decreases
with the expansion of the universe more slowly than the energy
density in matter. This idea grew in part out of the inflation
scenario, in part from ideas from particle physics. Early
examples are Weiss (1987) and Wetterich (1988).\footnote{
Other early examples include those cited in Ratra and Peebles
(1988) as well as
End\=o and Fukui (1977), Fujii (1982), Dolgov (1983), Nilles
(1985), Zee (1985), Wilczek (1985), Bertolami (1986),  Ford
(1987), Singh and Padmanabhan (1988), and Barr and Hochberg
(1988).}  
The former considers a quadratic potential with an ultralight 
effective mass, an idea that reappears in Frieman {\em et al.} (1995).
The latter considers the time variation
of the dark energy density in the case of the Lucchin and
Matarrese (1985a) exponential self-interaction potential 
(Eq.~[\ref{eq:exppot}]).\footnote{
For recent discussions of this model see Ferreira
and Joyce (1998), Ott (2001), Hwang and Noh (2001), and references
therein.}

In the exponential potential model the scalar field 
energy density varies with time in constant proportion to the
dominant energy density. 
The evidence is that radiation dominates at redshifts in the
range $10^3\la z\la 10^{10}$, 
from the success of the standard model for light element
formation, and matter dominates at $1\la z\la 10^3$, from the
success of the standard model for the gravitational growth of
structure. This would leave the dark energy
subdominant today, contrary to what is wanted. This led to the
proposal of the inverse power-law potential in
Eq.~(\ref{eq:powerlawV}) for a single real scalar field.\footnote{
In what follows we focus on this model, which was proposed by  
Peebles and Ratra (1988). The model assumes a conventionally 
normalized scalar field kinetic energy and spatial gradient term
in the action, and it assumes the scalar field is coupled only to
itself and gravity. The model is then completely characterized by
the form of the potential (in addition to all the other usual
cosmological parameters, including initial conditions). Models
based on other forms for $V(\Phi)$, with a more   
general kinetic energy and spatial gradient term, or with more general
couplings to gravity and other fields, are discussed in the Appendix.} 

We do not want the hypothetical field $\Phi$ to couple too
strongly to baryonic matter and fields, because that would
produce a ``fifth force'' that is not observed.\footnote{
The current value of the mass associated with spatial 
inhomogeneities in the field is $m_\phi(t_0) \sim H_0 
\sim 10^{-33}$ eV, as one would expect from the dimensions. More
explicitly, one arrives at this mass by 
writing the field as $\Phi (t, {\vec x}) = 
\langle \Phi \rangle (t) + \phi (t, {\vec x})$ and Taylor expanding
the scalar field potential energy density $V(\Phi)$ about the
homogeneous mean background $\langle \Phi \rangle$ to quadratic
order in the spatially inhomogeneous part $\phi$, to get
$m_\phi^2 = V^{\prime\prime}(\langle \Phi \rangle)$. Within the 
context of the inverse power-law model, the tiny value of the mass 
follows from the requirements that $V$ varies slowly with the field 
value and that the current value of $V$ be observationally
acceptable. The difference between the roles of $m_\phi$ and the
constant $m_q$ in the quadratic
potential model $V=m_q^2\Phi ^2/2$ is worth noting. The mass
$m_q$ has an assigned and arguably fine-tuned value. The
effective mass $m_\phi \sim H$ belonging to 
$V\propto\Phi ^{-\alpha}$ is a 
derived quantity, that evolves as the universe expands. The 
small value of $m_\phi (t_0)$ explains why the scalar field energy
cannot be concentrated with the non-relativistic mass in galaxies
and clusters of galaxies. Because of the tiny mass a scalar field
would mediate a new long-range fifth force if it were 
not weakly coupled to ordinary matter. Weak coupling
also ensures that the contributions to coupling constants (such
as the gravitational constant) from the exchange of dark energy 
bosons are small, so such coupling constants are not significantly 
time variable in this model. See, for example, 
Carroll (1998), Chiba (1999), Horvat (1999), Amendola (2000), Bartolo and 
Pietroni (2000), and Fujii (2000) for recent discussions of this 
and related issues.}${^{,}}$ \footnote{
Coupling between dark energy and
dark matter is not constrained by conventional fifth force
measurements. An example is discussed by Amendola and
Tocchini-Valentini (2002). Perhaps the first consideration is
that the fifth-force interaction between neighboring dark matter halos
must not be so strong as to shift regular galaxies of stars away
from the centers of their dark matter halos.} 
Within quantum field theory the inverse 
power-law scalar field potential makes the model non-renormalizable
and thus pathological. But the model is meant to describe what might 
emerge out of a more fundamental quantum theory, which maybe also
resolves the 
physicists' cosmological constant problem (Sec.~III.B), as the effective 
classical description of the dark energy.\footnote{
Of course, the zero-point energy of the quantum-mechanical 
fluctuations around the mean field value contributes to the 
physicists'
cosmological constant problem, and renormalization of the
potential could destroy the attractor solution (however, see
Doran and J\"ackel, 2002) and could generate couplings between 
the scalar field and other fields leading to an observationally 
inconsistent ``fifth force". The problems within quantum field 
theory with the idea that the energy of a classical scalar field 
is the dark energy, or drives inflation, are further discussed 
in the Appendix. The best we can hope is that the
effective classical model is a useful approximation to what
actually is happening, which might lead us to a more satisfactory
theory.}
The potential of this classical effective field is chosen {\it ad hoc}, 
to fit the scenario. But one can adduce analogs within supergravity, 
superstring/M, and brane theory, as reviewed in the Appendix.

The solution for the mass fraction in dark energy in the inverse 
power-law potential model (in Eq.~[\ref{eq:densityratio}] when
$\rho_\Phi \ll \rho$, and the numerical solution at lower
redshifts) is not unique, but it behaves as what has come to be 
termed an attractor or tracker: it is the asymptotic solution 
for a broad range of initial conditions.\footnote{
A recent discussion is in Brax and Martin (2000). Brax, Martin,
and Riazuelo (2000) present a thorough analysis of the evolution of 
spatial inhomogeneities in the inverse power-law scalar field potential 
model and confirm that these inhomogeneities do not destroy the homogeneous 
attractor solution. For other recent discussions of attractor solutions 
in a variety of contexts see Liddle and Scherrer (1999), 
Uzan (1999), de Ritis {\em et al.} (2000), Holden and Wands (2000), 
Baccigalupi, Matarrese, and Perrotta (2000), and Huey and Tavakol 
(2002).} 
The solution also has the property that $\rho _\Phi$ is
decreasing, but less rapidly than the mass densities in matter
and radiation. This may help alleviate two troubling aspects of
the cosmological constant. The coincidences issue is discussed 
in Sec.~III.B. The other is the characteristic energy scale set
by the value of $\Lambda$,
\begin{equation}
   \epsilon_\Lambda (t_0) = \rho _\Lambda (t_0)^{1/4} = 
	0.003 (1 - \OM)^{1/4} h^{1/2} {\rm eV} ,  
  \label{eq:lambdaenergy}
\end{equation} 
when $\OR$ and $\OK$ may be neglected. In the limit of constant
dark energy density, cosmology seems to indicate new
physics at an energy scale more typical of
chemistry. If $\rho _\Lambda$ is rolling toward zero the energy
scale might look more reasonable, as follows (Peebles and Ratra,
1988; Steinhardt {\em et al.}, 1999; Brax {\em et al.}, 2000).   

Suppose that as conventional inflation ends the scalar field
potential switches over to the inverse power-law form in
Eq.~(\ref{eq:powerlawV}). Let the energy scale at the end of
inflation be $\epsilon (t_{\rm I}) = \rho (t_{\rm I})^{1/4}$,
where $\rho (t_{\rm I})$ is the energy density in matter and
radiation at the end of
inflation, and let $\epsilon _\Lambda (t_{\rm I})$ be the energy 
scale of the dark energy at the end of inflation. Since the
present value $\epsilon _\Lambda (t_{\rm 0})$ of the dark
energy scale (Eq.~[\ref{eq:lambdaenergy}]) is comparable to the
present energy scale belonging to the matter, we have from
Eq.~(\ref{eq:densityratio})
\begin{equation}
   \epsilon_\Lambda (t_{\rm I}) \simeq \epsilon (t_{\rm I})
	\left( \epsilon_\Lambda (t_0)\over 
          \epsilon (t_{\rm I}) \right)^{2/(\alpha + 2)}. 
   \label{eq:energychemistry}
\end{equation}
For parameters of common inflation models, 
$\epsilon (t_{\rm I})\sim 10^{13}$~GeV, and 
$\epsilon_\Lambda (t_0)/\epsilon (t_{\rm I}) \sim 10^{-25}$.
If, say, $\alpha = 6$, then 
\begin{equation}
   \epsilon _\Lambda (t_{\rm I}) \sim 10^{-6} \epsilon (t_{\rm I})
	\sim 10^7\hbox{ GeV}.   
  \label{eq:thefactor}
\end{equation}
As this example illustrates, one can arrange the scalar field
model so it has a characteristic energy scale that exceeds the
energy $\sim 10^3$ GeV below which physics is thought to be
well understood: in this model cosmology does not
force upon us the idea that there is as yet undiscovered
physics at the very small energy in Eq.~(\ref{eq:lambdaenergy}).
Of course, where the factor $\sim 10^{-6}$ in Eq.~(\ref{eq:thefactor}) 
comes from still is an open question, but, as discussed in the 
Appendix, perhaps easier to resolve than the origin of
the factor $\sim 10^{-25}$ in the constant $\Lambda$ case.

When we can describe the dynamics of the departure from a
spatially  homogeneous field in linear perturbation theory,
a scalar field model generally is characterized by the 
time-dependent values of $w_{\rm X}$ (Eq.~[\ref{eq:wx}]) and
the speed of sound $c_{s{\rm X}}$ (e.g., Ratra, 1991;
Caldwell {\em et al.}, 1998). In the inverse power-law 
potential model the relation between the power-law index $\alpha$ 
and the equation of state parameter in the matter-dominated epoch is
independent of time (Ratra and Quillen, 1992),
\begin{equation}
  w_{\rm X} = - {2 \over \alpha + 2}.
  \label{eq:phieos}
\end{equation}
When the dark energy density starts to make an appreciable
contribution to the expansion rate the parameter $w_{\rm X}$
starts to evolve. The use of a constant value of $w_{\rm X}$ to
characterize the inverse power-law potential model thus can be
misleading. For example, Podariu and Ratra (2000, Fig. 2) show
that, when applied to the Type Ia supernova measurements, the
XCDM parametrization in Eq.~(\ref{eq:phieos}) leads to a
significantly tighter apparent upper limit 
on $w_{\rm X}$, at fixed $\OM$, than is warranted by the
results of a computation of the evolution of the dark energy
density in this scalar field model. Caldwell {\em et al.} (1998) deal
with the relation between scalar field models and the XCDM
parametrization by fixing $w_{\rm X}$, as a constant or some
function of redshift, deducing the scalar field potential
$V(\Phi)$ that produces this $w_{\rm X}$, and then computing the
gravitational response of $\Phi$ to the large-scale mass
distribution.  

\section{The cosmological tests}

Our intention is to supplement recent discussions of parameter
determinations within the standard relativistic cosmology\footnote{
See Bahcall {\em et al.} (1999), Schindler (2001), Sarkar (2002),
Freedman (2002), Plionis (2002), and references therein.}
with a broader consideration of the issues summarized in two
questions: what is the purpose of the cosmological tests, and how
well is the purpose addressed by recent advances and work in
progress?

The short answer to the first question used to be that we seek to
check the underlying physical theory, general
relativity, applied on the time and length scales of cosmology;
the model for the stress-energy tensor in Einstein's field
equation, suitably averaged over the rich small-scale structure
we cannot describe in any detail; and the boundary condition,
that the universe we can observe is close to homogeneous and
isotropic on the scale of the Hubble length. Recent advances make
use of the CDM prescription for the stress-energy tensor and the
boundary condition, so we must add the elements of the CDM model
to the physics to be checked.   

The short answer to the second question is that we now have
searching checks of the standard cosmology, which the 
model passes. But we believe it takes nothing away from the
remarkable advances of the tests, and the exemplary care in the
measurements, to note that there is a lot of room for
systematic errors. As we discussed in Sec.~I.A, the empirical
basis for the standard model for cosmology is not nearly as
substantial as is the empirical basis for the standard model for 
particle physics: in cosmology it is not yet a matter of
measuring parameters in a well-established physical theory.

We comment on the two main pieces of physics, general relativity
and the CDM  model, in Secs.~IV.A.1 and~IV.A.2. In
Sec.~IV.B we discuss the state of 13 cosmological tests,
proceeding roughly in order of increasing model dependence. We
conclude that there is a well-established scientific case for the
physical significance of the matter density parameter, and for
the result of the measurements, $0.15 \la \OM \la 0.4$ (in the 
sense of a two standard deviation range). Our reasoning is 
summarized in Sec.~IV.C, along with an explanation of why we are 
not so sure about the detection of $\Lambda$ or dark energy.

\subsection{The theories}

\subsubsection{General relativity}

Some early discussions of the cosmological tests, as in 
Robertson (1955) and Bondi (1960), make the point that
observationally important elements of a spatially homogeneous
cosmology follow by 
symmetry, independent of general relativity. This means some
empirical successes of the cosmology are not tests of
relativity. The point was important in the 1950s, because the
Steady State theory was a viable alternative to the 
Friedmann-Lema\^\i tre cosmology, and because the experimental tests
of relativity were quite limited.       

The tests of general relativity are much better now, but cosmology 
still is a considerable extrapolation. The length scales characteristic 
of the precision tests of general relativity in the Solar System and 
binary pulsar are $\la 10^{13}$~cm. An important scale for 
cosmology is the Hubble length,
$H_0^{-1}\sim 5000$~Mpc $\sim 10^{28}$~cm, fifteen orders
of magnitude larger. An extrapolation of fifteen
orders of magnitude in energy from that achieved at the 
largest accelerators,  
$\sim 10^{12}$~eV, brings us to the very different world of the
Planck energy. Why is the community not concerned about an
extrapolation of similar size in the opposite direction? One
reason is that the known open issues of physics have to do with
small  length scales; there is no credible reason to think
general relativity may fail on large scales. This is comforting,
to be sure, but, as indicated in footnote 7, not the same as a 
demonstration that we really know the physics of cosmology. 
Another reason is that if the physics of cosmology were very different 
from general relativity it surely would have already been
manifest in serious problems with the cosmological tests. This
also is encouraging, but we have to consider details, as follows. 

One sobering detail is that in the standard cosmology the two dominant 
contributions to the stress-energy tensor --- dark energy and
dark matter --- are hypothetical, introduced to make the theories 
fit the observations (Eq.~[\ref{eq:concordance}]). This need not
mean there is anything wrong 
with general relativity --- we have no reason to expect Nature to
have made all matter readily observable other than by its
gravity --- but it is a cautionary example of the challenges.
Milgrom's (1983) modified Newtonian dynamics 
(MOND) replaces the dark matter hypothesis with a hypothetical
modification of the gravitational force law. MOND gives
remarkably successful fits to observed motions within galaxies,
without dark matter (de Blok {\em et al.}, 2001). So why should we
believe there really is cosmologically significant mass in
nonbaryonic dark matter? Unless we are lucky enough to get a
laboratory detection, the demonstration must be through the tests
of the relativistic cosmology (and any other viable cosmological
models that may come along, perhaps including an extension of MOND).
This indirect chain of evidence for dark matter is becoming 
tight. A new example --- the prospect for a test of the inverse
square law for gravity on the length scales of cosmology --- is
striking enough for special mention here.\footnote{
Bin\'etruy and Silk (2001) and Uzan and Bernardeau (2001) pioneered 
this probe of the inverse square law. Related probes, based on the 
relativistic dynamics of gravitational lensing and the anisotropy of 
the 3~K thermal background, are discussed by these authors and White 
and Kochanek (2001).} 

Consider the equation of motion\footnote{
These relations are discussed in many books on cosmology, 
including Peebles (1980a).} 
of a freely moving 
test particle with nonrelativistic peculiar velocity $\vec v$ in
a universe with expansion factor $a(t)$, 
\begin{equation}
   {\partial\vec v\over\partial t} 
   +{\dot a\over a}\vec v = \vec g = -{1\over a}\nabla\varphi .
\label{eq:dynamics}
\end{equation}
The particle always is moving toward receding observers, which
produces the second term in the left-most expression. The 
peculiar gravitational acceleration
$\vec g$ relative to the homogeneous background model is computed 
from the Poisson equation for the gravitational potential $\varphi$, 
\begin{equation}
   \nabla ^2\varphi = 4\pi Ga^2[\rho (\vec x,t) -\langle\rho\rangle ].
\label{eq:poisson}
\end{equation}
The mean mass density $\langle\rho\rangle$ is subtracted because
$\vec g$ is computed relative to the homogeneous model. 
The equation of mass conservation expressed in terms of the
density contrast $\delta =\rho /\langle\rho\rangle - 1$ of the
mass distribution modeled as a continuous pressureless fluid is 
\begin{equation}
   {\partial\delta\over\partial t} + 
   {1\over a}\nabla\cdot (1+\delta )\vec v = 0.
\label{eq:mass}
\end{equation}
In linear perturbation theory in $\vec v$ and $\delta$ these
equations give 
\begin{equation}
   {\partial ^2\delta\over\partial t^2} +
   2{\dot a\over a}{\partial\delta\over\partial t} = 
	4\pi G\langle\rho\rangle\delta ,\qquad
  \delta (\vec x,t) = f(\vec x)D(t).
\label{eq:delta}
\end{equation}
Here $D(t)$ is the growing solution to the first
equation.\footnote{
The general solution is a sum of the growing and decaying solutions, 
but because the universe has expanded by a large factor since
nongravitational forces were last important on large scales we
can ignore the decaying part.} 
The velocity field belonging to the solution $D(t)$ is 
the inhomogeneous solution to Eq.~(\ref{eq:mass}) in linear
perturbation theory, 
\begin{equation}
   \vec v = {fH_oa\over 4\pi }\int {\vec y -\vec x\over 
   |\vec y -\vec x|^3}\delta (\vec y)\, d^3y,\qquad
   f\simeq\OM ^{0.6}.
\label{eq:linearpt}
\end{equation}
The factor $f=d\log D/d\log a$ depends on the cosmological model;
the second equation is a good approximation if $\Lambda= 0$ or
space curvature vanishes.\footnote{
This is illustrated in Fig.
13.14 in Peebles (1993). An analytic expression for spherical
symmetry is derived by Lightman and Schechter (1990).} 
One sees from Eq.~(\ref{eq:linearpt}) that the peculiar velocity
is proportional to the gravitational acceleration, as one would
expect in linear theory.   

The key point of Eq.~(\ref{eq:delta}) for the present purpose is
that the evolution of the density contrast $\delta$ at a given
position is not affected by the value of $\delta$ anywhere else. 
This is a consequence of the inverse square law. The mass 
fluctuation in a chosen volume element produces a peculiar 
gravitational acceleration $\delta\vec g$ that produces a peculiar 
velocity field $\delta\vec v\propto\vec g$ that has zero divergence 
and so the mass inside the volume element does not effect the exterior.   

For a ``toy'' model of the effect of a failure of the inverse
square law, suppose we adjust the expression for the
peculiar gravitational acceleration produced by a given mass
distribution to 
\begin{equation}
   \vec g = a^3R\int d^3y\,\delta (\vec y)
   {\vec y -\vec x\over |\vec y -\vec x|}Q(a|\vec y -\vec x|),
\label{eq:pseudoMOND}
\end{equation}
where $R$ is some function of world time only. 
In standard gravity physics $Q(w)=w^{-2}$. We have no basis in
fundamental physics for any other function of $w$. Although
Milgrom's (1983) MOND provides a motivation,
Eq.~(\ref{eq:pseudoMOND}) is not meant to be an extension of MOND
to large-scale flows. It is an {\it ad hoc}
model that illustrates an important property of the
inverse square law.  

We noted that in linear theory $\vec v\propto\vec g$. Thus we
find that the divergence of 
Eq.~(\ref{eq:pseudoMOND}), with the mass conservation
equation~(\ref{eq:mass}) in linear perturbation theory, gives
\begin{equation}
   {\partial ^2\delta _{\vec k}\over\partial t^2} + 
   2{\dot a\over a}{\partial\delta _{\vec k}\over\partial t} = 
   S(k,t)\delta _{\vec k},\qquad
   S(k,t) = (4\pi R k/a) \int _0^\infty w^2dw\, Q(w)j_1(kw/a),
\label{eq:nutty}
\end{equation}
where $\delta _{\vec k}(t)$ is the Fourier transform of the mass
density contrast $\delta (\vec x,t)$ and $j_1$ is a spherical
Bessel function. 
The inverse square law, $Q=w^{-2}$, makes the factor $S$ independent of 
the wavenumber $k$. This means all Fourier amplitudes grow by the
same factor in linear perturbation theory (when the growing mode
dominates), so the functional form of $\delta (\vec x,t)$ 
is preserved and the amplitude grows, as Eq.~(\ref{eq:delta})
says. When $Q$ is some other function,
the phases of the $\delta _{\vec k}$ are preserved
but the functional form of the 
power spectrum $|\delta _{\vec k}|^2$ evolves. 
For example, if $Q\propto w^{n-2}$ with $-2<n<1$ (so the integral 
in Eq.~[\ref{eq:nutty}] does not diverge) Eq.~(\ref{eq:nutty}) is
\begin{equation}
   {\partial ^2\delta _{\vec k}\over\partial t^2} + 
   2{\dot a\over a}{\partial\delta _{\vec k}\over\partial t} = 
   U(t)\left( a\over k\right)^n\delta _{\vec k},
\end{equation}
where $U$ is some function of world time. 

If $n>0$ density fluctuations grow faster on larger scales. 
If $Q(w)$ follows Newtonian gravity on the  scale of galaxies and
bends to $n>0$ on larger scales it reduces the mean mass density
needed to account for the measured large-scale galaxy flows, and
maybe reduces the need for dark matter. But there are testable
consequences: the apparent value of $\OM$ would vary with the
length scale of 
the measurement, and the form of the power spectrum of the
present mass distribution would not agree with the form at
redshift $z\sim 1000$ when it produced the observed angular power
spectrum of the 3~K cosmic microwave background. Thus we are very
interested in the evidence of consistency of these tests (as
discussed in Sec.~IV.B.13).  

\subsubsection{The cold dark matter model for structure formation}

Important cosmological tests assume the CDM model for structure
formation (Sec. III.C), so we must consider tests of the model.
The model has proved to be a useful basis for analyses of the
physics of formation of galaxies and clusters of galaxies 
(e.g., Kay {\em et al.}, 
2002; Colberg {\em et al.}, 2000; and references therein). There
are issues to consider, however; Sellwood and Kosowsky (2001)
give a useful survey of the situation. We remark on recent
developments and what seem to us to be critical issues. 

Numerical simulations of the dark mass distribution in the CDM
model predict that massive halos have many low mass satellites,
perhaps significantly more than the number observed around the
Milky Way galaxy (Klypin {\em et al.}, 1999; Moore {\em et al.},
1999a). The issue is of great interest but not yet a critical
test, because it is difficult to predict the nature of star
formation in a low mass dark halo: what does a dark halo look 
like when star formation or the neutral gas content makes it
visible? For recent discussions see Tully {\em et al.} (2002)
and Stoehr {\em et al.} (2002).

The nature of the dark mass distribution within galaxies is a
critical issue, because we know where to look for a distinctive
CDM feature: a cusp-like central mass distribution, the
density varying with radius $r$ as $\rho\propto r^{-\alpha}$ with
$\alpha\ga 1$. The power law is not unexpected, because there is
nothing in the CDM model to fix an astronomically interesting value 
for a core radius.\footnote{
Pioneering work on the theory of the central
mass distribution in a dark mass halo is in Dubinski and Carlberg
(1991). Moore (1994) and Flores and Primack (1994) are among the
first to point out the apparent disagreement between theory 
and observation.} 
A measure of the mass distribution in disk galaxies is the
rotation curve: the circular velocity as a function
of radius for matter supported by rotation. In some low surface 
brightness galaxies the observed rotation curves are close to solid
body, $v_c\propto r$, near the center, consistent with a near
homogeneous core, and inconsistent with the cusp-like CDM mass
distribution.\footnote{
The situation is reviewed by de Blok {\em et al.} (2001), and de
Blok and Bosma (2002). The galaxy NGC 3109 is a helpful example
because it is particularly close --- just outside the Local  
Group --- and so particularly well resolved. An optical image is 
in plate 39 in the {\it Hubble Atlas of Galaxies} (Sandage, 1961b). 
The radial 
velocity measurements across the face of the galaxy, in Figs.~1
and~2 in Blais-Ouellette, Amram, and Carignan (2001), are
consistent with circular motion with $v_c\propto r$ at 
$r\la 2$~kpc.} 

The circular velocity produced by the mass
distribution $\rho\propto r^{-1}$ is not very different from
solid body, or from the observations, and the difference might be
erased by gravitational rearrangement of the dark mass
by the fluctuations in the distribution of baryonic mass driven by
star formation, winds, or supernovae. This is too 
complicated to assess by current numerical simulations. But we do 
have a phenomenological hint: central solid body rotation
is most clearly seen in the disk-like galaxies with the lowest
surface brightnesses, the objects in which the baryon mass seems
least likely to have had a significant gravitational effect on 
the dark mass. This challenge to the CDM model is pressing. 

The challenge may be resolved in a warm dark matter
model, where the particles are assigned a primeval velocity 
dispersion that suppresses the initial power spectrum of density 
fluctuations on small scales (Moore {\em et al.}, 1999b; Sommer-Larsen
and Dolgov, 2001; Bode, Ostriker, and Turok, 2001). But
it seems to be difficult to reconcile the wanted suppression
of small-scale power with the observation of small-scale
clustering in the Lyman-$\alpha$ forest --- the neutral hydrogen
observed at $z\sim 3$ in the Lyman-$\alpha$ resonance absorption
lines in quasar spectra (Narayanan {\em et al.}, 2000; Knebe 
{\em et al.}, 2002). Spergel and Steinhardt (2000) 
point out that the scattering cross section of self-interacting
cold dark matter particles can be adjusted to suppress the
cusp-like core.\footnote{
In a power law halo with $\rho\propto r^{-\gamma}$, the velocity 
dispersion varies with
radius as  $\langle v^2\rangle\sim GM(<r)/r\propto r^{2-\gamma}$.
The particle scattering cross section must be adjusted to erase the
effective temperature gradient, thus lowering the mass density at
small radii, without promoting unacceptable core collapse.}
Dav\'e {\em et al.} (2001) demonstrate the effect in numerical
simulations. But Miralda-Escud\'e (2002) points out that the
collisions would tend to make the velocity distribution
isotropic, contrary to the evidence for ellipsoidal distributions
of dark matter in clusters of galaxies. For recent surveys of
the very active debate on these issues see Primack (2002) and
Tasitsiomi (2002); for references to still other possible fixes 
see Dav\'e {\em et al.} (2001). 

Another critical issue traces back to the biasing
picture discussed in Sec.~III.D. If $\OM$ is well below unity
there need not be significant mass in the voids defined by the
large galaxies. But the biasing process still operates, and might 
be expected to cause dwarf or irregular galaxies to trespass
into the voids outlined by the large regular galaxies. This seems to
happen in CDM model simulations to a greater extent than is
observed. Mathis and White (2002) discuss voids in $\Lambda$CDM 
simulations, but do not address the trespassing issue. The  
reader is invited to compare the relative distributions of big
and little galaxies in the simulation in Fig.~1 of Mathis and
White (2002) with the examples of observed distributions
in Figs. 1 and 2 in Peebles (1989b) and in Figs. 1 to 3 in
Peebles (2001). 

The community thought is that the trespassing issue need not be a 
problem for the CDM model: the low mass density in voids
disfavors formation of galaxies from the debris left in these
regions. But we have not seen an explanation of why local upward 
mass fluctuations, of the kind that produce normal galaxies in
populated regions, and appear also in the predicted debris in 
CDM voids, fail to  produce dwarf or irregular void galaxies. An 
easy explanation  
is that the voids contain no matter, having been gravitationally
emptied by the growth of primeval non-Gaussian mass density
fluctuations. The evidence in tests (10) and (11) in Sec.~IV.B is
that the initial conditions are close to Gaussian. But
non-Gaussian initial conditions that reproduce the character of
the galaxy distribution, including suppression of the trespassing
effect, would satisfy test (10) by construction. 

We mention finally the related issues of when the large elliptical
galaxies formed and when they  acquired the central compact
massive objects that are thought to be remnant quasar engines
(Lynden-Bell, 1969).  

In the CDM model large elliptical galaxies form in substantial
numbers at redshift $z<1$. Many astronomers do not see this as a
problem, because ellipticals do tend to contain relatively young
star populations, and some elliptical galaxies have grown by recent
mergers, as predicted in the CDM 
model.\footnote{
The classic merger example is also the 
nearest large elliptical galaxy, Centaurus A (NGC 5128). 
The elliptical image is crossed by a band of gas and dust that 
likely is the result of a merger with one of the spiral galaxies 
in the group around this elliptical. For a thorough
review of what is known about this galaxy see Israel (1998).} 
But prominent merger events are rare, and the young stars seen in
ellipticals generally seem to be a ``frosting'' (Trager {\em et
al.}, 2000) 
of recent star formation on a dominant old star population. The
straightforward reading of the evidence 
assembled in Peebles (2002) is that most of the large ellipticals
are present as assembled galaxies of stars at $z=2$.\footnote{
Papovich, Dickinson, and Ferguson (2002) find evidence that 
the comoving number density of all galaxies with star mass 
greater than $1\times 10^{10}M_\odot$, where $M_\odot$ is the 
mass of the Sun, is significantly less at redshift $z>1$ than now. 
This is at least roughly in line with the distribution of star 
ages in the Milky Way spiral galaxy: the bulge stars are old, 
while the stars in the thin disk have a broad range of ages. 
Thus if this galaxy evolved from $z = 2$ without significant 
growth by mergers its star mass at $z=2$ would be significantly 
less than the present value, which is about $5\times 10^{10}M_\odot$. 
Cimatti {\em et al.} (2002) show that the redshift distribution of 
faint galaxies selected at wavelenght $\lambda\sim 2~\mu$m is not 
inconsistent with the picture that galaxy evolution at $z<2$ is 
dominated by ongoing star formation rather than merging.}
The $\Lambda$CDM model prediction is uncertain because it depends on
the complex processes of star formation that are so difficult to
model. The reading of the situation by Thomas and Kauffmann
(1999) is that the predicted abundance of giant ellipticals at
$z=2$ is less than about one third of what it is now. Deciding
whether the gap between theory and observation can be closed is
not yet straightforward.  

A related issue is the significance of the observations of
quasars at redshift $z\sim 6$. By conventional estimates\footnote{
The quasars discovered in the Sloan  
Digital Sky Survey are discussed by Fan {\em et al.} (2001). If
the quasar radiation is not strongly beamed toward us, its
luminosity translates to an Eddington mass (the mass at which
the gravitational pull on unshielded plasma balances the
radiation pressure) $M_{\rm bh}\sim 10^{9.3}M_\odot$. In a 
present-day elliptical galaxy with this mass in the central 
compact object the line of sight velocity dispersion is 
$\sigma\simeq 350$ km~s$^{-1}$. This is close to the highest 
velocity dispersion observed in low redshift elliptical galaxies. 
For example, in the Faber {\em et al.} (1989) catalog of 500
ellipticals, 15 have $300<\sigma <400$ km~s$^{-1}$, and none has
a larger $\sigma$. From the present-day relation between $\sigma$
and luminosity, an elliptical galaxy with $\sigma = 350$ km~s$^{-1}$ 
has mass $\sim 10^{12.3}M_\odot$ in stars. The dark matter associated
with this many baryons is $M_{\rm DM}\sim 10^{13}M_\odot$. This
is a large mass to assemble at $z\sim 6$, but it helps that such
objects are rare. The present
number density of giant elliptical galaxies with $\sigma >300$
km~s$^{-1}$ is about $10^{-5}$ Mpc$^{-3}$, four
orders of magnitude more than the comoving number density of
quasars detected at $z\sim 6$.} 
these quasars are powered by black holes with masses at the upper end
of the range of masses of the central compact objects --- let us
call them black hole quasar remanants --- in the largest
present-day elliptical galaxies. Here are some options to
consider. First, the high redshift quasars may be in the few
large galaxies that have already formed at $z\sim 6$. Wyithe and
Loeb (2002), following Efstathiou and Rees (1988), show that this 
fits the $\Lambda$CDM model if the quasars  at $z\sim 6$ have black 
hole mass $\sim 10^9M_\odot$ in dark halos with mass 
$\sim 10^{12}M_\odot$. In the $\Lambda$CDM
picture these early galaxies would be considerably denser than
normal galaxies; to be checked is whether they would be rare
enough to be observationally acceptable. Second, the
quasars at $z\sim 6$ may be in more modest star clusters that
later grew by merging into giant ellipticals. To be established
is whether this growth would preserve the remarkably tight
correlation between the central black hole mass and the velocity
dispersion of the stars\footnote{
Ferrarese and Merritt (2000) and Gebhardt {\em et al.} (2000) 
show that the black hole mass correlates with the velocity 
dispersion of the stars in an elliptical galaxy and the velocity 
dispersion of the bulge stars in a spiral galaxy. This is not a 
direct gravitational effect: the black hole mass is less than 
1\%\ of the star mass in the bulge or the elliptical galaxy.},
and whether growth by merging would
produce an acceptable upper bound on black hole masses at the
present epoch. Third, large ellipticals might have grown by
accretion around pre-existing black holes, without a lot of
merging. This is explored by Danese {\em et al.} (2002). 

There does not seem to be a coherent pattern to the present list of 
challenges to the CDM model. The rotation curves of low surface 
brightness galaxies suggest we want to suppress the primeval density
fluctuations on 
small scales, but the observations of what seem to be mature
elliptical galaxies at high redshifts suggest we want to increase
small-scale fluctuations, or maybe postulate non-Gaussian
fluctuations that grow into the central engines for quasars at 
$z\sim 6$. We do not want these central engines to appear in low
surface brightness galaxies, of course. 

It would not be at all surprising if the confusion of challenges
proved to be at least in part due to the difficulty of comparing  
necessarily schematic analytic and numerical model analyses
to the limited and indirect empirical constraints. But it is also
easy to imagine that the CDM model has to be refined because the 
physics of the dark sector of matter and energy is more
complicated than $\Lambda$CDM, and maybe even more complicated
than any of the alternatives now under discussion. Perhaps some of
the structure formation ideas people were considering a decade
ago, which invoke good physics, also will prove to be significant  
factors in relieving the problems with structure formation. And
the important point for our purpose is that we do not know how
the relief might affect the cosmological tests.  

\subsection{The tests}

The literature on the cosmological tests is enormous compared to
what it was just a decade ago, and growing. Our references to
this literature are much sparser than in Sec.~III, on the
principle that no matter how complete the list 
it will be out of date by the time this review is published. For the
same reason, we do not attempt to present the best values of the
cosmological parameters based on their joint fit to the full
suite of present measurements. The situation will continue to
evolve as the measurements improve, and the state of the art is
best followed on astro-ph. We do take it to be our assignment to
consider what the tests are testing, and to assess the directions
the results seem to be leading us. The latter causes us to return
many times to two results that seem secure because they are so
well checked by independent lines of evidence, as follows.  

First, at the present state of the tests, optically selected 
galaxies are useful mass tracers. By that we mean the assumption 
that visible galaxies
trace mass does not seriously degrade the accuracy of analyses of
the observations. This will change as the measurements improve,
of course, but the case is good enough now that we suspect the
evidence will continue to be that optically selected galaxies are 
good indicators of where most of the mass is at the present
epoch. Second, the mass density in matter is significantly less
than the critical Einstein-de Sitter value. The case is
compelling because it is supported by so many different
lines of evidence (as summarized in Sec. IV.C). Each could be 
compromised by systematic error,
to be sure, but it seems quite unlikely the evidence could be so 
consistent yet  misleading. A judgement of the range of likely
values of the mass density is more difficult. Our estimate,
based on the measurements we most trust, is 
\begin{equation}
    0.15 \la \OM \la 0.4 , 
\label{eq:om}
\end{equation}
and we would put the central value at $\OM\simeq 0.25$. The
spread is meant in the sense of two standard deviations: we would
be surprised to find $\OM$ is outside this range. 

Several other policy decisions should be noted. The first is that
we do not comment on tests that have been considered but not yet
applied in a substantial campaign of measurements. A widely
discussed example is the Alcock and Paczy\'nski (1979) comparison
of the apparent depth and width of a system from its 
angular size and depth in redshift. 

In analyses of the tests of
models for evolving dark energy 
density, simplicity recommends the XCDM parametrization with a
single constant parameter $w_{\rm X}$, as is demonstrated by 
the large number of recent papers on this approach. But the more
complete physics recommends the scalar field model with an
inverse power-law potential. This includes the response of the
spatial distribution of the dark energy to the peculiar
gravitational field. Thus our comments on variable dark energy
density are more heavily weighted to the scalar field model than
is the case in the recent literature. 

The gravitational deflection of light appears not only as a
tool in cosmological tests, as gravitational lensing, but also as  
a source of systematic error. The gravitational
deflections caused by mass concentrations magnify the image of a
galaxy along a line of sight where the mass density is larger than 
the average, and reduce the solid angle of the image when the mass
density along the line of sight is low. 
The observed energy flux density is proportional to the solid
angle (because the surface brightness, erg cm$^{-2}$ s$^{-1}$
ster$^{-1}$ Hz$^{-1}$, is conserved at fixed redshift). 
Selection can be biased either way, by the magnification effect 
or by obscuration by the dust that tends to accompany
mass.\footnote{This was recognized by  Zel'dovich (1964), R.
Feynman, in 1964, and S. Refsdal, in 1965.
Feynman's comments in a colloquium are noted
by Gunn (1967). Peebles attended Refsdal's lecture at the
International Conference on General Relativity and Gravitation,
London, July 1965; Refsdal (1970) mentions the lecture.} 
When the tests are more precise we will have to correct them for 
these biases, through models for 
the mass distribution (as in Premadi {\em et al.}, 2001), and the 
measurements of the associated gravitational shear of
the shapes of the galaxy images. But the biases seem to be
small and will not be discussed here.  

And finally, as the cosmological tests improve a satisfactory
application will require a joint fit of all of the parameters to
all of the relevant measurements and constraints. 
%% a prototypical example is discussed at the end of test (12). 
Until recently it made sense to 
impose prior conditions, most famously the hope that if the
universe is not well described by the Einstein-de Sitter model
then surely it is the case either that $\Lambda$ is negligibly 
small or else that space curvature may be neglected. We
suspect the majority in the community still expect this is true,
on the basis of the coincidences argument in Sec.~II.B.2, but it
will be important to see what comes out of joint fits of both
$\OM$ and $\OL$, as well as all the other parameters, as is
becoming the current practice. Our test-by-test discussion is
useful for sorting out the physics and astronomy, we believe; it
is not the prototype for the coming generations of precision
application of the tests.  

Our remarks are ordered by our estimates of the model dependence.

\subsubsection{The thermal cosmic microwave background radiation} 

We are in a sea of radiation with spectrum  
very close to Planck at $T=2.73$~K, and isotropic to one part in
$10^5$ (after correction for a dipole term that usually is
interpreted as the result of our motion relative to the rest
frame defined by the radiation).\footnote{
The history of the discovery and measurement of this radiation,
and its relation to the light element abundances in test (2), is
presented in Peebles (1971, pp. 121-9 and 240-1), 
Wilkinson and Peebles (1990), 
and Alpher and Herman (2001). The precision spectrum measurements 
are summarized in Halpern, Gush, and Wishnow (1991) and Fixsen 
{\em et al.} (1996).} 
The thermal spectrum indicates thermal relaxation, for which the 
optical depth has to be large on the scale of the Hubble length 
$H_0^{-1}$. We know space now is close to transparent at the 
wavelengths of this 
radiation, because radio galaxies are observed at high redshift.
Thus the universe has to have expanded from a state quite
different from now, when it was hotter, denser, and optically
thick. This is strong evidence our universe is evolving.

This interpretation depends on, and checks, conventional local
physics with a single metric description of spacetime. Under
these assumptions the expansion of the universe preserves the
thermal spectrum and cools the temperature as\footnote{
To see this, recall the normal modes argument used to get
Eq.~(\ref{eq:z}). The occupation number in a normal mode with
wavelength $\lambda$ at temperature $T$ is the Planck form 
${\cal N} = [e^{\hbar c/k T\lambda} - 1]^{-1}$.
Adiabaticity says ${\cal N}$ is constant. Since the mode
wavelength varies as $\lambda\propto a(t)$, where $a$ is the
expansion factor in Eq.~(\ref{eq:scaling}), and ${\cal N}$ is
close to constant, the mode temperature
varies as $T\propto 1/a(t)$. Since the same
temperature scaling applies to each mode, an initially
thermal sea of radiation remains thermal in the absence of
interactions. We do not know the provenance of this argument; it
was familiar in Dicke's group when the 3~K cosmic microwave
background radiation was discovered.} 
\begin{equation}
   T\propto (1+z).
\end{equation}
Bahcall and Wolf (1968) point out that one can test this
temperature-redshift relation by measurements of the
excitation temperatures of fine-structure absorption line systems
in gas clouds along quasar lines of sight. The corrections for
excitations by collisions and the local radiation field are
subtle, however, and perhaps not yet fully sorted out (as
discussed by Molaro {\em et al.}, 2002, and references therein).

The 3~K thermal cosmic background radiation is a centerpiece of
modern cosmology, but its existence does not test general
relativity. 

\subsubsection{Light element abundances} 

The best evidence that the expansion and cooling of the universe
traces back to high redshift is the success of the standard model
for the origin of 
deuterium and isotopes of helium and lithium, by 
reactions among radiation, leptons, and atomic nuclei as the 
universe expands and cools through temperature $T\sim 1$~MeV at 
redshift $z\sim 10^{10}$. The free parameter in the
standard model is the present baryon number density. The model
assumes the baryons are uniformly distributed at high redshift,
so this parameter with the known present radiation temperature
fixes the baryon number density as a function of temperature and
the temperature as a function of time. The latter follows from
the expansion rate Eq.~(\ref{eq:f2}), which at the epoch of
light element formation may be written as 
\begin{equation}
   {\left(\dot a\over a\right)^2} = {8\over 3}\pi G\rho _r,
\label{eq:bbn}
\end{equation}
where the mass density $\rho _r$ counts radiation, which is
now at $T=2.73$~K, the associated neutrinos, and
$e^\pm$ pairs. The curvature and $\Lambda$ terms
are unimportant, unless the dark energy mass density varies
quite rapidly.  

Independent analyses of the fit to the measured element abundances,
corrected for synthesis and destruction in stars, by
Burles, Nollett, and Turner (2001), and Cyburt, Fields, and Olive
(2001), indicate
\begin{equation}
   0.018\leq \OB h^2 \leq 0.022, \hbox{ and } 
   0.006\leq \OB h^2 \leq 0.017, 
\label{eq:ob}
\end{equation}
both at 95\%\ confidence limits. Other analyses by Coc 
{\em et al.} (2002) and Thuan and Izotov (2002) result in ranges 
that lie between the two of Eq.~(\ref{eq:ob}). The difference in 
values may be a useful indication of remaining uncertainties; it
is mostly a consequence of the choice of isotopes used to
constrain $\OB h^2$. 
Burles {\em et al.} (2001) use the deuterium abundance, Cyburt
{\em et al.} (2001) favor the helium and lithium measurements,
and the other two groups use other combinations of abundances. 
Equation.~(\ref{eq:ob}) is consistent with the summary range,  
$0.0095\leq \OB h^2 \leq 0.023$ at 95\%\ confidence, of Fields 
and Sarkar (2002).

The baryons observed at low redshift, in stars and gas, amount to
(Fukugita, Hogan, and Peebles, 1998)
\begin{equation}
   \OB \sim 0.01.
\label{eq:fhp}
\end{equation}
It is plausible that the difference between Eqs.~(\ref{eq:ob})
and~(\ref{eq:fhp}) is in cool plasma, with temperature $T\sim 100$~eV, 
in groups of galaxies. It is difficult to observationally constrain the
idea that there is a good deal more cool plasma in the large voids 
between the concentrations of galaxies. A more indirect but eventually 
more precise constraint on $\OB$, from the anisotropy of the 3~K thermal 
cosmic microwave background radiation, is discussed in test~(11). 

It is easy to imagine complications, such as inhomogeneous entropy 
per baryon, or in the physics of neutrinos; examples may be traced back
through Abazajian, Fuller, and Patel (2001) and Giovannini, Keih\"anen, 
and Kurki-Suonio (2002). It seems difficult
to imagine that a more complicated theory would reproduce the 
successful predictions of the simple model, but Nature fools 
us on occasion. Thus before concluding that the
theory of the pre-stellar light element abundances is known,
apart from the addition of decimal places to the cross sections,
it is best to wait and see what advances in the physics of 
baryogenesis and of neutrinos teach us.  

How is general relativity probed? The only part of the
computation that depends specifically on this theory is
the pressure term in the active gravitational mass density, in
the expansion rate equation~(\ref{eq:f1}). If we did not have general
relativity, a simple Newtonian picture might have led us to write
down $\ddot a/a = -4\pi G\rho _r/3$ instead of Eq.~(\ref{eq:f1}). 
With $\rho _r\sim 1/a^4$, as appropriate since most of the mass
is fully relativistic at the redshifts of light element production,
this would predict the expansion time $a/\dot a$ is $2^{1/2}$
times the standard expression (that from Eq.~[\ref{eq:bbn}]). The larger
expansion time would hold the neutron to proton number density
ratio close to that at thermal equilibrium, $n/p = e^{-Q/kT}$, where $Q$
is the difference between the neutron and proton masses, to lower 
temperature. It would also allow more time for free decay of the
neutrons after thermal equilibrium is broken. 
Both effects decrease the final $^4$He abundance. The factor
$2^{1/2}$ increase in expansion time would reduce the helium
abundance by mass to $Y\sim 0.20$. This is significantly less
than what is observed in objects with the lowest heavy element
abundances, and so seems to be ruled out (Steigman, 2002).\footnote{
There is a long history of discussions of this probe of the 
expansion rate at the redshifts of light element production. 
The reduction of the helium abundance to $Y\sim 0.2$ if the 
expansion time is increased by the factor $2^{1/2}$ is seen 
in Figs.~1 and~2 in Peebles (1966). Dicke (1968) introduced
the constraint on evolution of the strength of the gravitational 
interaction; see Uzan (2002) for a recent review. 
The effect of the number of neutrino families on the expansion
rate and hence the helium abundance is noted by Hoyle and Tayler
(1964) and Shvartsman (1969). Steigman, Schramm, and Gunn (1977)
discuss the importance of this effect as a test of cosmology and
of the particle physics measures of the number of neutrino
families.} That is, we have positive evidence for the
relativistic expression for the active gravitational mass density
at redshift $z\sim 10^{10}$, a striking result. 

\subsubsection{Expansion times} 

The predicted time of expansion from the very early universe to 
redshift $z$ is 
\begin{equation}
t(z) = \int {da\over\dot a} 
= H_0^{-1}\int _z^\infty {dz\over (1+z)E(z)},
\label{eq:aitcht}
\end{equation}
where $E(z)$ is defined in Eq.~(\ref{eq:f2}). 
If $\Lambda =0$ the present age is $t_0 < H_0^{-1}$. In the
Einstein-de Sitter model the present age is $t_0 = 2/(3H_0)$. 
If the dark energy density is significant and evolving, we may
write $\rho_\Lambda = \rho_{\Lambda 0} f(z)$, where the function
of redshift is normalized to $f(0)=1$. Then $E(z)$ generalizes to  
\begin{equation}
   E(z) = \left(\OM (1+z)^3 + \OR (1+z)^4 + \OK (1+z)^2
   + \OL f(z)\right) ^{1/2}.
   \label{eq:eofz}
\end{equation}
In the XCDM parametrization with constant $w_{\rm X}$
(Eq.~[\ref{eq:xcdm}]), 
$f(z) = (1+z)^{3(1+w_{\rm X})}$. Olson and Jordan (1987) 
present the earliest discussion we have found of $H_0t_0$ 
in this picture (before it got the name).
In scalar field models, $f(z)$ generally must be evaluated
numerically; examples are in Peebles and Ratra (1988). 

The relativistic correction to the active gravitational mass
density (Eq.~[\ref{eq:f1}]) is not important at the redshifts at 
which galaxies can be observed and the ages of their star
populations estimated. At moderately high redshift, where the 
nonrelativistic matter term dominates, Eq.~(\ref{eq:aitcht}) is 
approximately 
\begin{equation}
   t(z) \simeq {2\over 3H_0\OM^{1/2}}(1+z)^{-3/2}.
   \label{eq:simt}
\end{equation}
That is, the ages of star populations at high redshift are an
interesting probe of $\OM$ but they are not
very sensitive to space curvature or to a near constant dark
energy density.\footnote{The predicted maximum age of star
populations in galaxies at redshifts $z \ga 1$ does still depend
on $\OL$ and $\OK$, and there is the advantage that the predicted
maximum age is a lot shorter than today. This variant of the
expansion time test is discussed by 
Nolan {\em et al.} (2001), Lima and Alcaniz (2001), and 
references therein.}

Recent analyses of the ages of old stars\footnote{
See Carretta {\em et al.} (2000), Krauss and Chaboyer (2001), 
Chaboyer and Krauss (2002), and references therein.}
indicate the expansion time is in the range 
\begin{equation}
   11\hbox{ Gyr} \la t_0 \la 17\hbox{ Gyr},
\end{equation}
at 95\%\ confidence, with central value $t_0 \simeq 13$ Gyr.
Following Krauss and Chaboyer (2001) these numbers add
0.8~Gyr to the star ages, under the assumption star formation
commenced no earlier than $z=6$ (Eq.~[\ref{eq:simt}]). A naive
addition in  
quadrature to the uncertainty in $H_0$ (Eq.~[\ref{eq:Ho}])
indicates the dimensionless age parameter is in the range
\begin{equation}
   0.72 \la H_0 t_0 \la 1.17 , 
\label{eq:Hoto}
\end{equation}
at 95\% confidence, with central value $H_0 t_0 \simeq 0.89$.
The uncertainty here is dominated by that in $t_0$. In the
spatially-flat $\Lambda$CDM model ($\OK = 0$), Eq.~(\ref{eq:Hoto})
translates to $0.15 \la \OM \la 0.8$, with central 
value $\OM \simeq 0.4$. In the open model with $\OL = 0$, the
constraint is $\OM \la 0.6$ with 
the central value $\OM \simeq 0.1$. In the inverse 
power-law scalar field dark energy case (Sec. II.C) with power-law index
$\alpha = 4$, the constraint is $0.05 \la \OM \la 0.8$.

We should pause to admire the unification of the theory and
measurements of 
stellar evolution in our galaxy, which yield the estimate of
$t_0$, and the measurements of the extragalactic distance scale,
which yield $H_0$, in the product in Eq.~(\ref{eq:Hoto}) that
agrees with the relativistic cosmology with dimensionless
parameters in the range now under discussion. As we indicated in 
Sec. III, there is a long history of discussion of the expansion
time as a constraint on cosmological models. The measurements now
are tantalizingly close to a check of consistency with the values
of $\OM$ and $\OL$ indicated by other cosmological tests. 
 
\subsubsection{The redshift-angular size and redshift-magnitude
relations}

An object at redshift $z$ with physical length $l$ perpendicular
to the line of sight subtends angle $\theta$ such that
\begin{equation}
   l = a(t)r(z)\theta = a_0 r(z)\theta /(1+z) ,
\label{eq:lofz}
\end{equation}
where $a_0 = a(t_0)$.
The angular size distance $r(z)$ is the coordinate position of
the object in the first line element in Eq.~(\ref{eq:rw}), with the
observer placed at the origin. The condition that light moves from
source to observer on a radial null geodesic is 
\begin{equation}
   \int _0^{r(z)}{dr\over\sqrt{1 + K r^2}} =\int {dt\over a(t)}, 
\end{equation}
which gives
\begin{equation}
   H_0 a_0 r(z) = {1\over\sqrt{\OK}}\sinh \left(\sqrt{\OK}\int_0^z
	{dz\over E(z)}\right), 
 \label{eq:rofz}
\end{equation}
where $E(z)$ is defined in Eqs.~(\ref{eq:f2})
and~(\ref{eq:eofz}).

In the Einstein-de Sitter model, the angular-size-redshift
relation is 
\begin{equation}
 \theta = {H_0l\over 2}{(1+z)^{3/2}\over\sqrt{1+z} - 1}.
 \label{eq:thedes}
\end{equation}
At $z\ll 1$, $\theta = H_0l/z$, consistent with the Hubble
redshift-distance relation. At $z\gg 1$ the image is
magnified,\footnote
{The earliest discussion we know of the
magnification effect is by Hoyle (1959). In the coordinate system
in Eq.~(\ref{eq:rw}), with the observer at the origin, light
rays from the object move to the observer along straight radial
lines. An image at high redshift is magnified because the light 
detected by the observer is emitted when the proper distance to 
the object measured at fixed world time is small. Because the proper
distance between the object and source is increasing faster than 
the speed of light, emitted light directed at the observer
is initially moving away from the observer.}
$\theta\propto 1+z$.

The relation between the luminosity of a galaxy and the energy
flux density received by an observer follows from Liouville's
theorem: the observed energy flux $i_{\nu_0}$ per unit time, area, solid
angle, and frequency satisfies
\begin{equation}
   i_{\nu_0} \delta\nu_0 = 
   i_{\nu _e}\delta \nu _e /(1+z)^4,
\label{eq:liou}
\end{equation}
with $i_{\nu_e}$ the emitted energy flux (surface brightness) 
at the source and $\delta\nu_e = \delta\nu_0 (1+z)$ the bandwidth 
at the source at redshift $z$. The redshift factor $(1+z)^4$ 
appears for the same reason as in the 3~K cosmic microwave 
background radiation energy density. With Eq.~(\ref{eq:lofz}) to 
fix the solid angle, Eq.~(\ref{eq:liou}) says 
the observed energy flux per unit area, time, and
frequency from a galaxy at redshift $z$ that 
has luminosity $L_{\nu_e}$ per frequency interval measured at the
source is  
\begin{equation}
   f_{\nu_0}= {L_{\nu _e}\over 4\pi a_0^2r(z)^2(1+z)}.
\label{eq:mofz}
\end{equation}

In conventional local physics with a single metric theory the 
redshift-angular size (Eq.~[\ref{eq:lofz}]) and
redshift-magnitude (Eq.~[\ref{eq:mofz}]) relations are physically
equivalent.\footnote{For a review of measurements of the
redshift-magnitude relation (and other cosmological tests) we 
recommend Sandage (1988). 
A recent application to the most luminous galaxies in
clusters is in Arag\'on-Salamanca, Baugh, and Kauffmann (1998).
The redshift-angular size relation is measured by Daly and Guerra
(2001) for radio galaxies, Buchalter {\em et al.} (1998) for
quasars, and Gurvits, Kellermann, and Frey (1999) for compact 
radio sources. Constraints on the cosmological parameters from
the Gurvits {\em et al.} data are discussed by 
Vishwakarma (2001), Lima and Alcaniz (2002), 
Chen and Ratra (2003), and references therein, and constraints
based on the radio galaxy data are discussed by
Daly and Guerra (2001), Podariu {\em et al.} (2003), and
references therein.}

The best present measurement of the redshift-magnitude relation 
uses supernovae of Type Ia.\footnote{
These supernovae are characterized by the absence of hydrogen lines 
in the spectra; they are thought to be the result of explosive nuclear 
burning of white dwarf stars.  
Pskovskii (1977) and Phillips (1993) pioneered the reduction of
the supernovae luminosities to a near universal standard candle.
For recent discussions of their use as a cosmological test see 
Goobar and Perlmutter (1995), Reiss {\em et al.}
(1998), Perlmutter {\em et al.} (1999a), Gott {\em et al.} (2001), and
Leibundgut (2001). We recommend Leibundgut's (2001)
cautionary discussion of astrophysical
uncertainties: the unknown nature of the trigger for
the nuclear burning, the possibility that the Phillips
correction to a fiducial luminosity actually depends on redshift
or environment within a galaxy, and possible obscuration by
intergalactic dust. There are also issues of physics that may
affect this test (and others): the 
strengths of the gravitational or electromagnetic interactions
may vary with time, and photon-axion conversion may reduce the 
number of photons reaching us. All of this is under active study.}  
The results are inconsistent with the Einstein-de
Sitter model, at enough standard deviations to make it clear 
that unless there is something quite substantially and
unexpectedly wrong with the measurements the Einstein-de Sitter
model is ruled out. The data 
require $\Lambda > 0$ at two to three standard deviations, 
depending on the choice of data and method of analysis (Leibundgut, 
2001; Gott {\em et al.}, 2001). The spatially-flat case with 
$\OM$ in the range of Eq.~(\ref{eq:om}) is a good fit for
constant $\Lambda$. The current data do not provide interesting
constraints on the models for evolving dark energy
density.\footnote{
Podariu and Ratra (2000) and Waga and Frieman (2000) discuss the 
redshift-magnitude relation in the inverse power-law scalar field
model, and Waga and Frieman (2000) and Ng and Wiltshire (2001)
discuss this relation in the massive scalar field model.}
Perlmutter {\em et al.} (http://snap.lbl.gov/) show that a
tighter constraint, from supernovae observations to redshift $z\sim 2$,
by the proposed SNAP satellite, is feasible and capable of giving
a significant detection of $\Lambda$ and maybe its 
evolution.\footnote{ 
Podariu, Nugent, and Ratra (2001), Weller and Albrecht (2002), Wang 
and Lovelace (2001), Gerke and Efstathiou (2002), Eriksson and Amanullah
(2002), and references therein, discuss constraints on cosmological
parameters from the proposed SNAP mission.}

\subsubsection{Galaxy counts}

Counts of galaxies --- or of other objects whose number density
as a function of redshift may be modeled --- probe the volume
element $(dV/dz)dz\delta\Omega$ defined by a solid angle
$\delta\Omega$ in the sky and a 
redshift interval $dz$. The volume is fixed by the angular size
distance (Eq.~[\ref{eq:lofz}]), which determines the area subtended
by the solid angle, in combination with the redshift-time relation
(Eq.~[\ref{eq:aitcht}]), which fixes the radial distance belonging
to the redshift interval. 

Sandage (1961a) and Brown and Tinsley (1974) showed that with the
technology then available galaxy counts are not a very sensitive
probe of the cosmological parameters. Loh and
Spillar (1986) opened the modern exploration of the galaxy 
count-redshift relation at redshifts near unity, where the
predicted counts are quite different in models with and without a 
cosmological constant (as illustrated in Figure 13.8 in Peebles,
1993).

The interpretation of galaxy counts requires an understanding of
the evolution of galaxy luminosities and the gain and loss of
galaxies by merging. Here is an example of the former in a
spatially-flat cosmological model with $\OM = 0.25$. The
expansion time from high redshift is $t_3=2.4$~Gyr at redshift 
$z=3$ and $t_0=15$~Gyr now. Consider a galaxy observed at $z = 3$.
Suppose the bulk of the stars in this galaxy formed at time $t_f$,
and the population then aged and faded without significant later
star formation. Then if $t_f \ll t_3$ the ratio of the observed
luminosity at $z=3$ to its present luminosity would be 
(Tinsley, 1972; Worthey, 1994)
\begin{equation}
  L_3/L_0\simeq (t_0/t_3)^{0.8}\simeq 4.
  \label{eq:passiveevolution}
\end{equation}
If $t_f$ were larger, but still less than $t_3$, this ratio
would be larger. If $t_f$ were greater than $t_3$ the galaxy
would not be seen, absent earlier generations of stars. 
In a more realistic picture significant
star formation may be distributed over a considerable range of
redshifts, and the effect on the typical galaxy luminosity at a 
given redshift accordingly more complicated. Since there are many
more galaxies with low luminosities than galaxies with
high luminosities, one has to know the luminosity evolution quite
well for a meaningful comparison of galaxy counts at high and low
redshifts. The present situation is illustrated by the rather 
different indications from studies by Phillipps {\em et al.}
(2000) and Totani {\em et al.} (2001). 

The understanding of galaxy evolution and the interpretation of 
galaxy counts will be improved by large samples of counts of
galaxies as a function of color, apparent magnitude, and
redshift. Newman and Davis (2000) point to a promising
alternative: count galaxies as a function of the internal velocity
dispersion that in spirals correlates with the dispersion in the
dark matter  halo. That could eliminate the need to understand
the evolution of star populations. There is still the issue of
evolution of the dark halos by merging and accretion, but that 
might be reliably modeled by numerical simulations within the CDM 
picture. Either way, with further work galaxy counts may provide
an important test for dark energy and its evolution (Newman and
Davis, 2000; Huterer and Turner, 2001; Podariu and Ratra, 2001).  

\subsubsection{The gravitational lensing rate} 

The probability of production of multiple images of a quasar or 
a radio source by
gravitational lensing by a foreground galaxy, or of strongly
lensed images of a galaxy by a foreground cluster of galaxies,
adds the relativistic expression for the deflection of light to
the physics of the homogeneous cosmological model. Fukugita,
Futamase, and Kasai (1990) and Turner (1990) point out the value
of this test: at small $\OM$ the predicted lensing rate is
considerably larger in a flat model with $\Lambda$ than in an 
open model with $\Lambda =0$ (as illustrated in Fig~13.12 in 
Peebles, 1993).  

The measurement problem for the analysis of quasar lensing
is that quasars that are not lensed are not magnified by
lensing, making them harder to find and the correction for
completeness of detection harder to establish. Present estimates
(Falco, Kochanek, and Mu\~noz, 1998; Helbig {\em et al.}, 1999) 
do not seriously constrain $\OM$ in an open model, and
in a flat model ($\OK = 0$) suggest $\OM > 0.36$ at
$2\sigma$. This is close to the upper bound in Eq.~(\ref{eq:om}). 
Earlier indications that the lensing rate in a flat model with
constant $\Lambda$ requires a larger value of $\OM$ than is
suggested by galaxy dynamics led Ratra and Quillen (1992) 
and Waga and Frieman (2000) to investigate the inverse power-law 
potential dark energy scalar field case. They showed this 
can significantly lower the predicted lensing rate at $\OK =0$
and  small $\OM$. The lensing rate still
is too uncertain to draw conclusions on this point, but advances
in the measurement certainly will be followed with interest. 

The main problem in the interpretation of the rate of strong
lensing of galaxies by foreground clusters as a cosmological test
is the sensitivity of the lensing cross section to the mass
distribution within the cluster (Wu and Hammer, 1993); for the
present still somewhat uncertain state of the art see Cooray
(1999) and references therein. 

\subsubsection{Dynamics and the mean mass density}

Estimates of the mean mass density from the relation
between the mass distribution and the resulting peculiar
velocities,\footnote{Early
estimates of the mean mass density, by 
Hubble (1936, p. 189) and Oort (1958), combine the galaxy number
density from galaxy counts with estimates of galaxy masses from
the internal motions of gas and stars. Hubble (1936, p. 180) was
quite aware that this misses mass between the 
galaxies, and that the motions of galaxies within clusters
suggests there is a lot more intergalactic mass (Zwicky, 1933;
Smith, 1936). For a recent review of this subject see
Bahcall {\em et. al.} (2000).}  and from the gravitational
deflection of light, probe 
gravity physics and constrain $\OM$. The former is not sensitive
to $\OK$, $\OL$, or the dynamics of the dark energy, the
latter only through the angular size distances. 

We begin with the redshift space of observed galaxy angular
positions and redshift distances $z/H_0$ in the radial
direction. The redshift $z$ has a contribution from the radial
peculiar  
velocity, which is a probe of the gravitational acceleration
produced by the inhomogeneous mass distribution.   
The two-point correlation function, $\xi_v$, in redshift space is 
defined by the probability that a randomly chosen galaxy has a
neighbor at distance $r_\parallel$ along the line of sight in
redshift space and perpendicular distance $r_\perp$, 
\begin{equation}
   dP = n\left( 1 + \xi _v(r_\parallel ,r_\perp )\right)
   dr_\parallel d^2r_\perp ,
\label{eq:xiv}
\end{equation}
where $n$ is the galaxy number density.
This is the usual definition of a reduced correlation function.
Peculiar velocities make the function anisotropic. On
small scales the random relative 
peculiar velocities of the galaxies broaden $\xi _v$ along the
line of sight. On large scales the streaming peculiar velocity of 
convergence to gravitationally growing mass concentrations
flattens $\xi _v$ along the line of sight.\footnote{
This approach grew out of the statistical method introduced by
Geller and Peebles (1973); it is derived in its present form
in Peebles (1980b) and first applied to a serious redshift sample 
in Davis and Peebles (1983b). These references give the theory 
for the second moment 
$\sigma ^2$ of $\xi _v$ in the radial direction --- the mean
square relative peculiar velocity --- in the small-scale stable 
clustering limit. The analysis of the anisotropy of $\xi _v$ in
the linear perturbation theory of large-scale flows
(Eq.~[\ref{eq:linearpt}]) is presented in Kaiser (1987).} 

At $10\hbox{ kpc}\la hr_\perp\la 1$~Mpc the measured
line-of-sight broadening is prominent, and indicates the
one-dimensional relative velocity
dispersion is close to independent of $r_\perp$ at 
$\sigma\sim 300$ km~s$^{-1}$.\footnote{
This measurement requires close attention to clusters that 
contribute little to the mean mass density but a broad and 
difficult to measure tail to the distribution of relative velocities. 
Details may be traced back through Padilla {\em et al.} (2001), 
Peacock {\em et al.} (2001), and Landy (2002).} 
This is about what would be expected if the mass two-and three-point
correlation functions were well approximated by the galaxy
correlation functions, the mass clustering on these scales were 
close to statistical equilibrium, and the density parameter were
in the range of Eq.~(\ref{eq:om}). 

We have a check from the motions of the galaxies in and around
the Local Group of galaxies, where the absolute errors in the 
measurements of
galaxy distances are least. The two largest group members are the 
Andromeda Nebula (M~31) and our Milky Way galaxy. If they contain most 
of the mass their relative motion is the classical two-body
problem in Newtonian mechanics (with minor corrections for
$\Lambda$, mass accretion at low redshifts, and the tidal torques
from neighboring galaxies). The two galaxies are 
separated by 800~kpc and approaching at 110 km~s$^{-1}$. In the  
minimum mass solution the galaxies have completed just over 
half an orbit in the cosmological expansion time 
$t_0\sim 10^{10}$~yr. By this argument Kahn and Woltjer (1959) 
find the sum of masses of the two galaxies has to be an order of 
magnitude larger than what is seen in the luminous parts. An 
extension to the analysis of the motions and distances of the 
galaxies within 4~Mpc distance from us, and
taking account of the gravitational effects of the galaxies out
to 20~Mpc distance, gives masses quite similar to what Kahn and
Woltjer found, and consistent with $\OM$ in the range of
Eq.~(\ref{eq:om}) (Peebles {\em et al.}, 2001). 

We have another check from weak lensing: the shear distortion of 
images of distant galaxies by the gravitational deflection by
the inhomogeneous mass distribution.\footnote{
Recent studies include Wilson, Kaiser, and Luppino (2001), 
Van Waerbeke {\em et al.} (2002), Refregier, Rhodes, and Groth (2002),
Bacon {\em et al.} (2002), and Hoekstra, Yee, and Gladders (2002).
See Munshi and Wang (2002) and references therein for discussions 
of how weak lensing might probe dark energy.} 
If galaxies trace mass these measurements say the
matter density parameter measured on scales from about 1~Mpc to
10~Mpc is in the range of Eq.~(\ref{eq:om}). It
will be interesting to see whether these measurements can check
the factor of two difference between the relativistic
gravitational deflection of light and the naive Newtonian
deflection angle. 

The redshift space correlation function 
$\xi _v$ (Eq.~[\ref{eq:xiv}]) is measured 
well enough at $hr_\perp\sim 10$~Mpc to demonstrate the
flattening effect, again consistent with $\OM$ in the range of
Eq.~(\ref{eq:om}), if galaxies trace mass. Similar numbers follow
from galaxies selected as far infrared IRAS sources 
(Tadros {\em et al.}, 1999) and from optically selected galaxies 
(Padilla {\em et al.}, 2001; Peacock {\em et al.}, 2001).
The same physics, applied to estimates of the mean relative
peculiar velocity of galaxies at separations $\sim 10$~Mpc,
yet again indicates a similar density parameter (Juszkiewicz {\em
et al.}, 2000). 

Other methods of analysis of the distributions of astronomical
objects and peculiar velocities smoothed over scales $\ga 10$~Mpc
give a variety of results for the mass density, some above the 
range in Eq.~(\ref{eq:om}),\footnote{
The methods and results may be traced through Fisher, Scharf, 
and Lahav (1994), Sigad {\em et al.} (1998), and Branchini
{\em et al.} (2000).}
others towards the bottom end of the range (Branchini {\em et al.}, 
2001). The measurement of $\OM$ from large-scale streaming
velocities thus remains open. But we are impressed by an apparently 
simple local situation, the peculiar motion of the Local Group 
toward the Virgo cluster of galaxies.
This is the nearest known large mass concentration, 
at distance $\sim 20$~Mpc. Burstein (2000) 
finds that our virgocentric velocity is $v_v=220$ km~s$^{-1}$, 
indicating $\OM\simeq 0.2$ (Davis and Peebles, 1983a, Fig.~1). 
This leads us to conclude that the weight of the evidence from
dynamics on scales $\sim 10$~Mpc favors low $\OM$, in the range
of Eq.~(\ref{eq:om}).
 
None of these measurements is precise. But many have been under 
discussion for a long time and seem to us to be reliably
understood. Weak lensing is new, but the measurements are 
checked by several independent groups. The result, in our
opinion, is a well checked and believable network of evidence
that over two decades of well-sampled length scales, 100~kpc to
10~Mpc, the apparent value of $\OM$ is constant to a factor 
of three or so, in the range $0.15\la\OM\la 0.4$. The key
point for the purpose of this review is that this result is
contrary to what might have been expected from biasing, or from a
failure of the inverse square law (as will be discussed in test
[13]). 

\subsubsection{The baryon mass fraction in clusters of
galaxies} 

Abell (1958) made the first useful catalog of the rich clusters 
considered here and in the next test. A typical value of the
Abell cluster mass 
within the Abell radius $r_a=1.5h^{-1}$~Mpc is 
$3\times 10^{14}h^{-1}M_\odot$. The cluster masses are reliably
measured (within Newtonian gravity) from consistent results from
the velocities of the galaxies, the pressure of the intracluster
plasma, and the gravitational deflection of light from background
galaxies.  

White (1992) and White {\em et al.} (1993) point
out that rich clusters likely are large enough to contain
a close to fair sample of baryons and dark matter, meaning the 
ratio of baryonic to total mass in a cluster is a good measure of  
$\OB /\OM$. With $\OB$ from the model for light elements
(Eqs.~[\ref{eq:ob}]), this gives a measure of the mean mass  
density. The baryon mass fraction in clusters is still under
discussion.\footnote
{See Hradecky {\em et al.} (2000), Roussel, Sadat, and Blanchard 
(2000), Allen, Schmidt, and Fabian (2002), and references therein.} 
We adopt as the most direct and so maybe most reliable approach 
the measurement of the baryonic gas mass fraction of clusters, 
$f_{\rm gas}$, through the Sunyaev-Zel'dovich microwave decrement 
caused by Thomson-Compton scattering of cosmic microwave background 
radiation by the intracluster plasma. The Carlstrom {\em et al.} 
(2001) value for $f_{\rm gas}$ gives $\OM\sim 0.25$,\footnote{
This assumes $\OB h^2 = 0.014$ from Eqs.~(\ref{eq:ob}). For the full
range of values in Eqs.~(\ref{eq:Ho}) and~(\ref{eq:ob}), $0.1 \la
\OM \la 0.4$ at two standard deviations.}
in the range of Eq.~(\ref{eq:om}). This test does not directly constrain
$\OK$, $\OL$, or the dynamics of the dark energy. 

\subsubsection{The cluster mass function}

In the CDM model rich clusters of galaxies grow out of the rare
peak upward fluctuations in the primeval Gaussian mass
distribution. Within this model one can adjust the amplitude
of the mass fluctuations to match the abundance of clusters at
one epoch. In the Einstein-de Sitter model it is difficult to see
how this one free adjustment can account for the abundance of
rich clusters now and at redshifts near unity.\footnote{Early
discussions of this problem include Evrard (1989), Peebles,
Daly, and Juszkiewicz (1989), and Oukbir and Blanchard (1992).} 

Most authors now agree that the low density flat $\Lambda$CDM  
model can give a reasonable fit to the cluster abundances
as a function of redshift. The constraint on $\OM$ from the
present cluster abundance still is under discussion, but
generally is found to be close to $\OM\sim 0.3$ if galaxies 
trace mass.\footnote{
For recent discussions see Pierpaoli, Scott, and White (2001),
Seljak (2001), Viana, Nichol, and Liddle (2002), Ikebe {\em et al.}
(2002), Bahcall {\em et al.} (2002), and references therein. Wang 
and Steinhardt (1998) consider this test in the context of the 
XCDM parametrization; to our knowledge it 
has not been studied in the scalar field dark energy case.}
The constraint from the evolution of the cluster number density also 
is under discussion.\footnote{
Examples include Blanchard {\em et al.} (2000), and Borgani 
{\em et al.} (2001).}
The predicted evolution is slower in
a lower density universe, and at given $\OM$ the evolution is
slower in an open model with $\Lambda =0$ than in a 
spatially-flat model with $\Lambda$ (for the reasons discussed in
Sec.~III.D). Bahcall and Fan (1998) emphasize that we have good
evidence for the presence of some massive clusters at $z\sim 1$,
and that this is exceedingly difficult to understand in the  
CDM model in the Einstein-de Sitter cosmology (when biasing is 
adjusted to get a reasonable present number density). 
Low density models with or without $\Lambda$ can account for the
existence of some massive clusters at high redshift.
Distinguishing between the predictions of the spatially curved
and flat low density cases awaits better measurements. 

\subsubsection{Biasing and the development of nonlinear mass
density fluctuations}  

Elements of the physics of cluster formation in test (9) appear
in this test of the early stages in the nonlinear growth of 
departures from homogeneity. An initially
Gaussian mass distribution becomes skew as low density
fluctuations start to bottom out and high density fluctuations
start to develop into prominent mass peaks. The early 
signature of this nonlinear evolution is the disconnected
three-point mass autocorrelation function, 
$\langle\delta (\vec x,t)\delta (\vec y,t) \delta (\vec z,t)\rangle$, 
where $\delta (\vec x,t)=\delta\rho /\rho$ is the dimensionless
mass contrast. If galaxies are useful mass tracers the galaxy
three-point function is a good measure of this mass function. 

The form for the mass three-point function, for Gaussian initial
conditions at high redshift, in lowest nonzero order in
perturbation theory, is worked out in Fry (1984), and Fry (1994)
makes the point that measurements of the galaxy three-point
function test how well galaxies trace mass.\footnote{
Other notable contributions to the development of this point include
Bernardeau and Schaeffer (1992), Fry and Gazta\~naga (1993),
and Hivon {\em et al.} (1995).}
There are now two sets of measurements of the galaxy three-point
function on scales $\sim 10$ to 20~Mpc, where the
density fluctuations are not far from Gaussian. One uses  
infrared-selected IRAS galaxies,\footnote{
Two sub-samples of IRAS 
galaxies are analyzed by Scoccimarro {\em et al.} (2001) and by Feldman 
{\em et al.} (2001).} 
the other optically-selected galaxies (Verde {\em et al.}, 2002). 
The latter is consistent with the
perturbative computation of the mass three-point function for
Gaussian initial conditions. The former says infrared-selected
galaxies are adequate mass tracers apart from the densest
regions, which IRAS galaxies avoid. That has a simple
interpretation in astrophysics: galaxies in dense
regions tend to be swept clear of the gas and dust that make
galaxies luminous in the infrared. 

This test gives evidence of consistency of three ideas: 
galaxies are useful mass tracers on scales $\sim 10$~Mpc, the
initial conditions are close to Gaussian, and conventional
gravity physics gives an adequate description of this aspect of
the growth of structure. It is in principle sensitive to $\OL$,
through the suppression of the growth of small departures from
homogeneity at low redshift, but the effect is small. 

\subsubsection{The anisotropy of the cosmic microwave background
radiation} 

The wonderfully successful CDM prediction of the power spectrum
of the angular distribution of the temperature of the 3~K
cosmic microwave background radiation has converted many of the 
remaining skeptics in the
cosmology community to the belief that the CDM model likely does
capture important elements of reality. 

Efstathiou (2002) provides a useful measure of the information in
the present measurements\footnote{
Recent measurements are presented in Lee {\em et al.} (2001), 
Netterfield {\em et al.} (2002), Halverson {\em et al.} (2002), 
Miller {\em et al.} (2002a), Coble {\em et al.} (2001), Scott 
{\em et al.} (2002), and Mason {\em et al.} (2002).}:
the fit to the CDM model significantly  
constrains three linear combinations of the free parameters. 
We shall present three sets of considerations that roughly follow
Efstathiou's constraints. We begin with reviews of the standard
measure of the temperature anisotropy and of the conditions at
redshift $z\sim 1000$ that are thought to produce the observed
anisotropy.  

The 3~K cosmic microwave background temperature $T(\theta ,\phi)$ 
as a function of position in the sky usually is expressed as an 
expansion in spherical harmonics, 
\begin{equation}
   \delta T(\theta ,\phi )  = T(\theta ,\phi ) -\langle T\rangle 
   = \sum _{l,m} a_l^nY_l^m(\theta ,\phi ).
\end{equation}
The square of $\delta T$ averaged over the sky is
\begin{equation}
   \langle\delta T^2\rangle = {1\over 4\pi }\sum _l
   (2l+1)\langle |a_l^m|\rangle ^2,
\end{equation}
where $|a_l^m|^{2}$ is statistically independent of $m$. This may be 
rewritten as 
\begin{equation}
   \langle\delta T^2\rangle = \sum _l{1\over l}\delta T_l^2,
   \qquad\delta T_l^2 = {l(2l+1)\over 4\pi }\langle |a_l^m|^2\rangle .
\label{eq:Tvariance}
\end{equation}
Since $\sum l^{-1}$ is close to $\int d\ln l$, $\delta T_l^2$ is
the variance of the temperature per logarithmic interval of $l$.
A measure of the angular scale belonging to the multipole index 
$l$ is that the minimum distance between 
zeros of the spherical harmonic $Y_l^m$, in longitude or 
latitude, is $\theta =\pi/l$, except close to the poles, where
$Y_l^m$ approaches zero.\footnote{
A more careful analysis
distinguishes averages across the sky from ensemble averages. By
historical accident the conventional normalization replaces $2l+1$
with $2(l+1)$ in Eq.~(\ref{eq:Tvariance}). Kosowsky (2002)
reviews the physics of the polarization of the radiation.}

Now let us consider the main elements of the physics that
determines the 3~K cosmic microwave background anisotropy.\footnote{
The physics is worked out in Peebles and Yu (1970) and Peebles (1982). 
Important analytic considerations are in Sunyaev and Zel'dovich (1970). 
The relation of the cosmic microwave background anisotropy to the
cosmological parameters is explored in many papers; examples of
the development of ideas include Bond (1988), Bond {\em et al.}
(1994), Hu and Sugiyama (1996), Ratra {\em et al.} (1997, 1999), 
Zaldarriaga, Spergel, and Seljak (1997), and references therein.} 
At redshift $z_{\rm dec}\sim 1000$ the
temperature reaches the critical value at which the primeval
plasma combines to atomic hydrogen (and slightly earlier to
neutral helium). This removes the coupling between baryons and
radiation by Thomson scattering, leaving the radiation to
propagate nearly freely (apart from residual gravitational
perturbations). Ratios of mass densities near the epoch $z_{\rm dec}$
when matter and radiation decouple are worth 
noting. At redshift $z_{\rm eq}=2.4\times 10^4\OM h^2$ the mass
density in matter --- including the baryonic and nonbaryonic
components --- is equal to the relativistic mass density in
radiation and neutrinos assumed to have low masses. At decoupling
the ratio of mass densities is 
\begin{equation}
   {\rho_{\rm M}(z_{\rm dec}) \over \rho_{\rm R}(z_{\rm dec})} 
   = {z_{\rm eq}\over z_{\rm dec}}
   \sim 20\OM h^2 \sim 2,
\end{equation}
at the central values of the parameters in Eqs.~(\ref{eq:Ho}) 
and~(\ref{eq:om}). The ratio of mass densities in baryons and in 
thermal cosmic microwave background radiation --- not counting 
neutrinos --- is 
\begin{equation}
  {\rho _{\rm B}(z_{\rm dec})\over\rho _{\rm CBR}(z_{\rm dec})} 
  = {4\times 10^4\OB h^2 \over z_{\rm dec}}
  \sim 0.5.
\end{equation}
That is, the baryons and radiation decouple just as the expansion
rate has become dominated by nonrelativistic matter and the
baryons are starting to lower the velocity of sound in the
coupled baryon-radiation fluid (presenting us with still more
cosmic coincidences). 

The acoustic peaks in the spectrum of angular fluctuations of the
3~K cosmic microwave background radiation come from the Fourier 
modes of the coupled
baryon-radiation fluid that have reached maximum or minimum
amplitude at decoupling. Since all Fourier components start
at zero amplitude at high redshift --- in the growing density
perturbation mode --- this condition is 
\begin{equation}
   \int _0^{t_{\rm dec}}kc_sdt/a \simeq n\pi /2,
   \label{eq:eigen}
\end{equation}
where $c_s$ is the velocity of sound in the baryon-radiation
fluid. Before decoupling the mass density in radiation is greater
than that of the baryons, so the velocity of sound is close 
to $c/\sqrt{3}$. The proper wavelength at the first acoustic peak
thus is  
\begin{equation}
   \lambda _{\rm peak}\sim t_{\rm dec}\propto h^{-1}\OM ^{-1/2}.
   \label{eq:peak}
\end{equation}
The parameter dependence comes from Eq.~(\ref{eq:simt}).
The observed angle subtended by $\lambda _{\rm peak}$ is
set by the angular size distance $r$ computed from $z_{\rm eq}$
to the present (Eq.~[\ref{eq:rofz}]). If
$\OK =0$ or $\OL =0$ the angular size distance is 
\begin{equation}
   H_0 a_0 r \simeq 2\OM ^{-m}, \quad m = 1\ \hbox{if}\ \OL =0, 
   \quad m \simeq 0.4\ \hbox{if}\ \OK =0.
\end{equation} 
If $\Lambda =0$ this expression is analytic at large $z_{\rm eq}$.
The expression for $\OK =0$ is a reasonable approximation to the
numerical solution. So the angular scale of the peak varies with
the matter density parameter as 
\begin{eqnarray}
   \theta _{\rm peak}  \sim  z_{\rm dec}\lambda _{\rm peak}/a_0 r 
     & \propto & \OM ^{1/2} \ \hbox{ if }\ \OL =0, \label{eq:openpeak}\\
     & \propto & \OM ^{-0.1} \ \hbox{ if }\ \OK =0. \label{eq:flatpeak}
\end{eqnarray}

The key point from these considerations is that the angle defined
by the first peak in the fluctuation power spectrum is sensitive
to $\OM$ if $\Lambda =0$ (Eq.~[\ref{eq:openpeak}]), but not
if $\OK =0$ (Eq.~[\ref{eq:flatpeak}]).\footnote{
This ``geometrical degeneracy" is discussed by Efstathiou and Bond 
(1999). Marriage (2002) presents a closer analysis of the effect.
Sugiyama and Gouda (1992), Kamionkowski, Spergel, and Sugiyama 
(1994b), and Kamionkowski {\em et al.} (1994a) are early discussions
of the cosmic microwave background anisotropy in an open model.} 
We have ignored the sensitivity of $z_{\rm dec}$ and $t_{\rm dec}$ 
to $\OM$, but the effect is weak. More detailed computations, which 
are needed for a precise comparison with the data, show that the 
CDM model predicts that the first and largest peak of $\delta T_l$ 
appears at multipole index $l_{\rm peak}\simeq 220\OM ^{-1/2}$ if 
$\Lambda =0$, and at $l_{\rm peak}\simeq 220$ if $\OK =0$ and $0.1 \la 
\OM \la 1$.\footnote{
Brax {\em et al.} (2000) and Baccigalupi {\em et al.} (2000) compute
the angular spectrum of the cosmic microwave background anisotropy in the 
dark energy scalar field model. Doran {\em et al.} (2001) discuss the angular
scale of the peaks in this case, and Corasaniti and Copeland (2002),
Baccigalupi {\em et al.} (2002), and references therein, compare 
model predictions and observations --- it is too early to draw profound
conclusions about model viability, and new data are eagerly anticipated.
Wasserman (2002) notes that the cosmic microwave background anisotropy data 
could help discriminate between different dark energy scalar field models 
whose predictions do not differ significantly at low redshift.}  

The measured spectrum\footnote{
For analyses see Knox and Page (2000), Podariu {\em et al.} 
(2001), Wang, Tegmark, and Zaldarriaga (2002), Durrer,
Novosyadlyj, and  Apunevych (2002), Miller {\em et al.} (2002b), and
references therein.} peaks at $\delta T_l\sim 80~\mu$K at 
$l\sim 200$, thus requiring small space curvature in the CDM
model. This is the first of Efstathiou's constraints. Because of
the geometric degeneracy this measurement does not yet seriously
constrain $\OM$ if $\OK =0$.  

The second constraint comes from the spectrum of temperature
fluctuations on large scales, $l\la 30$, where pressure gradient
forces never were very important. Under the scale-invariant initial 
conditions discussed in Sec.~III.C the Einstein-de Sitter model
predicts $\delta T_l$ is nearly independent of 
$l$ on large scales. A spatially-flat model with $\OM\sim 0.3$,
predicts $\delta T_l$ decreases slowly with increasing $l$
at small $l$.\footnote{
The physics was first demonstrated by Sachs and Wolfe (1967) and
applied in the modern context by Peebles (1982). The intermediate
Sachs-Wolfe effect that applies if the universe is  
not Einstein-de Sitter is shown in Eq.~(93.26) in
Peebles (1980). This part of the Sachs-Wolfe effect receives a
contribution from  
the low redshift matter distribution, so cross-correlating the 
observed large-scale cosmic microwave background anisotropy with the
low redshift matter distribution could provide another test of the
world model (Boughn and Crittenden, 2001, and references therein).} 
The measured spectrum is close to flat at 
$\delta T_l\sim 30~\mu$K, but not well enough constrained for a
useful measure of the parameters $\OM$ and $\OL$.\footnote{
See, e.g., G\'orski {\em et al.} (1998). This ignores the ``low" value
of the cosmological quadrupole ($ l = 2 $) moment, whose value depends
on the model used to remove foreground Galactic emission (see, e.g., Kogut
{\em et al.}, 1996). Contamination due to non-cosmic microwave background
emission is an issue for some of the anisotropy data sets (see, e.g., 
de Oliveira-Costa {\em et al.}, 1998; Hamilton and Ganga, 2001; Mukherjee
{\em et al.}, 2002, and references therein). Other issues that need care
in such analyses include accounting for the uncertainty in the calibration 
of the experiment (see, e.g., Ganga {\em et al.}, 1997; Bridle {\em et al.}, 
2002), and accounting for the shape of the antenna pattern (see, e.g., Wu
{\em et al.}, 2001a; Souradeep and Ratra, 2001; Fosalba, Dore, and Bouchet,
2002).}
Because of the simplicity of the physics on large angular scales, this  
provides the most direct and so perhaps most reliable normalization of the
CDM model power spectrum (that is, the parameter $A$ in 
Eqs.~[\ref{eq:scaleinvariantpk}] and~[\ref{eq:exponentialV}]). 

The third constraint is the baryon mass density. It affects the
speed of sound $c_s$ (Eq.~[\ref{eq:eigen}]) in the 
baryon-radiation fluid prior to decoupling, and the mean free
path for the radiation at $z\sim z_{\rm dec}$. These in turn
affect the predicted sequence of acoustic peaks 
(see, e.g., Hu and Sugiyama, 1996). The detected peaks are consistent with 
a value for the baryon density parameter $\OB$ in a range that
includes what is derived from the light elements abundances
(Eqs.~[\ref{eq:ob}]).\footnote{ 
The $\OB h^2$ values estimated from the cosmic microwave background
anisotropy measured by Netterfield {\em et al.} (2002), Pryke {\em
et al.} (2002), and Stompor {\em et al.} (2001), are more consistent
with the higher, deuterium based, Burles {\em et al.} (2001) range
in Eqs.~(\ref{eq:ob}).}
This impressive check may be much improved by the measurements 
of $\delta T_l$ in progress.   

The measurements of $\delta T_l$ are consistent with a near
scale-invariant power spectrum (Eq.~[\ref{eq:exponentialV}] 
with $n \simeq 1$) with negligible contribution from gravity 
wave or isocurvature fluctuations (Sec.~III.C.1). 
The 3~K cosmic microwave background temperature fluctuations  
show no departure from a Gaussian random process.\footnote{
Colley, Gott, and Park (1996) present an early discussion of the 
situation on large angular scales; more recent discussions are in 
Mukherjee, Hobson, and Lasenby (2000), Phillips and Kogut (2001), and
Komatsu {\em et al.} (2002). Degree and sub-degree angular scale 
anisotropy data are studied in Park {\em et al.} (2001), Wu {\em et al.}
(2001b), Shandarin {\em et al.} (2002), and Polenta {\em et al.} (2002).} 
This agrees with the picture in test (10) for the
nonlinear growth of structure out of Gaussian initial mass
density fluctuations. 

The interpretation of the cosmic microwave background temperature 
anisotropy measurements assumes and tests general relativity and 
the CDM model. One can write down other models for 
structure formation that put the peak of $\delta T_l$ at about
the observed angular scale --- an example is Hu and Peebles 
(2000) --- but we have seen none so far that seem likely to fit 
the present measurements of $\delta T_l$. Delayed recombination
of the primeval plasma in an low density $\Lambda =0$ CDM model
can shift the peak of $\delta T_l$ to the observed
scale.\footnote{The model in Peebles, Seager, and Hu (2000)
assumes stellar ionizing radiation at $z\sim 1000$ produces
recombination  Lyman~$\alpha$ photons. These resonance photons
promote photoionization from the $n=2$ level of atomic hydrogen.
That allows delayed recombination with a rapid transition to
neutral atomic hydrogen, as required to get the shape of $\delta
T_l$ about right.} The physics is valid, but the scenario 
is speculative and arguably quite improbable. On the other hand,
we cannot be sure a fix of the challenges to CDM reviewed in Sec.
IV.A.2 will not affect our assessments of such issues, and hence
of this cosmological test. 

\subsubsection{The mass autocorrelation function and nonbaryonic
matter} 

If the bulk of the nonrelativistic matter, with density parameter 
$\OM\sim 0.25$, were baryonic, then under adiabatic initial
conditions the most immediate problem would be the strong
dissipation of primeval mass density fluctuations on the scale of
galaxies by diffusion of radiation through the baryons at
redshifts near decoupling.\footnote{
Early analyses of this effect are in Peebles (1965), and Silk 
(1967, 1968).}
Galaxies could form by fragmentation of the 
first generation of protocluster ``pancakes,'' as Zel'dovich
(1978) proposed, but this picture is seriously challenged by the
evidence that the galaxies formed before clusters of
galaxies.\footnote{
For example, our Milky Way galaxy is in the Local Group, which 
seems to be just forming, because the time for a group member
to cross the Local Group is comparable 
to the Hubble time. The Local Group is on the outskirts of the 
concentration of galaxies around the Virgo cluster. We and
neighboring galaxies are moving away from the cluster, but
at about 80 percent of the mean Hubble flow, as if the local
mass concentration were slowing the local expansion. That is, 
our galaxy, which is old, is starting to cluster with
other galaxies, in a ``bottom up'' hierarchical growth of
structure, as opposed to the ``top down'' evolution of the pancake
picture.} 
In a baryonic dark matter model we could accommodate the
observations of 
galaxies already present at $z\sim 3$ by tilting the primeval
mass fluctuation spectrum to favor large fluctuations on small 
scales, but that would mess up the cosmic microwave background
anisotropy. The search for isocurvature initial
conditions that might fit both in a baryonic dark matter model
has borne no fruit so far (Peebles, 1987).

The most important point of this test is 
the great difficulty of reconciling the power spectra of matter 
and radiation without the postulate of nonbaryonic dark matter.  
The CDM model allows hierarchical growth of structure, from
galaxies up, which is what seems to be observed, because the
nonbaryonic dark matter interacts with 
baryons and radiation only by gravity; the dark matter
distribution is not smoothed by the dissipation of density
fluctuations in the baryon-radiation fluid at redshifts 
$z\ga z_{\rm eq}$. 

As discussed in Sec.~III.D, in the CDM model the small scale part
of the dark matter power spectrum bends from the primeval
scale-invariant form  $P(k) \propto k$ to $P(k) \propto k^{-3}$, and
the characteristic length at the break scales inversely with
$\OM$ (Eq.~[\ref{eq:lambdabreak}]). Evidence of such a break in
the galaxy power spectrum $P_g(k)$ has been known for more than a 
decade\footnote{
The first good evidence is discussed in Efstathiou {\em et al.} (1990); 
for recent examples see Sutherland {\em et al.} (1999), 
Percival {\em et al.} (2001), and Dodelson {\em et al.} (2002).};
it is consistent with a value of $\OM$ in the range of
Eq.~(\ref{eq:om}). 

\subsubsection{The gravitational inverse square law}

The inverse square law for gravity determines the relation
between the mass distribution and the gravitationally-driven
peculiar velocities that enter estimates of the  
matter density parameter $\OM$. The peculiar velocities also
figure in the evolution of the mass distribution, and hence the
relation between the present mass fluctuation spectrum and the
spectrum of cosmic microwave background temperature fluctuations 
imprinted at redshift 
$z\sim 1000$. We are starting to see demanding tests of
both aspects of the inverse square law.

We have a reasonably well checked set of measurements of the 
apparent value of $\OM$ on scales ranging from 
100~kpc to 10~Mpc (as reviewed under test [7]). Most agree with a 
constant value of the apparent $\OM$, within a factor of three or
so. This is not the precision one would like, but the subject
has been under discussion for a long time, and, we believe, is
now pretty reliably understood, within the factor of three or so.
If galaxies were biased tracers of mass  
one might have expected to have seen that $\OM$ increases
with increasing length scale, as the increasing scale includes
the outer parts of extended massive halos. Maybe that is 
masked by a gravitational force law that decreases more rapidly
than the inverse square law at large distance. But the much more 
straightforward reading is that the slow variation of $\OM$
sampled over two orders of magnitude in length scale agrees with
the evidence from tests~(7) to (10) that galaxies are useful mass
tracers, and that the inverse square law therefore is a useful
approximation on these scales.

The toy model in Eq.~(\ref{eq:nutty}) illustrates how a failure
of the inverse square law would affect the evolution of 
the shape of the mass fluctuation power spectrum $P(k,t)$ as a
function of the comoving wavenumber $k$, in linear perturbation 
theory. This is tested by the measurements of the mass and 
cosmic microwave background temperature fluctuation power spectra. 
The galaxy power spectrum $P_g(k)$ varies 
with wavenumber at $k\sim 0.1h$~Mpc$^{-1}$ about as expected
under the assumptions that the mass distribution grew by gravity
out of adiabatic scale-invariant initial conditions, the mass is
dominated by dark matter that does not suffer radiation drag at
high redshift, the galaxies are useful
tracers of the present mass distribution, the matter density
parameter is $\OM\sim 0.3$, and, of course, the evolution is
adequately described by conventional physics (Hamilton and
Tegmark, 2002, and references therein). If the inverse
square law were significantly wrong at $k\sim 0.1h$ Mpc$^{-1}$,
the near scale-invariant form would have to be an accidental
effect of some failure in this rather long list of assumptions.
This seems unlikely, but a check certainly is desirable. We
have one, from the cosmic microwave background anisotropy  
measurements. They also are consistent with near scale-invariant  
initial conditions applied at redshift $z\sim 1000$. This
preliminary check on the effect of the gravitational inverse
square law applied on cosmological length scales and back to 
redshift $z\sim 1000$ will be improved by better understanding of the 
effect on $\delta T_l$ of primeval tensor perturbations to
spacetime, and of the dynamical response of the dark energy
distribution to the large-scale mass distribution.   

Another aspect of this check is the comparison of values of the
large-scale rms fluctuations in the present distributions of mass 
and the cosmic microwave background radiation. The
latter is largely set at decoupling, after which the former grows
by a factor of about $10^3$ to the present epoch, in the standard
relativistic cosmological model. If space curvature is negligible the
growth factor agrees with the observations to 
about 30\%, assuming galaxies trace mass.   
In a low density universe with $\Lambda =0$ the standard model
requires that mass is more smoothly distributed than galaxies, 
$\delta N/N\sim 3\delta M/M$, or that the
gravitational growth factor since decoupling is a factor of three
off the predicted factor $\sim 1000$; this factor of three is about 
as large a deviation from unity as is viable. We are not proposing this
interpretation of the data, rather we are impressed by the modest size 
of the allowed adjustment to the inverse square law.

\subsection{The state of the cosmological tests}

Precision cosmology is not very interesting if it is based on
faulty physics or astronomy. That is why we have emphasized the
tests of the standard gravity physics and structure formation
model, and the checks of consistency of measures based on
different aspects of the astronomy. 

There are now five main lines of evidence that significantly
constrain the value of $\OM$ to the range of Eq.~(\ref{eq:om}):
the redshift-magnitude   
relation (test~[4]), gravitational dynamics and weak lensing (test
[7]), the baryon mass fraction in clusters of galaxies (test
[8]), the abundance of clusters as a function of mass and redshift
(test~[9]), and the large-scale galaxy distribution (test~[12]).
There are indications for larger values of 
$\OM$, from analyses of the rate of strong lensing of
quasars by foreground galaxies (test [6]) and some analyses of
large-scale flows (test [7]), though we know of no well-developed
line of evidence that points to the Einstein-de Sitter value 
$\OM =1$.  
Each of these measures of $\OM$ may suffer from systematic
errors: we must bear in mind the tantalus principle
mentioned in Sec. I.A, and we have to remember that the
interpretations could be corrupted by a failure of standard
physics. But the general pattern of results from a  considerable
variety of independent approaches seems so close to consistent as
to be persuasive. Thus we conclude that there is a well-checked
scientific case for the proposition that the measures of the mean
mass density of matter in forms capable of clustering are physically 
meaningful, and that the mass density parameter almost certainly 
is in the range $0.15\la\OM\la 0.4$. 

In the standard cosmology the masses of the galaxies are
dominated by dark matter, with mass density parameter
$\ODM \sim 0.2$, that is not baryonic (or acts that way).
We do not have the direct evidence of a 
laboratory detection; this is based on two indirect lines of 
argument. First, the successful model for the  
origin of the light elements (test~[2]) requires baryon density
$\OB \sim 0.05$. It is 
difficult to see how to reconcile a mass density this small with
the mass estimates from dynamics and lensing; the
hypothesis that $\OM$ is dominated by matter that is not baryonic
allows us to account for the difference.
Second, the
nonbaryonic matter allows us to reconcile the theory of the 
anisotropy of the cosmic microwave background radiation with 
the distributions of galaxies and groups and clusters of galaxies, 
and the presence of galaxies at $z\sim 3$ (tests [11] and [12]). 
This interpretation requires a value for $\OB$ that is in line
with test~(2). The consistency is impressive. But
the case is not yet as convincing as the larger network of
evidence that $\OM$ is well below unity.  

The subject of this review is Einstein's cosmological constant
$\Lambda$, or its equivalent in dark energy. The evidence for
detection of $\Lambda$ by the redshift-magnitude relation for
type Ia supernovae is checked by the angular distribution of the
3~K cosmic microwave background temperature together with the 
constraints on $\OM$.
This certainly makes a serious case for dark energy. But we keep
accounts by the number of significant independent checks, 
and by this reckoning the case is not yet as strong as for
nonbaryonic dark matter.     

\section{Concluding remarks}

We cannot demonstrate that there is not some other physics,
applied to some other cosmology, that equally well agrees with
the cosmological tests. The same applies to the whole enterprise
of physical science, of course. Parts of physics are so densely
checked as to be quite convincing approximations to the way the
world really is. The web of tests is much less dense in
cosmology, but, we have tried to demonstrate, by no means
negligible, and growing tighter.  

A decade ago there was not much discussion of how to test
general relativity theory on the scales of cosmology. That was in
part because the theory seems so logically compelling, and
certainly in part also because there was not much evidence to work
with. The empirical situation is much better now. We mentioned
two tests, of the relativistic active gravitational 
mass density and of the gravitational inverse square 
law. The consistency of constraints on $\OM$ from dynamics and
the redshift-magnitude relation adds a test of the effect of
space curvature on the expansion rate. These tests are
developing; they will be greatly improved by work in progress. 
There certainly may be surprises in the gravity physics of
cosmology at redshifts $z\la 10^{10}$, but it already is clear
that if so the surprises are subtly hidden.  

A decade ago it was not at all clear which direction the theory
of large-scale structure formation would take. Now the simple CDM
model has proved to be successful enough that there is good
reason to expect the standard model ten years from now will
resemble CDM. We have listed challenges to this structure
formation model. Some may well be only apparent, a result of the 
complications of interpreting the theory and observations. Others
may prove to be real and drive adjustments of the model.
Fixes certainly will include one element from the ideas of
structure formation that were under debate a decade ago:
explosions that rearrange matter in ways that are difficult to
compute. Fixes may 
also include primeval isocurvature departures from homogeneity,
as in spacetime curvature fluctuations frozen in during
inflation, and maybe in new cosmic fields. It would not be 
surprising if cosmic field defects, that have such a good
pedigree from particle physics, also find a role in structure
formation. And a central point to bear in mind is that fixes,
which do not seem unlikely, could affect the cosmological tests.   

A decade ago we had significant results from the cosmological tests. 
For example, estimates of the product $H_0t_0$ suggested we might
need positive $\Lambda$, though the precision was not quite
adequate for a convincing case. That still is so; the
community will be watching for further advances. We had pretty
good constraints on $\OB$ from the theory of the origin of the
light elements. The abundance measurements are improving; an 
important recent development is the detection of deuterium 
in gas clouds at redshifts $z\sim 3$. Ten years ago we had useful
estimates of masses from peculiar motions on relatively small
scales, but more mixed messages from larger scales. The story
seems more coherent now, but there still is room for improved
consistency. We had a 
case for nonbaryonic dark matter, from the constraint on $\OB$
and from the CDM model for structure formation. The case is
tighter now, most notably due to the successful fit of the CDM
model prediction to the measurements of the power spectrum of the 
anisotropy of the 3~K cosmic microwave background radiation.  
In 1990 there were believable observations of galaxies identified
as radio sources at $z\sim 3$. Now the distributions of galaxies
and the intergalactic medium at $z\sim 3$ are mapped out in
impressive detail, and we are seeing the development of a
semi-empirical picture of galaxy formation and evolution. Perhaps
that will lead us back to the old dream of using galaxies as
markers for the cosmological tests.  

Until recently it made sense to consider the constraints on one
or two of the parameters of the cosmology while holding all the
rest at ``reasonable'' values. That helps us understand what the  
measurements are probing; it is the path we have followed in
Sec.~IV.B. But the modern and very sensible trend is to consider
joint fits of large numbers of parameters to the full suite of  
observations.\footnote{
Examples are Cole {\em et al.} (1997), Jenkins {\em et al.} (1998), 
Lineweaver (2001), and Percival {\em et al.} (2002).} 
This includes a measure of the
biasing or antibiasing of the distribution of galaxies relative
to mass, rather than our qualitative argument that one usefully
approximates the other. In a fully satisfactory cosmological test
the parameter set will also include parametrized departures from
standard physics extrapolated to the scales of cosmology. 

Community responses to advances in empirical evidence are not
always close to linear. The popularity of the Einstein-de Sitter
model continued longer into the 1990s than seems logical to us,
and the switch to the now standard $\Lambda$CDM cosmological model 
--- with flat space sections, nonbaryonic cold dark matter, and 
dark energy --- arguably is more abrupt than warranted 
by the advances in the evidence. Our review leads us to conclude
that there is now a good scientific case that the matter density
parameter is $\OM \sim 0.25$, and a pretty good case that
about three quarters of that is not baryonic. The cases for dark 
energy and for the $\Lambda$CDM model are significant, too,
though beclouded by observational issues of whether we have an
adequate picture of structure formation. But we expect the rapid
advances in the observations of structure formation will soon
dissipate these clouds, and, considering the record, likely
reveal new clouds on the standard model for cosmology a decade
from now. 

A decade ago the high energy physics community had a well-defined
challenge: show why the dark energy density vanishes. Now there
seems to be a new challenge and clue: show why the dark energy  
density is exceedingly small but not zero. The present
state of ideas can 
be compared to the state of research on structure formation a
decade ago: in both situations there are many lines of thought
but not a clear picture of which is the best direction to take.
The big difference is that a decade ago we could be reasonably
sure that observations in progress would guide us to a better
understanding of how structure formed. Untangling the physics  
of dark matter and dark energy and their role in gravity physics is a
much more subtle challenge, but, we may hope, will also be guided
by advances in the exploration of the phenomenology.  
Perhaps in another ten years that will include detection of
evolution of the dark energy, and maybe detection of the
gravitational response of the dark energy distribution to the
large-scale mass distribution. There may be three unrelated
phenomena to deal with: dark energy, dark matter, and a vanishing
sum of zero-point energies and whatever goes with them. Or
the phenomena may be related. Because our only evidence of
dark matter and dark energy is from their gravity it is a natural 
and efficient first step to suppose their properties are as 
simple as allowed by the phenomenology. But it makes sense to
watch for hints of more complex physics within the dark sector. 

The past eight decades have seen steady advances in the
technology of application of the cosmological tests, from
telescopes to computers; advances in the
theoretical concepts underlying the tests; and progress through
the learning curves on how to apply the concepts and technology. 
We see the results: the basis for cosmology is much firmer
than a decade ago. And the basis surely will be a lot more solid
a decade from now. 

Einstein's cosmological constant, and the modern variant, dark
energy, have figured in a broad range of topics in physics and
astronomy that have been under discussion, in at least some
circles, much of the time for the past eight decades. Many of
these issues undoubtedly have been discovered more than once. But
in our experience such ideas tend to persist for a long time at 
low visibility and sometimes low fidelity. Thus the community has
been very well prepared for the present evidence for detection of
dark energy. And for the same reason we believe that dark
energy, whether constant, or rolling toward zero, or maybe even
increasing, still will be an active topic of research, in at
least some circles, a decade from now, whatever the outcome of
the present work on the cosmological tests. Though this much is
clear, we see no basis for a prediction of whether the standard 
cosmology a decade from now will be a straightforward elaboration 
of $\Lambda$CDM, or whether there will be more substantial
changes of direction. 

\section*{Acknowledgments}

We are indebted to Pia Mukherjee, Michael Peskin, and Larry Weaver
for detailed comments on drafts of this review. We thank Uwe 
Thumm for help in translating and discussing papers written in
German. We have also benefited from discussions with Neta Bahcall, 
Robert Caldwell, Gang Chen, Andrea Cimatti, Mark Dickinson, Michael Dine, 
Masataka Fukugita, Salman Habib, David Hogg, Avi Loeb, Stacy McGaugh, 
Paul Schechter, Chris Smeenk, Gary Steigman, Ed Turner, Michael Turner, 
Jean-Philippe Uzan, David Weinberg, and Simon White. BR acknowledges 
support from NSF CAREER grant AST-9875031, and PJEP acknowledges support 
in part by the NSF.   

\section{Appendix: Recent dark energy scalar field research} 

Many dark energy models are characterized by attractor or
tracker behavior. The goal also is to design the model so that
the field energy density is subdominant at high redshift, where
it is not wanted, and dominant at low redshift, where that is
what seems to be observed.  

In the simplest such scalar field models the action has
a conventionally normalized scalar field kinetic and spatial
gradient term, and the real scalar field is coupled only to
itself and gravity. Then the scalar field part of the model is
fully characterized by  
the scalar field potential (along with some broad constraints on
the initial conditions for the field, if the attractor behavior is
realized). The inverse power law potential model is 
described in Secs.~II.C and III.E, and used in some of the
cosmological tests in Sec.~IV. Here we list other
scalar field potentials now 
under discussion for these minimal dark energy models,
modifications of the kinetic part of the action, possible
guidance from high energy particle physics ideas, and constraints
on these ideas from cosmological observations in the context of
dark energy models. 

To those of us not active in this field the models may
seem baroque in their complexity, but that may be the way Nature
is. And as we accumulate more and better data it will be possible
to test and constrain these models.

This is an active area of research, with frequent introduction of
new models, so our discussion must be somewhat sketchy.\footnote{
In particular, we do not discuss variable mass term models, complex or
multiple scalar field models, or repulsive matter. We also omit 
non-scalar field aspects of brane models, Kaluza-Klein models, bimetric 
theories of gravitation, quantum mechanical running of the cosmological
constant, the Chaplygin gas, and the superstring tachyon.}
We limit discussion to a cosmological model with space
sections that are flat thanks to the presence of the dark energy
density.  
    
As mentioned in Sec~III.E, the simplest exponential potential scalar 
field model is unacceptable because it cannot produce the wanted
transition from sub-dominant to dominant energy density. Ratra
and Peebles (1988) consider more complex potentials, such as
powers of linear combinations  
of exponentials of the field $\Phi$. Related models are under active 
investigation. These include potentials that are powers of 
cosh($\Phi$) and/or sinh($\Phi$).\footnote{
For examples see Chimento and Jakubi (1996), Starobinsky (1998),
Kruger and Norbury (2000), Di Pietro and Demaret (2001),
Ure\~na-L\'opez and Matos (2000), Gonz\'alez-D\'{\i}az (2000), 
and Johri (2001).}
Sahni and Wang (2000) present a detailed discussion of a specific example,
$V(\Phi) \propto ({\rm cosh}\lambda\Phi - 1)^p$, where $\lambda$ and $p$ are
constants, emphasizing that this potential interpolates from $V \propto
e^{-p\lambda\Phi}$ to $V \propto (\lambda\Phi)^{2p}$ as $|\lambda\Phi|$
increases, thus preserving some of the desirable properties of
the simplest exponential potential case. 
De la Macorra and Piccinelli (2000) consider potentials that are 
exponentials of more complicated functions of $\Phi$, such as $\Phi^2$
and $e^\Phi$. Skordis and Albrecht (2002) discuss a model with 
$V(\Phi) \propto [1 + (\Phi -A)^2] {\rm exp}[-\Phi\sqrt{q/2}]$, while
Dodelson, Kaplinghat, and Stewart (2000) study a potential 
$V(\Phi) \propto [1 + A\sin(B\Phi)] {\rm exp}[-\Phi\sqrt{q/2}]$, 
and Ng, Nunes, and Rosati (2001) consider a model with 
$V(\Phi) \propto \Phi^A {\rm exp}[B\Phi^C]$, a simple example of a 
class of supergravity-inspired potentials studied by Copeland, Nunes,
and Rosati (2000). Here $A$, $B$, and $C$ are parameters.
Steinhardt {\em et al.} (1999) consider more 
complicated functions of inverse powers of $\Phi$, such as $V(\Phi) \propto
{\rm exp}(1/\Phi)$ and linear combinations of inverse powers of $\Phi$.
Brax and Martin (2000) consider a supergravity-inspired generalization with
$V(\Phi) = \kappa \Phi^{-\alpha} e^{\Phi^2/2}$.\footnote{
For still more examples see Green and Lidsey (2000), Barreiro,
Copeland, and Nunes (2000), Rubano and  Scudellaro (2002), Sen
and Sethi (2002), and references therein.} 

It will be quite a challenge to select from this wide range of
functional forms for the potential those that have particular
theoretical merit and some chance of being observationally
acceptable. 

Following Dolgov (1983), there have been discussions of
non-minimally coupled dark energy scalar field models.\footnote{
See Uzan (1999), Chiba (1999), Amendola (1999, 2000), Perrotta, Baccigalupi,
and Matarrese (2000), Bartolo and Pietroni (2000), Bertolami and Martins 
(2000), Fujii (2000), Faraoni (2000), Baccigalupi {\em et al.} (2000), 
Chen, Scherrer, and Steigman (2001), and references therein.}
These Jordan-Brans-Dicke type models have an explicit coupling
between the Ricci scalar and a function of the scalar field. This
causes the effective gravitational ``constant" $G$ (in units
where masses are constant) to vary with time,
which may be an observationally interesting effect and a useful 
constraint.\footnote{
Nucleosynthesis constraints on these and related models are discussed by
Arai, Hashimoto, and Fukui (1987), Etoh {\em et al.} (1997), Perrotta 
{\em et al.} (2000), Chen {\em et al.} (2001), and Yahiro {\em et al.}
(2002).} 

In yet another approach, people have considered modifying the
form of the dark energy scalar field kinetic and spatial
gradient term in the action.\footnote{
See Fujii and Nishioka
(1990), Chiba, Okabe, and Yamaguchi (2000), Armendariz-Picon,
Mukhanov, and Steinhardt (2001), Hebecker and Wetterich (2001),
and references therein.} 

We discussed in Sec.~III.E the idea that at the end of inflation the
dark energy scalar field potential might patch on to the part of
the scalar field potential responsible for inflation. In an
inverse power law model for the dark energy potential function
this requires an abrupt drop in $V(\Phi)$ at the end of
inflation. More sophisticated models, now dubbed quintessential
inflation, attempt to smooth out this 
drop by constructing scalar field potentials that interpolate
smoothly between the part responsible for inflation and the low
redshift, dark energy, part. These models assume either minimally
or non-minimally (Jordan-Brans-Dicke) coupled scalar
fields.\footnote{ 
Early work includes Frewin and Lidsey (1993), Spokoiny (1993), Joyce and
Prokopec (1998), and Peebles and Vilenkin (1999); more recent discussions
are given by Kaganovich (2001), Huey and Lidsey (2001), Majumdar (2001), 
Sahni, Sami, and Souradeep (2002), and Dimopoulos and Valle (2001).}  

The dark energy scalar field models we have reviewed here are meant to be
classical, effective, descriptions of what might come out of a more 
fundamental quantum mechanical theory. The effective dark energy scalar
field is coupled to itself and gravity, and is supposed to be
coupled to the other fields in the universe only by gravity. This
might be what Nature chooses, but we lack an understanding of why
the coupling of dark and ordinary fields that are allowed by the 
symmetries are not present, or have coupling strengths that are
well below what might be expected by naive dimensional analysis
(e.g., Kolda and Lyth, 1999). A satisfactory solution remains elusive. 
Perhaps this is not unexpected, because it likely requires a proper
understanding of how to reconcile general relativity and quantum
mechanics.

We turn now to scalar field dark energy models that arguably are
inspired by particle physics. Inverse power law scalar field
potentials are generated non-perturbatively 
in models of dynamical supersymmetry breaking.
In supersymmetric non-Abelian gauge theories, the resulting scalar 
field potential may be viewed as being generated by instantons, 
the potential being proportional to a power of
$e^{-1/g^2(\Phi)}$, where $g(\Phi)$ is the gauge coupling
constant which evolves logarithmically with the scalar 
field through the renormalization group equation. Depending on
the parameters  
of the model, an inverse power-law scalar field potential can result.
This mechanism may be embedded in supergravity and superstring/M
theory models.\footnote{
In the superstring/M theory case, since the coupling constant
is an exponential function of the dilaton scalar field, the
resulting potential is  usually
not of the inverse power-law form. However, it is perhaps not unreasonable
to think that after the dilaton has been stabilized, it or one of the
other scalar fields in superstring/M theory might be able to play the role 
of dark energy. Gasperini, Piazza, and Veneziano (2002) and    
Townsend (2001) consider other ways of using the dilaton as a dark energy 
scalar field candidate.}
This has not yet led to a model that might be compared to 
the observations.\footnote{
See, Davis, Dine, and Seiberg (1983), and Rossi and 
Veneziano (1984) for early discussions of supersymmetry breaking, and 
Quevedo (1996), Dine (1996), Peskin (1997), and Giudice and Rattazzi (1999)
for reviews. Applications of dynamical non-perturbative supersymmetry
breaking directly relevant to the dark energy scalar field model are
discussed in Bin\'etruy (1999), Masiero, Pietroni, and Rosati (2000), 
Copeland {\em et al.} (2000), Brax, Martin, and Riazuelo (2001),
de la Macorra and Stephan-Otto (2001), and references therein.}

In the model considered by Weiss (1987), Frieman {\em
et al.} (1995), and others the dark energy field potential is of
the form $V(\Phi) = M^4[\cos(\Phi/f) + 1]$, where $M$ and $f$ are
mass scales and the mass of the inhomogeneous scalar field fluctuation
$\sim M^2/f$ is on the order of the present value of the Hubble parameter. 
For discussions of how this model might be more firmly placed on a
particle physics foundation see Kim (2000), Choi (1999),
Nomura, Watari, and Yanagida (2000), and Barr and Seckel (2001).

There has been much recent interest in the idea of inflation in  
the brane scenario. Dvali and Tye (1999) note that the potential
of the scalar field which describes the relative separation
between branes can be of a form that leads to inflation, and will
include some inverse  power-law scalar field terms.\footnote{ 
Recent discussions of this setup include Halyo (2001b), Shiu and Tye (2001),
Burgess {\em et al.} (2002), Kyae and Shafi (2002), Garc\'{\i}a-Bellido,
Rabad\'an, and Zamora (2002), Blumenhagen {\em et al.} (2002), and Dasgupta
{\em et al.} (2002).} It will be interesting to learn whether
these considerations can lead
to a viable dark energy scalar field model. Brane models allow
for a number 
of other possibilities for dark energy scalar fields,\footnote{ 
See Uzawa and Soda (2001), Huey and Lidsey (2001), Majumdar (2001), Chen and
Lin (2002), Mizuno and Maeda (2001), Myung (2001), Steinhardt and
Turok (2002) and references therein.}
but it is too early to decide whether any of these options
give rise to observationally acceptable dark energy scalar field
models. 
 
Building on earlier work\footnote{
See Maldacena and Nu\~nez (2001), Bousso (2000), Banks and
Fischler (2001), and references therein.},
Hellerman, Kaloper, and Susskind (2001), and Fischler {\em et al.} (2001)
note that dark energy scalar field cosmological models have 
future event horizons characteristic of the de Sitter
model. This means some events have causal
futures that do not share any common events. In these dark energy scalar
field models, some correlations are therefore unmeasurable, which destroys
the observational meaning of the S-matrix. This indicates that it is not
straightforward to bring superstring/M theory into consistency with dark
energy models in which the expansion of the universe is
accelerating\footnote{
Other early discussions of this issue may be found in He (2001), Moffat
(2001), Deffayet, Dvali, and Gabadadze (2002), Halyo (2001a),
and Kolda and Lahneman (2001). The more recent literature may be 
accessed from Larsen, van der Schaar, and Leigh (2002), and Medved (2002).}.

At the time of writing, while there has been much work, it appears 
that the dark energy scalar field scenario still lacks a firm, high 
energy physics based, theoretical foundation. While this is a significant
drawback, the recent flurry of activity prompted by developments in 
superstring/M and brane theories appears to hold significant promise 
for shedding light on dark energy. Whether this happens before the 
observations rule out or ``confirm" dark energy is an intriguing
question.

\end{document}